\documentclass[fleqn,useAMS,usenatbib]{mn2e}

\usepackage[T1]{fontenc}
\usepackage{ae,aecompl}

\usepackage{graphicx}
\usepackage{amsmath}
\usepackage{amssymb}
\usepackage{amsfonts}
\usepackage{natbib}
\usepackage{graphicx}
\usepackage{epsfig}
\include{journaldefs}
\usepackage[caption=false]{subfig}
\usepackage{url}
\usepackage{textcomp}
\usepackage[squaren]{SIunits}
\usepackage{float}
\usepackage[T1]{fontenc} 
\usepackage{aecompl}
\usepackage{amsmath}
\usepackage{multicol}
\usepackage{amssymb}
\usepackage{savesym}
\usepackage{color}
\usepackage{tipa}

\newcommand{\mnras}{MNRAS}
\newcommand{\apjl}{ApJL}
\newcommand{\apj}{ApJ}
\newcommand{\apjs}{ApJS}

\newcommand{\nat}{Nature}
\newcommand{\aap}{A\&A}

\newcommand{\pasj}{PASJ}

\newcommand{\apss}{Ap\&SS}
\newcommand{\araa}{ARA\&A}
\newcommand{\pasp}{PASP}

\usepackage{multirow}
\usepackage{epic,eepic}
\usepackage{graphicx}
\usepackage{amssymb}
\usepackage{threeparttable}

\title[Eclipsing X-ray binaries in M\,51]
  {Discovery of two eclipsing X-ray binaries in M\,51}
\author[S.~Wang et al.]{Song Wang$^{1}$\thanks{E-mail: songw@bao.ac.cn},
  Roberto Soria$^{2,3,4}$\thanks{E-mail: roberto.soria@curtin.edu.au},
  Ryan Urquhart$^{3}$\thanks{E-mail: ryan.urquhart@icrar.org},
  and Jifeng Liu$^{1,2}$\thanks{E-mail: jfliu@bao.ac.cn}\\
$^{1}$Key Laboratory of Optical Astronomy, National Astronomical Observatories,
Chinese Academy of Sciences, Beijing 100012, China\\
$^{2}$College of Astronomy and Space Sciences, University of Chinese Academy of Sciences, Beijing 100049, China\\
$^{3}$International Centre for Radio Astronomy Research, Curtin University, GPO Box U1987, Perth, WA 6845, Australia\\
$^{4}$Sydney Institute for Astronomy, School of Physics A28, The University of Sydney, Sydney, NSW 2006, Australia}

\begin{document}




\maketitle
\begin{abstract}

We discovered eclipses and dips in two luminous (and highly variable) X-ray sources in M\,51. One (CXOM51 J132943.3$+$471135) is an ultraluminous supersoft source, with a thermal spectrum at a temperature of about 0.1 keV and characteristic blackbody radius of about $10^4$ km. The other (CXOM51 J132946.1$+$471042) has a two-component spectrum with additional thermal-plasma emission; it approached an X-ray luminosity of $10^{39}$erg s$^{-1}$ during outbursts in 2005 and 2012. From the timing of three eclipses in a series of {\it Chandra} observations, we determine the binary period ($52.75 \pm 0.63$ hr) and eclipse fraction ($22\% \pm 0.1\%$) of CXOM51 J132946.1$+$471042. We also identify a blue optical counterpart in archival {\it Hubble Space Telescope} images, consistent with a massive donor star (mass of $\sim$20--$35 M_{\odot}$). By combining the X-ray lightcurve parameters with the optical constraints on the donor star, we show that the mass ratio in the system must be $M_2/M_1 \ga 18$, and therefore the compact object is most likely a neutron star (exceeding its Eddington limit in outburst). The general significance of our result is that we illustrate one method (applicable to high-inclination sources) of identifying luminous neutron star X-ray binaries, in the absence of X-ray pulsations or phase-resolved optical spectroscopy. Finally, we discuss the different X-ray spectral appearance expected from super-Eddington neutron stars and black holes at high viewing angles.

\end{abstract}


\section{Introduction}
\label{intro.sec}

Ultraluminous X-ray sources (ULXs) are extra-nuclear, accreting compact objects with an observed luminosity in excess of $10^{39}$ erg s$^{-1}$,
which is the Eddington limit for a typical stellar-mass black hole (BH) with a mass of $\approx$10 $M_{\odot}$.
Hundreds of ULXs have been discovered in nearby galaxies
\citep{Liu2005, Swartz2011, Walton2011}
and all types of galaxies contain ULXs \citep{Mushotzky2006}.
Although the first examples of ULXs were already discovered by the {\it Einstein} satellite more than 30 years ago \citep{Long1983}, the nature of these sources remains an unsolved fundamental question.
The most popular explanation for the majority of ULXs is that they are the high-luminosity end of the X-ray binary (XRB) population \citep{Gladstone2009, Feng2011}. They may include: neutron stars (NSs) accreting at highly super-Eddington rates onto a magnetized surface \citep{Bachetti2014,Furst2016,Israel2017a}; ordinary stellar-mass BHs ($M_{\rm{BH}} \la 20~M_{\odot}$) accreting at super-Eddington rates; more massive BHs ($20 \la M_{BH} \la 80~M_{\odot}$) formed from the collapse of metal-poor stars, accreting around their Eddington limit; in a few rare cases \citep{Farrell2009,Zolotukhin2016}, intermediate-mass black holes (IMBHs), with $M \sim 10^3$--$10^4~M_{\odot}$, accreting below their Eddington limit.

In parallel with the uncertainty on the nature of the compact objects in ULXs, the physical interpretation of their phenomenological X-ray spectral states remains unclear \citep{Soria2007, Gladstone2009, Sutton2013b, Urquhart2016a,Pintore2017,Kaaret2017}. ULX spectra are often classified into three empirical regimes \citep{Sutton2013b}, characterized by either a single-component curved spectrum (broadened disk regime), or a two-component spectrum peaking in the soft band, below 1 keV (soft ultraluminous regime), or in the hard band, around 5 keV (hard ultraluminous regime). It is not clear how those empirical regimes quantitatively depend on the nature of the compact object, the accretion rate, and/or the viewing angle. The most plausible scenario is that in super-critical accretion, a massive radiatively driven disk outflow forms a lower-density polar funnel around the central regions. In this scenario, those three different regimes correspond to different amounts of scattering and absorption of the X-ray photons along our line of sight, function of (mass-scaled) accretion rate and viewing angle; softer X-ray emission is mostly emerging through the down-scattering wind, while harder X-ray emission from the innermost regions can only be directly seen for relatively low (face-on) viewing angles, as we look into the funnel.

A fourth spectral class has recently been added to the three ULX regimes mentioned above: that of ultraluminous supersoft sources (ULSs), characterized by thermal spectra with $kT \approx 50$--100 eV and little or no emission above 1 keV \citep{Kong2003,Kong2005}. Although their observed X-ray luminosity barely reaches the ULX threshold at $L_{\rm X} \approx 10^{39}$ erg s$^{-1}$, their accretion rate may be highly super-Eddington, and their ultrasoft thermal spectra may result from reprocessing of the emitted photons in an optically thick wind \citep{Urquhart2016a,Soria2016,Shen2015,Poutanen2007}. ULSs had previously been interpreted as IMBHs in the sub-Eddington high/soft state, or as white dwarfs with nuclear burning of accreted materials on their surfaces (analogous to Galactic super-soft sources). However, serious difficulties with both the IMBH scenario and the white dwarf scenario were highlighted by \cite{Urquhart2016a} and \cite{Liu2008b}, respectively.

The main reason for the continuing uncertainty in the ULX nature, accretion flow geometry, and spectral state classification is the scarcity of reliable measurements of mass and viewing angle. Dynamical measurements of BH masses in ULXs with  phase-resolved optical spectroscopy are particularly challenging, given the faintness of their optical counterparts in external galaxies. Moreover, some of the line emission may come from a wind, in which case the observed velocity shifts would not be reliable for dynamical measurements. Independent measurements of the viewing angle ({\it {i.e.}}, not based on spectral appearance) are also not available in most cases.

To make progress on those two issues, we searched for eclipsing ULXs and/or ULSs in nearby galaxies. Eclipsing XRBs give us two advantages. Firstly, we know by default that their orbital plane is seen nearly edge-on from our line of sight. Secondly, mass measurements are made relatively simpler by the presence of eclipses. It is obvious that the projected orbital velocities of the two components in an edge-on binary system, and therefore the Doppler shifts of their emission and absorption lines, are higher than in systems seen face-on, and therefore it is easier to determine a mass function from phase-resolved spectroscopy; this led to spectroscopic mass measurements for example in M\,33 X-7 \citep{Pietsch2006, Orosz2007} and M\,101 X-1 \citep{Liu2013}. More importantly, even in the absence of phase-resolved optical spectroscopic data (which is the case for almost all ULXs/ULSs), the mass ratio and inclination are constrained by the observed eclipse fraction ({\it {i.e.}}, the relative fraction of time spent in eclipse); thus, if the mass of the donor star is also known or constrained, we can constrain the mass of the compact object from photometry alone, as we discuss in this paper.


Here, we report on our search for eclipsing X-ray sources in the grand-design spiral galaxy M\,51, located at a distance of 8.0 $\pm$ 0.6 Mpc \citep{Bose2014}.
We have already found \citep{Urquhart2016b} two eclipsing ULXs in M\,51 and have noted the small probability to find two luminous, eclipsing sources so close to each other in the same {\it Chandra}/ACIS  field of view.
In this paper, we illustrate the discovery of two new eclipsing binaries in the same {\it Chandra} field.
The two sources were catalogued as CXOM51\,J132943.3$+$471135 (hereafter S1) and CXOM51\,J132946.1$+$471042 (hereafter S2) in \citet{Terashima2004}; based on the full stacked dataset of {\it Chandra} observations,
the most accurate positions for the two sources are R.A. $=$ 13$^{h}$29$^{m}$43$^{s}$.32 and Dec. $=$ $+47^{\circ}11'34".9$ for S1, and
R.A. $=$ 13$^{h}$29$^{m}$46$^{s}$.13 and Dec. $=$ $+47^{\circ}10'42".3$ for S2 \citep{Wang2016}.
We used archival data from {\it Chandra} and {\it XMM-Newton} to analyze their X-ray timing and spectral properties. For one of the two sources (S2), we determine the orbital period, identify a candidate optical counterpart, and show that the compact object is most likely a neutron star, exceeding its Eddington limit in outburst.

\begin{table*}
\scriptsize
\begin{center}
\caption[]{Key parameters for each {\it Chandra} and {\it XMM-Newton} observation of S1 and S2.}
\label{Xdata.tab}
\begin{tabular}{llccccccccc}
\hline\noalign{\smallskip}
ObsID  & Detector & Exp.~Time  &    MJD   &  OAA                & VigF    &   Net Cts  &  Bkg Cts  &  ${M-S \over H+M+S}$  &  ${H-M \over H+M+S}$  & Class \\
	   &          & (s)   &          & ($^{\prime\prime}$) &         &           &        &                       &                       &     \\
(1)    &   (2)    &    (3)  &    (4)   &   (5)               & (6)     &     (7)   &    (8) & (9)                   & (10)                  & (11)\\
\hline\noalign{\tiny}\\[-5pt]
\multicolumn{11}{c}{S1}\\[3pt]
\hline\noalign{\tiny}
414    & ACIS-S3 &    1137.4  & 51566.259  &  419  & 0.750 &   24.7(5.4)   &   0.5  & -0.62(0.32)   & -0.12(0.70)  & SS \\
354    & ACIS-S3 &   14862.2  & 51715.336  &   69  & 0.995 &  268.0(16.5)  &   0.4  & -0.92(0.04)   & -0.04(0.41)  & SS  \\
1622   & ACIS-S3 &   26808.1  & 52083.783  &   71  & 0.996 &  365.0(19.2)  &   0.7  & -0.88(0.04)   & -0.06(0.13)  & SS  \\
3932   & ACIS-S3 &   47969.8  & 52858.605  &  153  & 0.980 &  982.0(31.6)  &   2.5  & -0.74(0.03)   & -0.13(0.06)  & SS  \\
12562  & ACIS-S3 & 9630.1      & 55724.286 & 246   & 0.674 &  4.5(4.2)      & -      & -             & -            &  d   \\
13813  & ACIS-S3 &  179196.4  & 56179.742  &  201  & 0.954 & 1095.0(33.8)  &   8.7  & -0.82(0.03)   & -0.09(0.07)  & SS  \\
13812  & ACIS-S3 &  157457.0  & 56182.767  &  202  & 0.954 & 1230.0(35.7)  &   6.4  & -0.73(0.03)   & -0.12(0.08)  & SS  \\
15496  & ACIS-S3 &   40965.5  & 56189.390  &  216  & 0.963 &  322.0(18.4)  &   2.3  & -0.80(0.05)   & -0.10(0.11)  & SS  \\
13814  & ACIS-S3 &  189849.8  & 56190.308  &  216  & 0.962 & 2334.0(49.4)  &  11.8  & -0.69(0.02)   & -0.15(0.04)  & SS  \\
13815  & ACIS-S3 &   67183.5  & 56193.343  &  221  & 0.961 &  692.0(26.8)  &   3.3  & -0.74(0.04)   & -0.13(0.05)  & SS  \\
13816  & ACIS-S3 &   73103.6  & 56196.217  &  220  & 0.963 &   24.4(6.4)   &  17.9  & -0.55(0.42)   & -0.52(0.50)  & SS  \\
15553  & ACIS-S3 &   37573.4  & 56210.031  &  235  & 0.966 &   19.3(5.3)   &   2.3  & -0.95(0.21)   &  0.22(0.50)  & QS  \\
0112840201 & EPIC/MOS1 & 20457.1 & 52654.553 & 94 & 0.935 & 192.0(15.6) & 17.0 & -0.88(0.05) & -0.08(0.03) & SS \\
0112840201 & EPIC/MOS2 & 20474.9 & 52654.553 & 94 & 0.733 & 152.0(14.5) & 24.0 & -0.78(0.07) & -0.11(0.04) & SS \\
0212480801 & EPIC/MOS2 & 48427.2 & 53552.276 & 113 & 0.955 & 491.3(26.6) & 130.8 & -0.77(0.04) & -0.1(0.03) & SS \\
0212480801 & EPIC/pn & 41252.6 & 53552.291 & 113 & 0.985 & 1932.3(55.9) & 861.8 & -0.9(0.02) & -0.04(0.02) & SS \\
0303420101 & EPIC/MOS1 & 49299.0 & 53875.274 & 107 & 0.987 & 23.0(20.3) & 232.0 & 0.13(0.68) & -0.3(0.9) & QS \\
0303420101 & EPIC/MOS2 & 48738.7 & 53875.274 & 107 & 0.967 & 39.5(19.3) & 196.5 & -0.59(0.66) & -0.52(0.78) & QS \\
0303420101 & EPIC/PN & 41798.1 & 53875.289 & 107 & 0.941 & 108.0(39.7) & 926.0 & -0.83(0.37) & -0.13(0.37) & SS \\
0303420201 & EPIC/MOS1 & 36063.7 & 53879.469 & 107 & 0.983 & 254.3(19.9) & 81.8 & -0.83(0.06) & -0.08(0.04) & SS \\
0303420201 & EPIC/MOS2 & 36075.4 & 53879.469 & 107 & 0.967 & 298.3(20.8) & 75.8 & -0.81(0.05) & -0.06(0.04) & SS \\
0303420201 & EPIC/pn & 30674.2 & 53879.484 & 107 & 0.92 & 1094.5(40.0) & 342.5 & -0.9(0.02) & -0.06(0.02) & SS \\
0677980701 & EPIC/MOS1 & 10641.2 & 55719.207 & 237 & 0.654 & 20.3(7.5) & 16.8 & -0.31(0.36) & -0.31(0.31) & QS \\
0677980701 & EPIC/MOS2 & 10657.2 & 55719.207 & 237 & 0.896 & 22.8(7.3) & 13.3 & -0.75(0.32) & -0.01(0.23) & SS \\
0677980701 & EPIC/pn & 9796.9 & 55719.223 & 237 & 0.862 & 111.5(14.7) & 59.5 & -1.08(0.1) & -0.0(0.06) & SS \\
0677980801 & EPIC/MOS1 & 2493.3 & 55723.197 & 237 & 0.903 & $< 9.2$ & - & - & - & u \\
0677980801 & EPIC/MOS2 & 2450.9 & 55723.197 & 237 & 0.9 & $< 12.7$ & - & - & - & u \\
0677980801 & EPIC/pn & 8689.8 & 55723.212 & 237 & 0.852 & 45.5(23.1) & 353.5 & -0.6(0.43) & -0.6(0.67) & QS \\
\hline\noalign{\tiny}\\[-5pt]
\multicolumn{11}{c}{S2}\\[3pt]
\hline\noalign{\tiny}
354    & ACIS-S3 &    14862.2 & 51715.336  &   60  & 0.994 &   29.9(5.6)    &   0.3  & -0.30(0.34)   & -0.20(0.54)  &  H \\
1622   & ACIS-S3 & 26808.1    & 52083.783  &   61  & 0.994 &   $< 8.1$       & -      & -             & -             &  u  \\
3932   & ACIS-S3 &    47969.8 & 52858.605  &  113  & 0.901 &   11.5(3.7)    &   2.3  & -0.41(0.74)   &  0.03(0.96)  &  H \\
12562  & ACIS-S3 & 9630.1     & 55724.286  &  201  & 0.889 & $< 2.7$       & -      & -             & -             &  u  \\
12668  & ACIS-S3 & 9989.9     & 55745.439  &  206  & 0.861 & $< 3.0$       & -      & -             & -             &  u  \\
13813  & ACIS-S3 &   179196.4 & 56179.742  &  167  & 0.959 & 1484.0(39.1)   &   5.7  &  0.14(0.04)   & -0.29(0.04)  & H \\
13812  & ACIS-S3 &   157457.0 & 56182.767  &  167  & 0.960 & 1632.0(40.9)   &   6.1  &  0.13(0.04)   & -0.29(0.04)  & H \\
15496  & ACIS-S3 &    40965.5 & 56189.390  &  184  & 0.854 &  341.0(18.7)   &   1.6  &  0.12(0.09)   & -0.29(0.09)  & H \\
13814  & ACIS-S3 &   189849.8 & 56190.308  &  184  & 0.800 &  826.0(29.4)   &   6.7  &  0.07(0.06)   & -0.26(0.06)  & H \\
13815  & ACIS-S3 &    67183.5 & 56193.343  &  190  & 0.778 &  175.0(13.6)   &   4.4  &  0.03(0.14)   & -0.21(0.15)  & H \\
13816  & ACIS-S3 &    73103.6 & 56196.217  &  190  & 0.772 &   42.9(6.9)    &   2.6  &  0.09(0.29)   & -0.44(0.28)  &  QS  \\
15553  & ACIS-S3 & 37573.4    & 56210.031  &  207  & 0.966 & 7.5(5.7)       & -      & -             & -             &  d  \\
0112840201 & EPIC/MOS2 & 20474.9 & 52654.553 & 95 & 0.939 & 15.8(6.9) & 14.3 & -0.57(0.43) & -0.43(0.32) & SS \\
0112840201 & EPIC/pn & 17120.7 & 52654.568 & 95 & 0.915 & 47.0(11.8) & 55.0 & -0.46(0.21) & -0.32(0.15) & QS \\
0212480801 & EPIC/MOS1 & 48366.7 & 53552.276 & 68 & 0.996 & 289.5(22.6) & 141.5 & -0.17(0.07) & -0.27(0.06) & H \\
0212480801 & EPIC/MOS2 & 48427.2 & 53552.276 & 68 & 0.982 & 373.5(23.8) & 115.5 & -0.16(0.06) & -0.18(0.05) & H \\
0212480801 & EPIC/pn & 41252.6 & 53552.291 & 68 & 0.999 & 1037.5(45.5) & 753.5 & -0.4(0.04) & -0.14(0.03) & H \\
0303420101 & EPIC/MOS1 & 49299.0 & 53875.274 & 60 & 0.992 & 77.5(18.4) & 179.5 & -0.47(0.2) & -0.31(0.23) & QS \\
0303420101 & EPIC/pn & 41798.1 & 53875.289 & 60 & 0.95 & 196.8(36.4) & 844.3 & -0.67(0.16) & -0.07(0.17) & SS \\
0303420201 & EPIC/MOS1 & 36063.7 & 53879.469 & 60 & 0.997 & 26.0(11.6) & 69.0 & -0.79(0.44) & 0.22(0.35) & SS \\
0303420201 & EPIC/pn & 30674.2 & 53879.484 & 60 & 0.93 & 59.0(22.9) & 336.0 & -1.19(0.52) & -0.1(0.4) & SS \\
0677980701 & EPIC/MOS1 & 10641.2 & 55719.207 & 196 & 0.928 & 7.0(5.9) & - & - & - & d \\
0677980701 & EPIC/MOS2 & 10657.2 & 55719.207 & 196 & 0.927 & 13.3(6.5) & 12.8 & -0.49(0.48) & -0.17(0.4) & QS \\
0677980801 & EPIC/MOS1 & 2493.3 & 55723.197 & 196 & 0.93 & $< 6.4$ & -  &  - &  - &  u \\
0677980801 & EPIC/MOS2 & 2450.9 & 55723.197 & 196 & 0.94 & $< 5.2$ &  - &  - &  - &  u \\
0677980801 & EPIC/pn & 8689.8 & 55723.212 & 196 & 0.875 &  $< 38.0$ &  - &  - &  - &  u \\
\hline\noalign{\smallskip}
\end{tabular}
\end{center}
\begin{flushleft}
{The columns are:
(1) observation ID;
(2) instrument and detector.
(3) source exposure time after deadtime correction;
(4) observation date;
(5) off-axis angle in arcseconds;
(6) vignetting factor;
(7) background-subtracted photon counts with their uncertainty in brackets;
(8) background counts within the source region;
(9) X-ray color index $C_{MS} = (M-S)/(H+M+S)$ with its uncertainty in brackets;
(10) X-ray color index $C_{HM} = (H-M)/(H+M+S)$ with its uncertainty in brackets.
(11) Spectral hardness classification: SS  = supersoft;
      QS  = quasi-soft; H = hard; d = dim ($0 < C_N - C_E \leq 10$); u = not significantly detected ($C_N - C_E \leq 0$).}
\end{flushleft}
\end{table*}

\section{X-ray data analysis}
\label{data.sec}

\begin{figure}
\center
\includegraphics[width=0.47\textwidth]{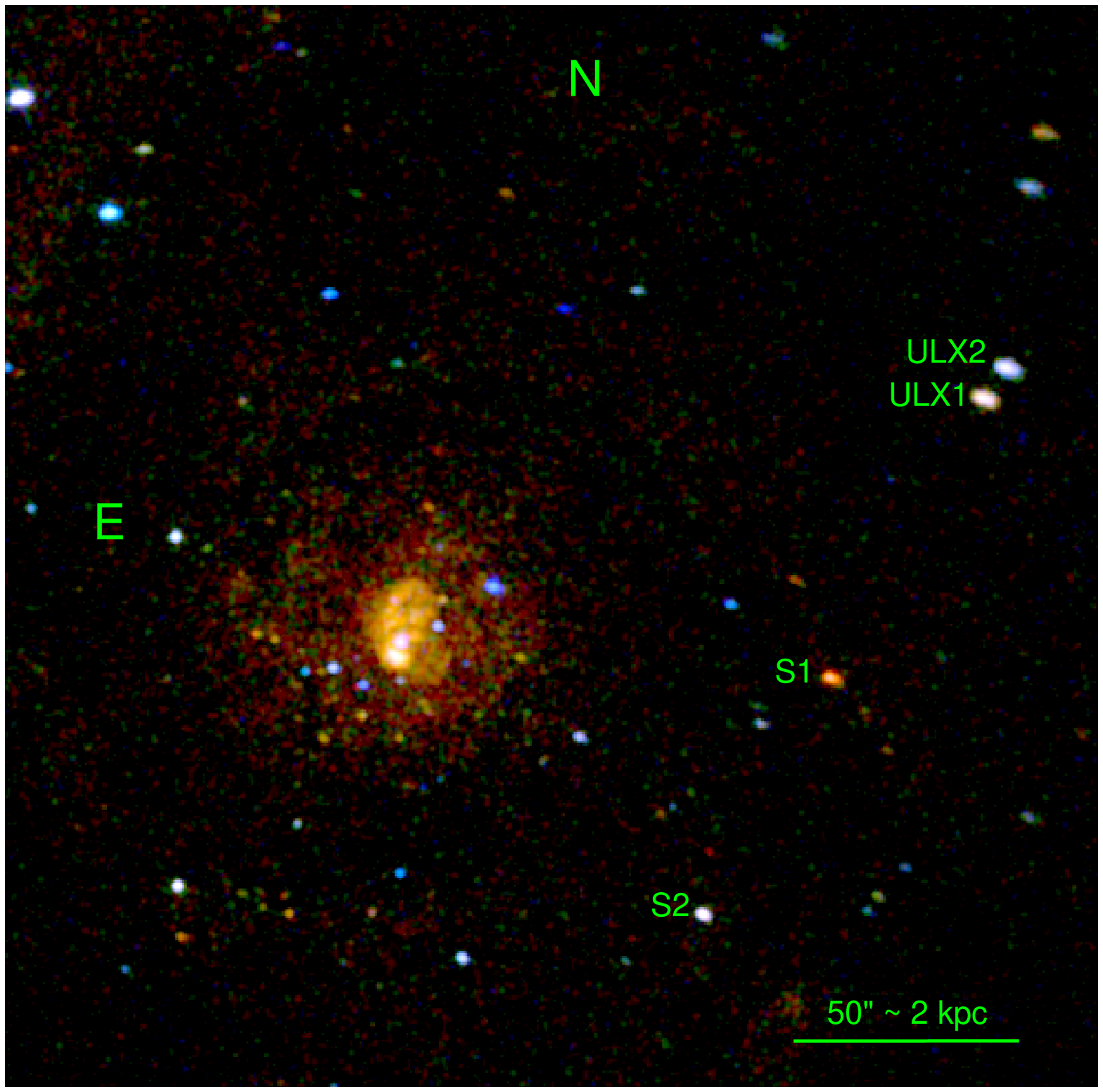}
\caption{Combined {\it Chandra}/ACIS true-color map of M\,51,
with red = 0.3--1 keV, green = 1--2 keV, and blue = 2--8 keV.
The two new eclipsing sources are labelled S1 and S2, while ULX1 and ULX2 are the eclipsing sources discussed in \citet{Urquhart2016b}.}
\label{chandra.fig}
\end{figure}

M\,51 has been observed many times by the {\it Chandra}/ACIS and {\it XMM-Newton}/EPIC detectors.
We listed (Table \ref{Xdata.tab}) the observations that cover the location of S1 and S2, including
Obsid, instrument, exposure time, observation date, off-axis angle, vignetting factor. There are twelve {\it Chandra} observations between 2000 June and 2012 October, and six {\it XMM-Newton} observations between 2003 January and 2011 June.
The two sources are displayed and labelled in Figure \ref{chandra.fig},
together with the two eclipsing ULXs studied by \citet{Urquhart2016b}.

The {\it Chandra} data were downloaded from the public archive and reprocessed with the
{\it Chandra} Interactive Analysis of Observations software ({\small{CIAO}}), version 4.6 \citep{Fruscione2006}. We used the {\it wavdetect} tool \citep{Freeman2002}
to determine whether S1 and S2 are significantly detected in each of the {\it Chandra}/ACIS images,
measure the position of their centroids, and define their $3\sigma$ elliptical source regions for subsequent extraction of lightcurves and spectra.
For both sources, local background regions were taken as elliptical annuli around the source regions, with the length of the inner and outer major axes fixed at 2 and 4 times the length of the major axes of the source ellipses.
Using {\it mkexpmap} task, we created an exposure map and calculated the vignetting factor for each observation, that is the ratio between the local effective and the nominal exposure time.
For our timing analysis, we first applied {\it axbary} to all observations for the barycenter correction, and then used {\it dmextract} to produce light-curves.
Background-subtracted source spectra and their corresponding response matrices were created with {\it specextract}.

We processed the {\it XMM-Newton}/EPIC data with the Science Analysis System (SAS), version 15.0.0.
A circular region of 13$''$ radius was used as the source region,
and a neighboring circular region of 26$''$ radius was used to estimate the local background.
The event patterns were selected in the 0 to 12 range for the MOS detectors
and in the 0 to 4 range for the pn detector.
The flagging criteria \#XMMEA\_EM and \#XMMEA\_EP were also applied for the MOS and pn detectors respectively,
together with a ``FLAG=0'' filter for the pn.
To account for vignetting, we created an exposure map with {\it evselect} and {\it eexpmap}.
For our timing study, we applied the barycenter correction with {\it barycen},
and extracted the light-curves with {\it evselect} and {\it epiclccorr}.
We created background-subtracted MOS and pn spectra and their corresponding response files, using {\it evselect}, {\it backscale}, {\it rmfgen}, and {\it arfgen}.
To create a weighted-average EPIC spectrum, we combined
the MOS1, MOS2, and pn spectra of each observation with {\it epicspeccombine}.
Finally, we grouped the spectra with {\it specgroup} so that the number of channels does not oversample the spectral resolution.

For both sources, we used {\sc xspec} \citep{Arnaud1996} version 12.09.0 to perform spectral fitting to the data from each observation with more than 200 counts.

\section{Long-term X-ray monitoring}
\label{longterm.sec}

For both sources in each observation (excluding observations when either source unfortunately fell onto a chip gap), we computed the net counts $C_N$ from aperture photometry: $C_N = C_s - C_b\times A_s/A_b$, where $C_s$ are the raw counts, $C_b$ the background counts,
$A_s$ the source region area, and $A_b$ the background region area. We also computed the corresponding net count errors $C_E$, defined as $C_E = 1 + \sqrt{0.75 + C_s + C_b \times (A_s/A_b)^2}$ \citep{Gehrels1986}.
In observations where a source is not detected ($C_N -C_E \leq$ 0), we report the upper limit ($C_N +C_E$) to their net counts.
We defined X-ray colours for both {\it Chandra} and {\it XMM-Newton} data using the standard soft band $S$ from 0.3 keV to 1.0 keV, medium band $M$ from 1.0 keV to 2.0 keV, and hard band $H$ from 2.0 keV to 8.0 keV \citep{Prestwich2003}. We applied the hierarchical classification of \cite{Di Stefano2003a, Di Stefano2003b} to classify the observed X-ray colours in the various epochs as supersoft, quasi-soft, or hard.
Table \ref{Xdata.tab} lists the background-subtracted photon counts with error, expected background counts in the source region, X-ray colors, and source classification.

\begin{figure}
\center
\hspace*{-20pt}
\includegraphics[width=0.42\textwidth]{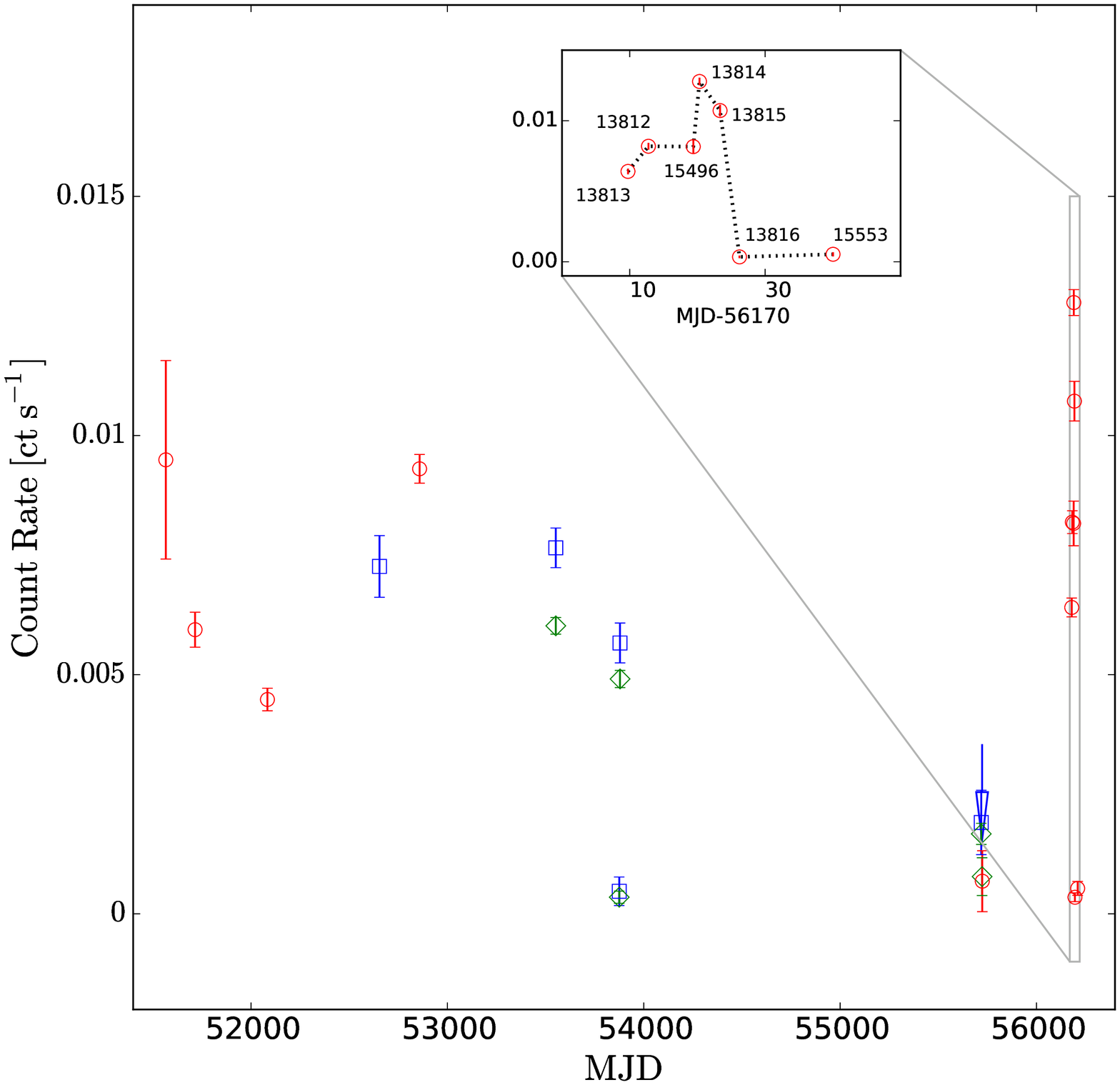}\\
\includegraphics[width=0.48\textwidth]{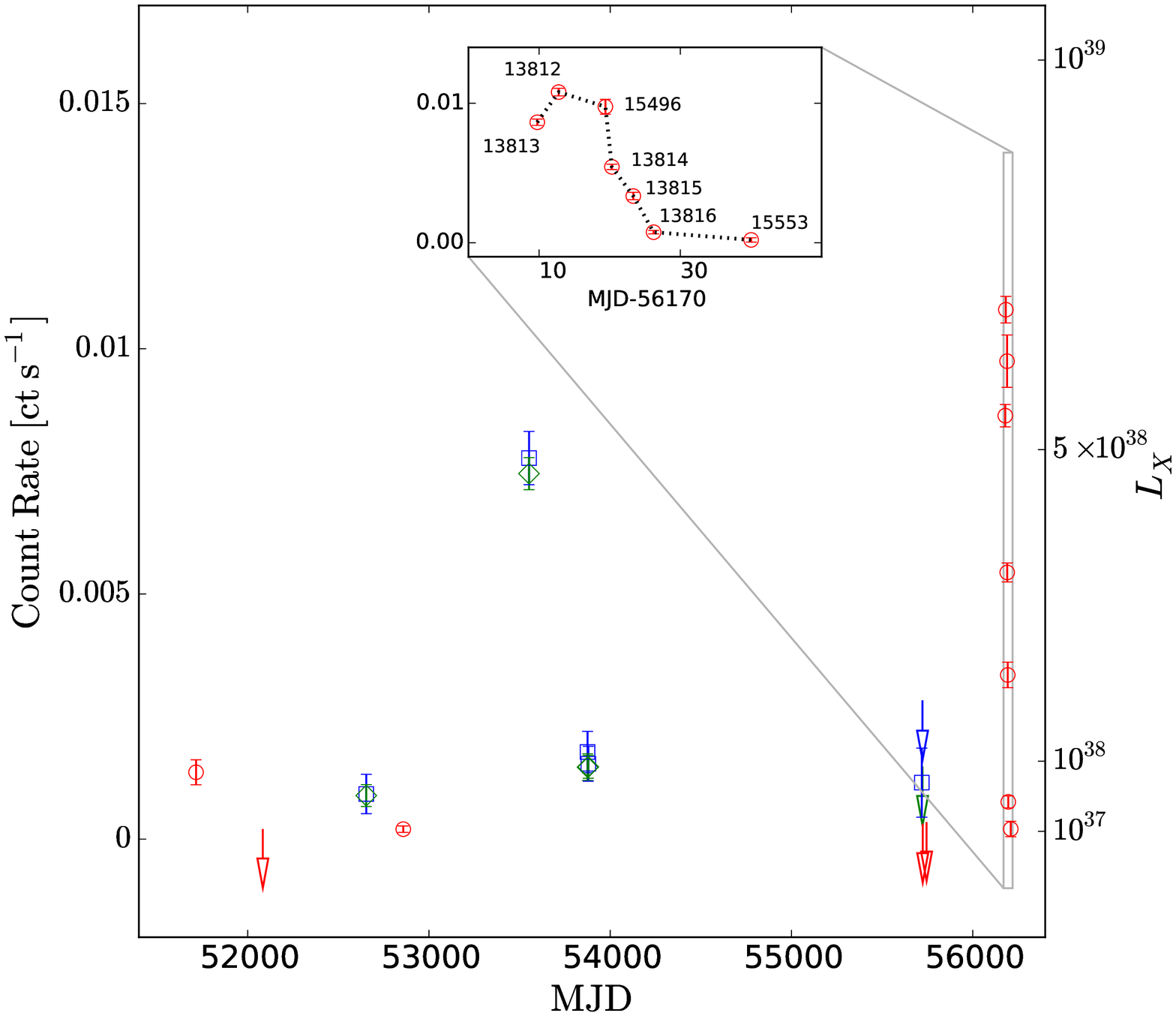}\\
\caption{
Top panel: long-term X-ray light-curve of S1. Red symbols are datapoints from {\it Chandra}/ACIS observations, blue and green symbols represent {\it XMM-Newton}/EPIC MOS and pn datapoints, respectively; arrows indicate upper limits. We converted the observed count rates to an equivalent {\it Chandra}/ACIS-S Cycle-13 count rate, using PIMMS with a blackbody spectrum at $kT = 0.1$ keV and Galactic line-of-sight absorption ($n_{\rm H} \approx 2\times 10^{-20}~{\rm cm}^{-2}$).
Bottom panel: long-term X-ray light-curve of S2. Symbols are defined as in the top panel. We converted all count rates to an equivalent {\it Chandra}/ACIS-S Cycle 13 count rate, using PIMMS with a power-law model (photon index $\Gamma = 1.7$) and Galactic line-of-sight absorption ($n_{\rm H} \approx 2\times 10^{-20}~{\rm cm}^{-2}$).
The X-ray luminosities for S2 are calculated using $L_{\rm X} \equiv 4\pi \, d^2 \, f^{\rm em}_X$, with the de-absorbed fluxes $f^{\rm em}_X$ estimated from count rates and line-of-sight Galactic absorption with PIMMS.
}
\label{fullseries.fig}
\end{figure}

S1 shows strong long-term variability in its observed count rate over the almost 13 years of sporadic monitoring (Figure \ref{fullseries.fig}, and Table \ref{XlumS2.tab}), including a couple of epochs when it is below the detection limit. 
We will argue (Section~\ref{xspec.sec}) that the apparent luminosity variability is due to changes in the optical depth of the thick outflow surrounding the compact object, rather than changes in the accretion rate or in the geometry of the system. Changes in the radius and temperature of the outflow photosphere have been invoked to explain the spectral properties and evolution of ULSs \citep{Soria2016,Urquhart2016a}, and other sources where the primary X-ray photons are reprocessed in an optically thick, variable wind \citep{Shidatsu2016}.

S2 behaves like a standard XRB, with a broadband X-ray spectrum; therefore, we can use its net count rates as a rough proxy for X-ray luminosity, before doing any detailed spectral modelling. Assuming a power-law spectrum with photon index $\Gamma = 1.7$ and line-of-sight Galactic absorption, we find that in most of the observations, its luminosity (Figure \ref{fullseries.fig} and Table \ref{XlumS2.tab}) hovers at $L_{\rm X} \approx$0.5--1 $\times 10^{38}$ erg s$^{-1}$; it reached $L_{\rm X} \approx 5$--7  $\times 10^{38}$ erg s$^{-1}$ during the 2005 and 2012 outbursts. The long-term average as well as the peak luminosity suggest disk accretion via Roche-lobe overflow, rather than wind accretion. Instead, a non-detection in the {\it Chandra} ObsID 1622 (2001 June 23) means that it must have been fainter than $L_{\rm X} \approx 10^{37}$ erg s$^{-1}$ at that time. There are several physical mechanisms for XRB outbursts, depending on the nature of the primary and secondary stars and on the system parameters. Therefore, we need to determine such parameters before we can favour any scenario.

\begin{table}
  \renewcommand\arraystretch{0.9}
\begin{center}
\caption[]{Count rates and X-ray luminosities for S1 and S2.}
\label{XlumS2.tab}
\begin{tabular}{llcc}
\hline
ObsID  & Detector & Count Rate$^{a}$  &  $L_X$ [0.3--8 keV]$^{b}$  \\
	   &          & (ks$^{-1}$)        &   (10$^{39}$ erg s$^{-1}$)        \\
\hline
\multicolumn{4}{c}{S1}\\[0pt]
\hline
414  & ACIS-S3 &       9.49$\pm$      2.08 & -\\
354  & ACIS-S3 &       5.94$\pm$      0.37 & -\\
1622  & ACIS-S3 &       4.48$\pm$      0.24 & -\\
0112840201  & EPIC/MOS &       7.26$\pm$      0.64 & -\\
3932  & ACIS-S3 &       9.30$\pm$      0.30 & -\\
0212480801  & EPIC/MOS &       7.65$\pm$      0.41 & -\\
0212480801  & EPIC/pn &       6.02$\pm$      0.17 & -\\
0303420101  & EPIC/MOS &       0.47$\pm$      0.30 & -\\
0303420101  & EPIC/pn &       0.35$\pm$      0.13 & -\\
0303420201  & EPIC/MOS &       5.66$\pm$      0.42 & -\\
0303420201  & EPIC/pn &       4.91$\pm$      0.18 & -\\
0677980701  & EPIC/MOS &       1.91$\pm$      0.67 & -\\
0677980701  & EPIC/pn &       1.67$\pm$      0.22 & -\\
0677980801  & EPIC/MOS & $<$      3.55 & -\\
0677980801  & EPIC/pn &       0.78$\pm$      0.39 & -\\
12562  & ACIS-S3 &       0.68$\pm$      0.64 & -\\
13813  & ACIS-S3 &       6.41$\pm$      0.20 & -\\
13812  & ACIS-S3 &       8.19$\pm$      0.24 & -\\
15496  & ACIS-S3 &       8.16$\pm$      0.47 & -\\
13814  & ACIS-S3 &      12.78$\pm$      0.27 & -\\
13815  & ACIS-S3 &      10.72$\pm$      0.42 & -\\
13816  & ACIS-S3 &       0.35$\pm$      0.09 & -\\
15553  & ACIS-S3 &       0.53$\pm$      0.15 & -\\
\hline
\multicolumn{4}{c}{S2}\\[0pt]
\hline
354  & ACIS-S3 &       1.36$\pm$      0.26 &       0.09$\pm$      0.02\\
1622  & ACIS-S3 & $<$      0.20 & $<$      0.01\\
0112840201  & EPIC/MOS &       0.92$\pm$      0.40 &       0.06$\pm$      0.03\\
0112840201  & EPIC/pn &       0.89$\pm$      0.22 &       0.06$\pm$      0.01\\
3932  & ACIS-S3 &       0.20$\pm$      0.07 &      0.013$\pm$     0.004\\
0212480801  & EPIC/MOS &       7.77$\pm$      0.54 &       0.49$\pm$      0.03\\
0212480801  & EPIC/pn &       7.45$\pm$      0.33 &       0.47$\pm$      0.02\\
0303420101  & EPIC/MOS &       1.78$\pm$      0.42 &       0.11$\pm$      0.03\\
0303420101  & EPIC/pn &       1.47$\pm$      0.27 &       0.09$\pm$      0.02\\
0303420201  & EPIC/MOS &       0.81$\pm$      0.36 &       0.05$\pm$      0.02\\
0303420201  & EPIC/pn &       0.61$\pm$      0.24 &       0.04$\pm$      0.01\\
0677980701  & EPIC/MOS &       1.15$\pm$      0.70 &       0.07$\pm$      0.04\\
0677980801  & EPIC/MOS & $<$      2.81 & $<$      0.18\\
0677980801  & EPIC/pn & $<$      1.48 & $<$      0.09\\
12562  & ACIS-S3 & $<$      0.31 & $<$      0.02\\
12668  & ACIS-S3 & $<$      0.35 & $<$      0.02\\
13813  & ACIS-S3 &       8.64$\pm$      0.23 &       0.54$\pm$      0.01\\
13812  & ACIS-S3 &      10.80$\pm$      0.27 &       0.68$\pm$      0.02\\
15496  & ACIS-S3 &       9.75$\pm$      0.53 &       0.61$\pm$      0.03\\
13814  & ACIS-S3 &       5.44$\pm$      0.19 &       0.34$\pm$      0.01\\
13815  & ACIS-S3 &       3.35$\pm$      0.26 &       0.21$\pm$      0.02\\
13816  & ACIS-S3 &       0.76$\pm$      0.12 &       0.05$\pm$      0.01\\
15553  & ACIS-S3 &       0.21$\pm$      0.16 &      0.013$\pm$     0.010\\
\hline
\end{tabular}
\end{center}
\begin{flushleft}
$^a$For S1: we converted the observed count rates to an equivalent {\it Chandra}/ACIS-S Cycle-13 count rate, using PIMMS with a blackbody spectrum at $kT = 0.1$ keV and Galactic line-of-sight absorption ($n_{\rm H} \approx 2\times 10^{-20}~{\rm cm}^{-2}$). For S2: we converted to an equivalent Cycle-13 count rate using a power-law model with photon index $\Gamma = 1.7$ and line-of-sight absorption.\\
$^b$For S1: we did not convert count rates to a luminosity, because at such low temperatures, the conversion is strongly model dependent and would produce spurious results. For S2: we converted count rates to de-absorbed fluxes $f^{\rm em}_{\rm X}$, using PIMMS with a power-law model ($\Gamma = 1.7$) and line-of-sight absorption; then, we converted the de-absorbed fluxes to luminosities with the relation $L_{\rm X} \equiv 4\pi \, d^2 \, f^{\rm em}_X$, at the assumed distance $d = 8.0$ Mpc.
\end{flushleft}
\end{table}

\section{X-ray Timing Results}
\label{xtime.sec}

\subsection{Eclipse and dips for S1}
\label{xtimeS1.sec}

Source S1 was discussed by \cite{Urquhart2016a} in the context of its supersoft spectrum, and high-inclination viewing angle. They noted the presence of a deep dip in the light-curve from {\it Chandra} ObsID 13814, and of a very low flux state at the beginning of {\it Chandra} ObsID 13815. The light-curves of other X-ray sources in the same observation and same ACIS chip do not display similar variability: this rules out instrumental problems.

\begin{figure}
\center
\includegraphics[width=0.40\textwidth]{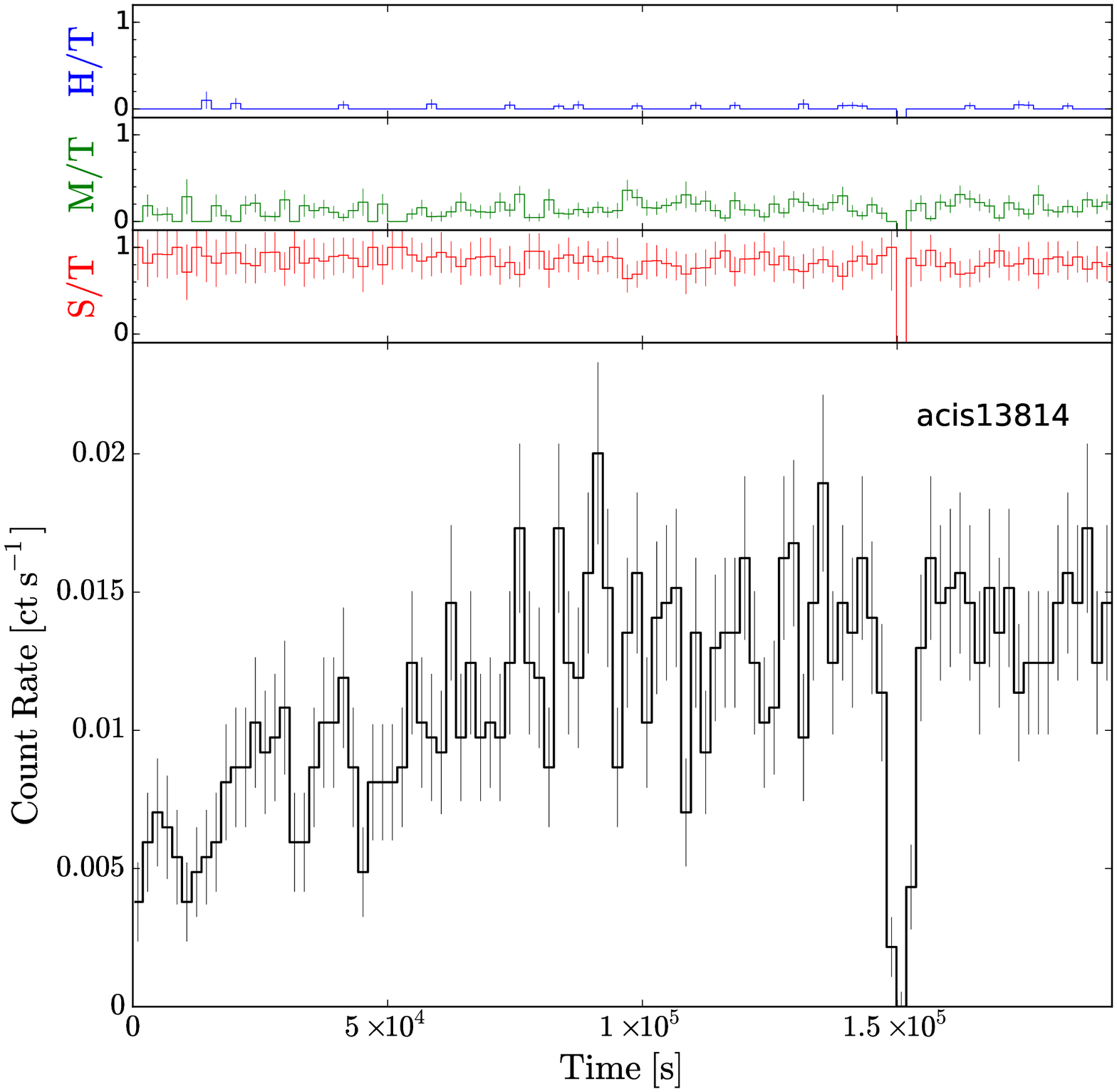}\\[-5pt]
\includegraphics[width=0.40\textwidth]{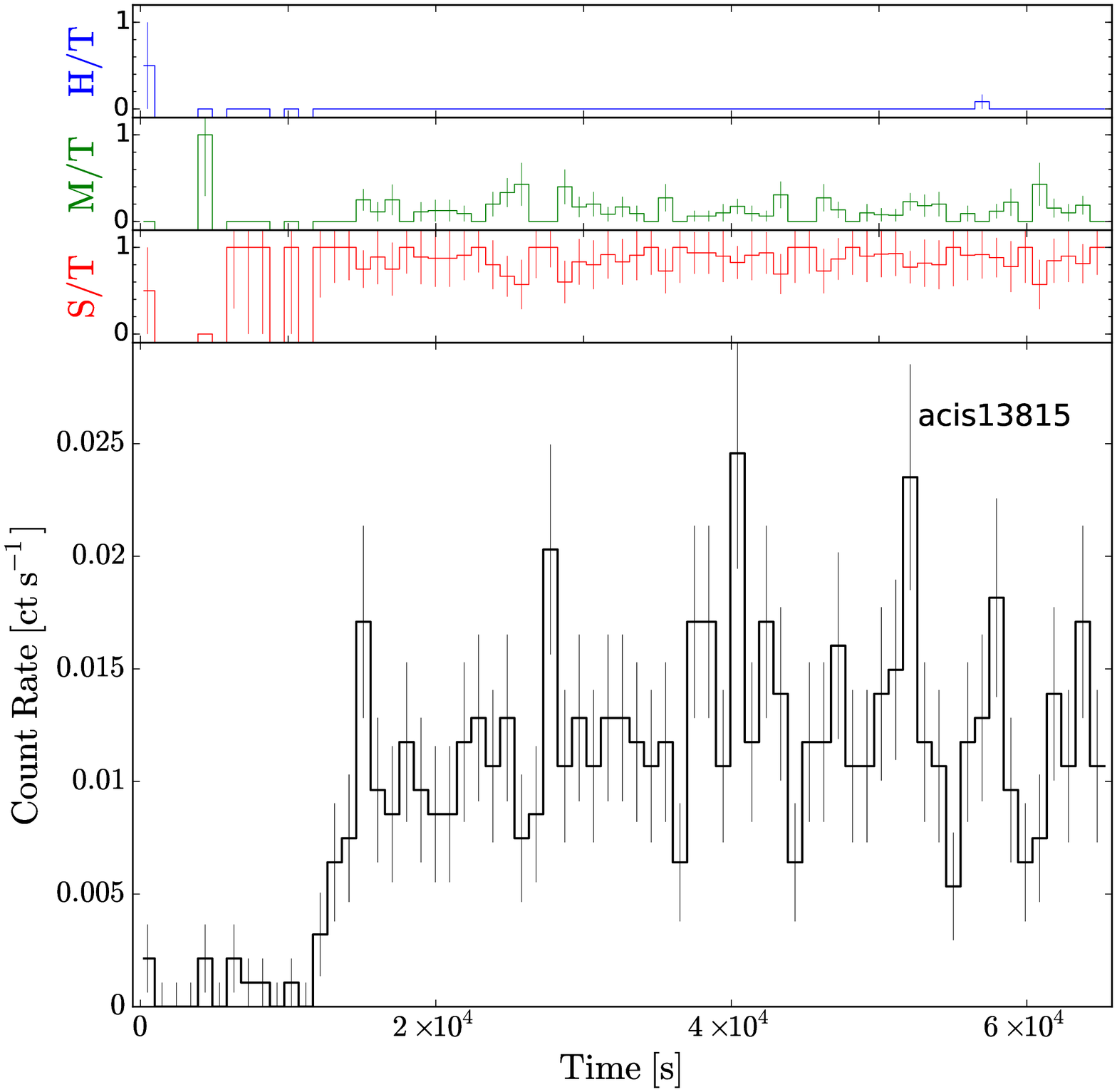}\\[-5pt]
\includegraphics[width=0.40\textwidth]{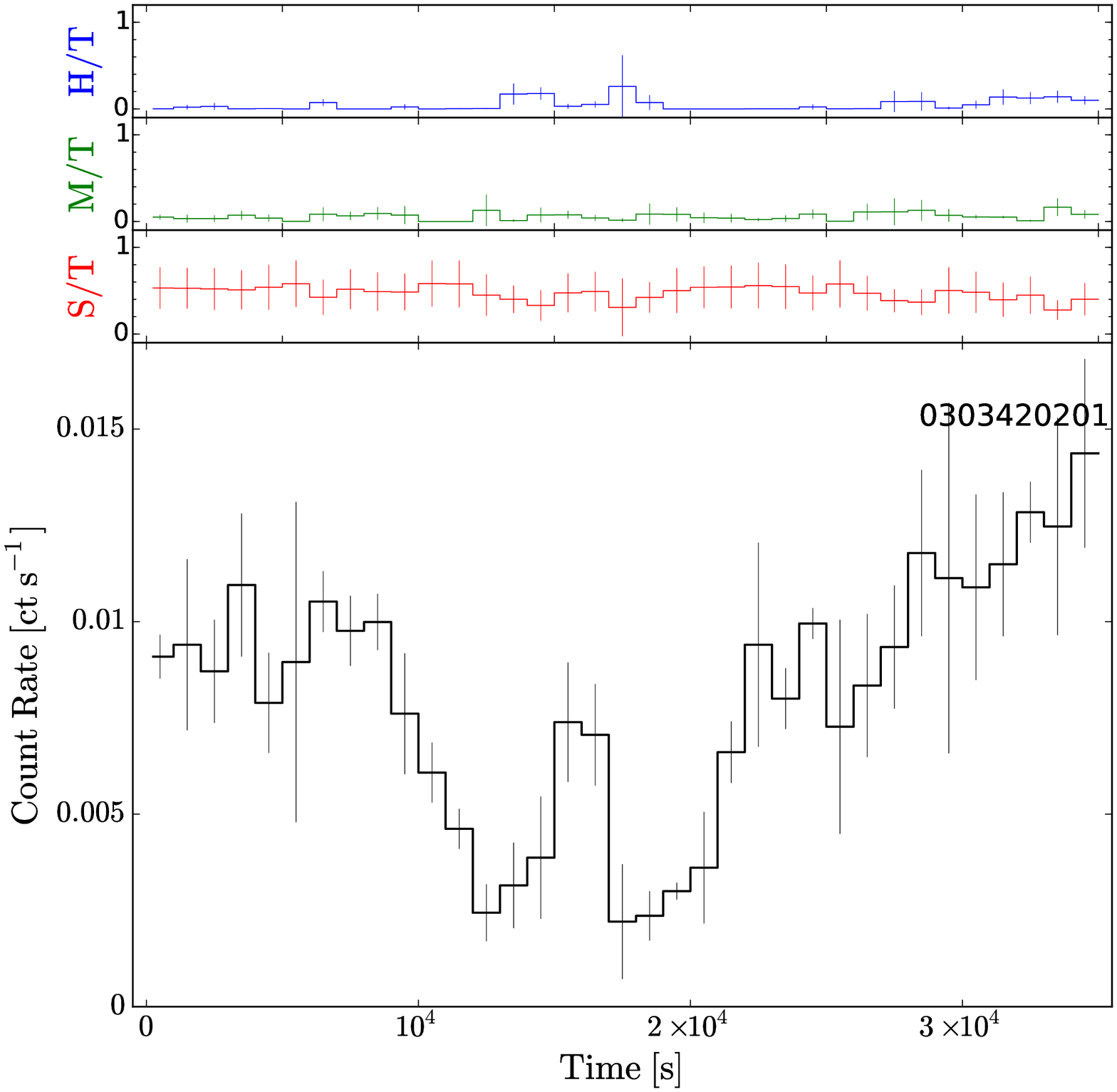}
\caption{Top panel: {\it Chandra}/ACIS light-curve for S1 in ObsID 13814, binned to 1000-s intervals.
The three upper sub-panels show the fractions of photon counts in the soft (0.3--1 keV), medium (1--2 keV), and hard (2--8 keV) bands.
Middle panel: as in the top panel, for ObsID 13815.
Bottom panel: as in the top panel, for the {\it XMM-Newton}/EPIC ObsID 0303420201; we plotted the combined EPIC pn $+$ MOS light-curve in the 0.3--8 keV band.
}
\label{eachS1.fig}
\end{figure}

We re-examined the data and confirm the findings of \cite{Urquhart2016a}. We interpret the step-like flux behaviour at the beginning of {\it Chandra} ObsID 13815 (Figure \ref{eachS1.fig}, top right) as a strong candidate for an eclipse of the X-ray source by the donor star. The quick transition from low to high count rates all but rules out other explanations such as state transitions. The eclipse was already in progress at the beginning of the observation, so we can only place a lower limit on its duration, $\gtrsim$ 12 ks.
We also recover the detection of the sharp dip (Figure \ref{eachS1.fig}, top left); the count rate falls from an average baseline to zero over a time-scale of $\approx$4000 s,
and rises back in a similar short time.
In addition, we found another dip in the EPIC light-curve from {\it XMM-Newton} Obsid 0303420201 (Figure \ref{eachS1.fig}, bottom left) .
The duration from ingress to egress is slightly longer than 10 ks.
Unlike the other two episodes, in this case the count rate does not decrease to zero; there is also a bump during the dip. We do not see a repeat of such dipping profile in any other observation of this source.


\subsection{Eclipses and orbital period for S2}
\label{xtimeS2.sec}

\begin{figure}
\center
\includegraphics[width=0.4\textwidth]{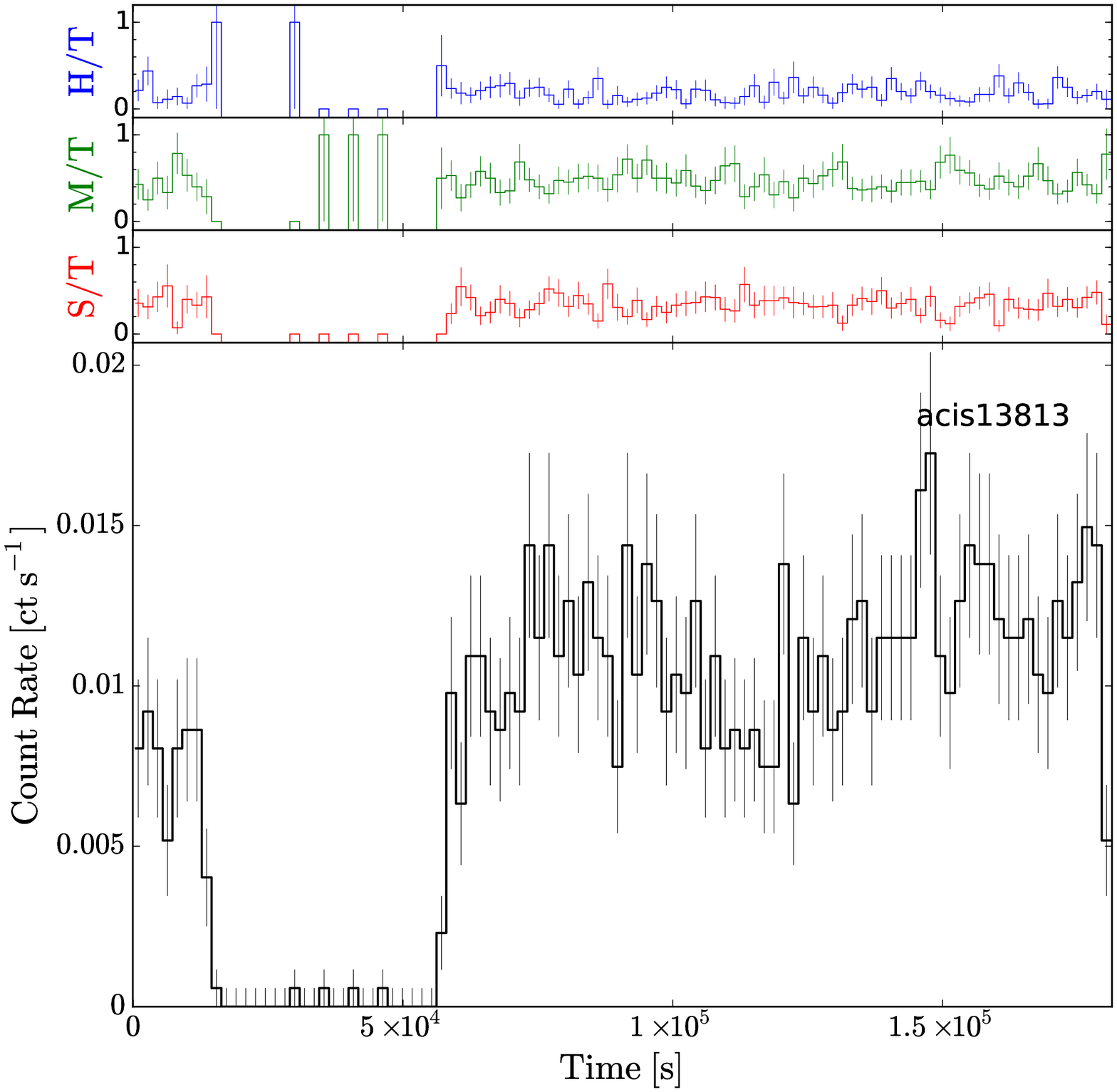}\\[-5pt]
\includegraphics[width=0.4\textwidth]{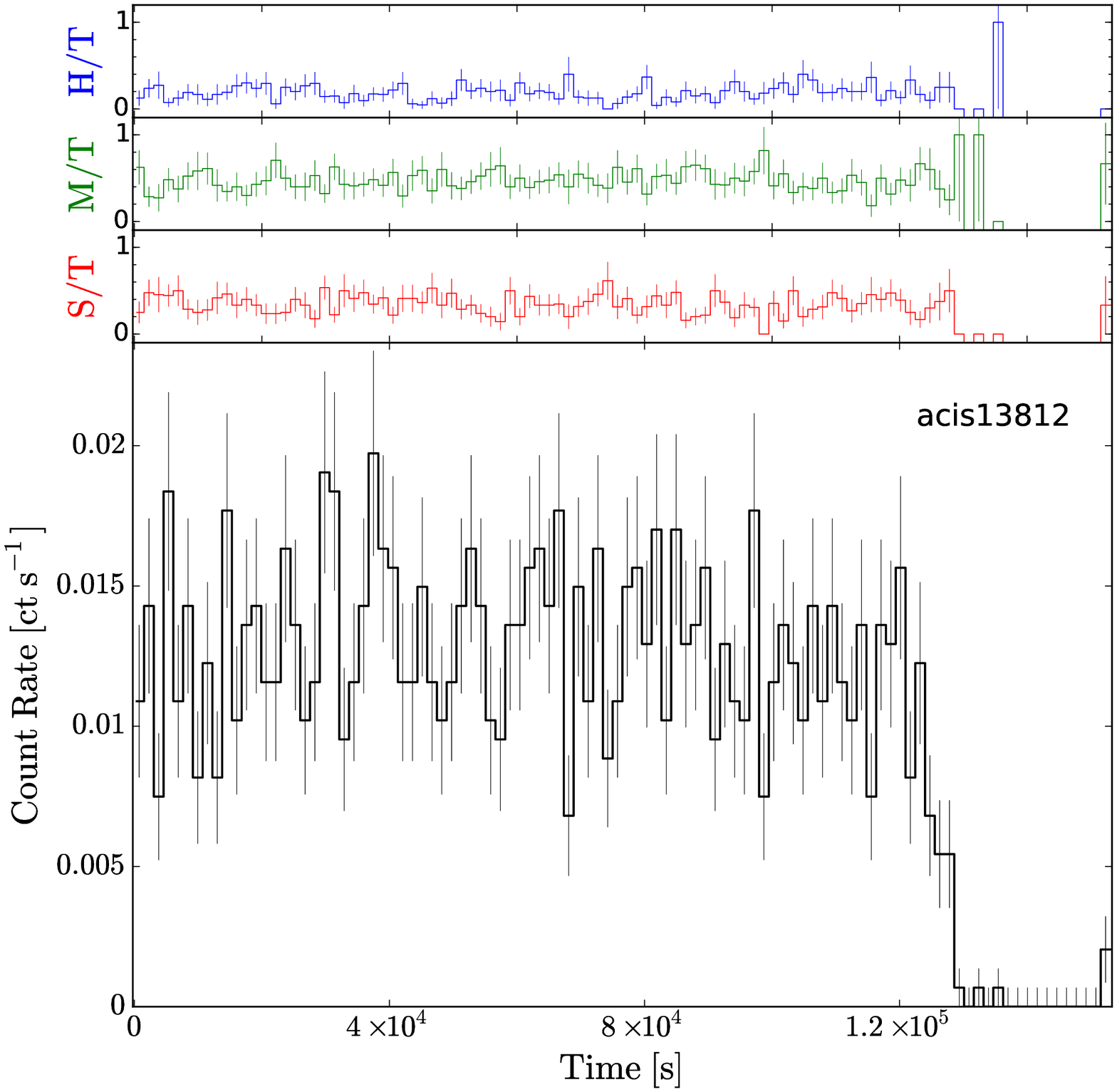}\\[-5pt]
\includegraphics[width=0.4\textwidth]{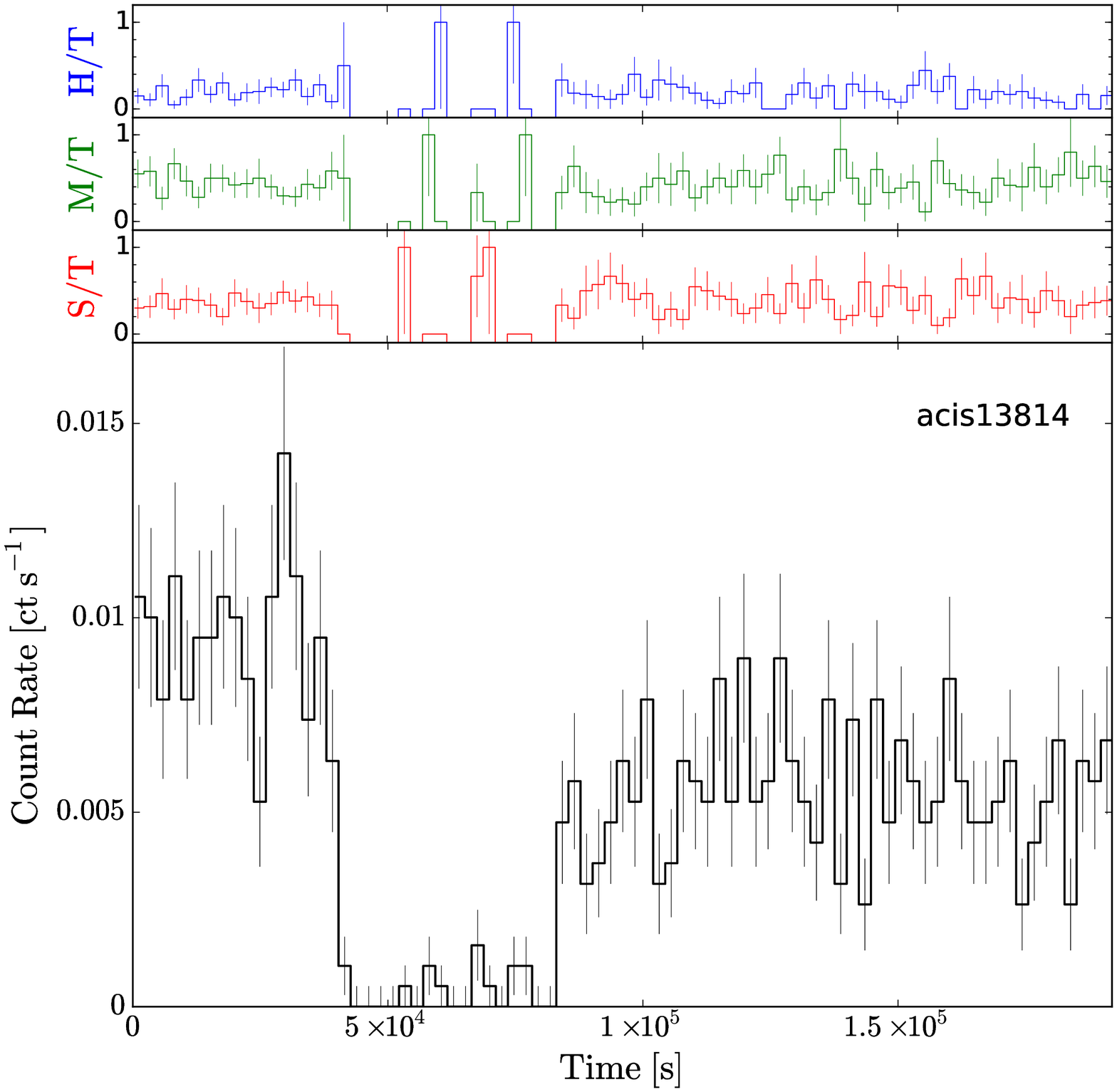}
\caption{Top panel: {\it Chandra}/ACIS light-curve for S2 in ObsID 13813, binned to 1000-s intervals.
The three upper sub-panels show the fractions of photon counts in the soft (0.3--1 keV), medium (1--2 keV), and hard (2--8 keV) bands.
Middle panel: as in the top panel, for ObsID 13812.
Bottom panel: as in the top panel, for ObsID 13814.
}
\label{eachS2.fig}
\end{figure}

If the interpretation of the flux dips in S1 is somewhat uncertain, the behaviour of S2 is much clearer. The long-term monitoring of this source shows that it is most often in quiescence or in a low state, but went into outburst for a few weeks during the 2012 {\it Chandra} observations (Figure \ref{eachS2.fig}, bottom right panel). It is during this outburst that we discovered three unambiguous stellar eclipses, in ObsIDs 13813, 13812, and 13814\footnote{Notice that the order of the three observations is not a typo, that is ObsID 13813 did precede ObsID 13812.} (Figure \ref{eachS2.fig}). In two of those three cases, we could measure the eclipse duration as $T_{\rm {ecl}} \approx 40$ ks $\approx 11$ hr.
Luckily, the scheduling of the {\it Chandra} observations enabled us to catch the moment of ingress of all three eclipses, only a few days from each other.
The interval between the ingress times in ObsIDs 13813 and 13812 is $\sim$ 104.02 hr,
while the interval between the ingress times in ObsIDs 13812 and 13814 is $\sim$ 156.75 hr.
Therefore, the binary period should be
\begin{equation}
P \approx 104.02/n_1 \approx 156.75/n_2,
\label{P1.equ}
\end{equation}
where $n_1$ and $n_2$ are integer numbers $\geq 1$ and $n1 = (2n_2)/3$. Thus, we can define $n_2 \equiv 3n$ where $n$ is an integer. For $n=1$, $P \approx 52$ hr; this is an acceptable solution. For $n=2$, $P \approx 26$ hr; however this is not an acceptable solution, because it would produce two eclipses during each of the long observations (ObsIDs 13813, 13812 and 13814, with a duration of $\approx$2 days each; see Table 1) as well as one eclipse (or part of) during ObsIDs 13815 and 13816, which is not observed. Even shorter periods ($n > 2$) are ruled out for the same reason.
In summary, the only possible binary period is $P \approx 52$ hr, with a rather long eclipse fraction $T_{\rm {ecl}}/P \approx 0.2$.


We then analyzed our X-ray light-curves with two independent techniques, to derive an even more precise and accurate value of the orbital period.
The first technique is the Lomb-Scargle method \citep[astropy.stats.LombScargle;][]{Press1989}
which is devised for unevenly spaced data.
We used the five observations with the highest number of counts (Obsid 13813, 13812, 15496, 13814, 13815) to compute the Lomb periodogram.
We found three significant peaks (true period and/or aliases) at $\approx$161.55 hr, $\approx$53.84 hr, and $\approx$69.31 hr, with a probability $<10^{-10}$ that any of those peaks are due to random fluctuations of photon counts.
The searching is on the assumption of the null hypothesis of no signal.
The second technique is the phase dispersion minimization (PDM) analysis \citep{Stellingwerf1978}.
A first search of periods in the range of 50--80 hr with steps of 0.01 hr
returns a group of phase dispersion minima at approximately
52.29 hr, 52.75 hr, 52.79 hr, 52.84 hr, 52.89 hr, 53.02 hr, 53.09 hr, 53.21 hr, 53.23 hr, 53.25 hr, 53.34 hr, and 53.36 hr. We then performed a second search with finer steps of 1 s around each of those minima, with an interval width of $\pm$ 30 s.

\begin{figure*}
\center
\includegraphics[width=0.49\textwidth]{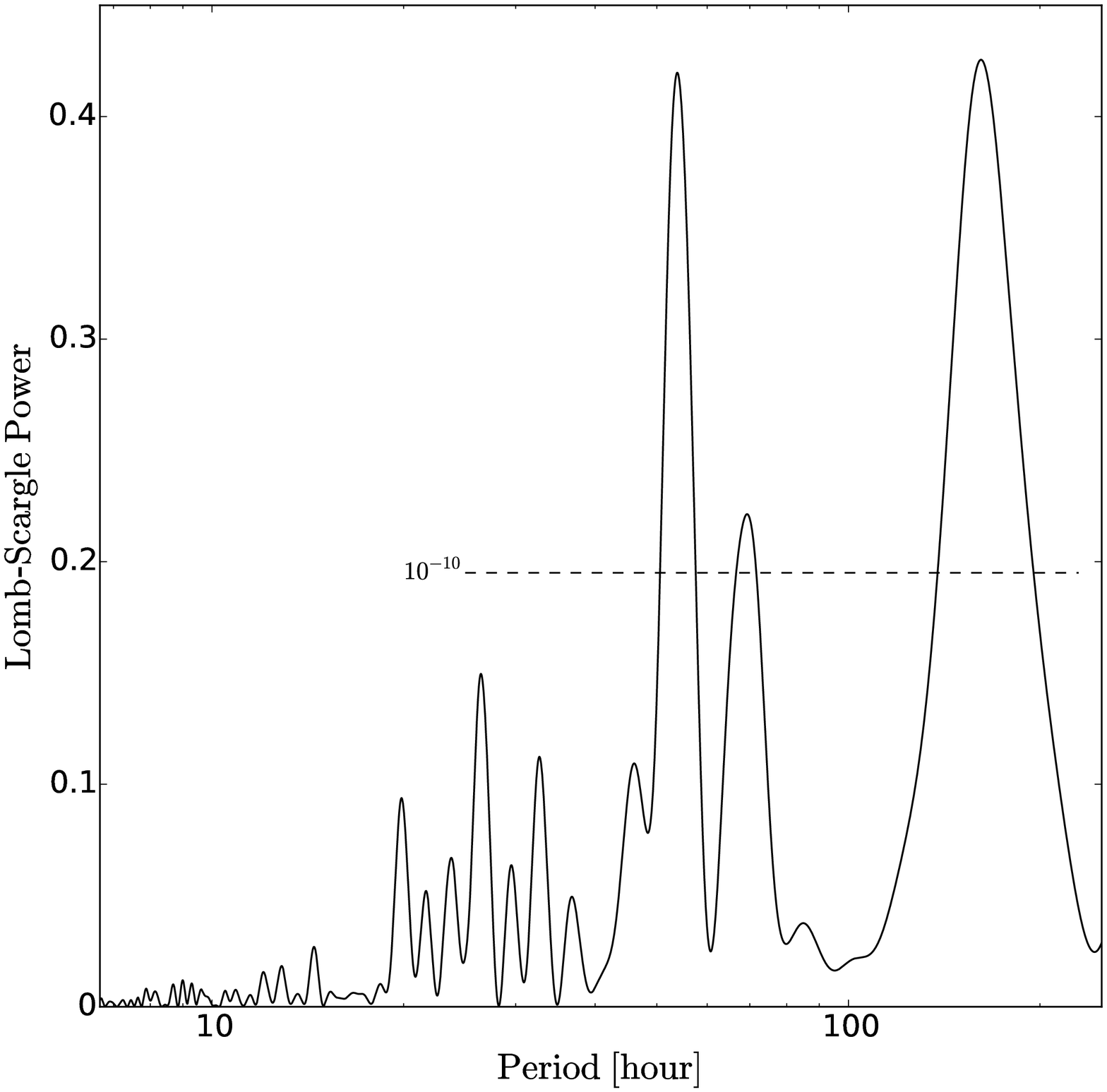}
\includegraphics[width=0.49\textwidth]{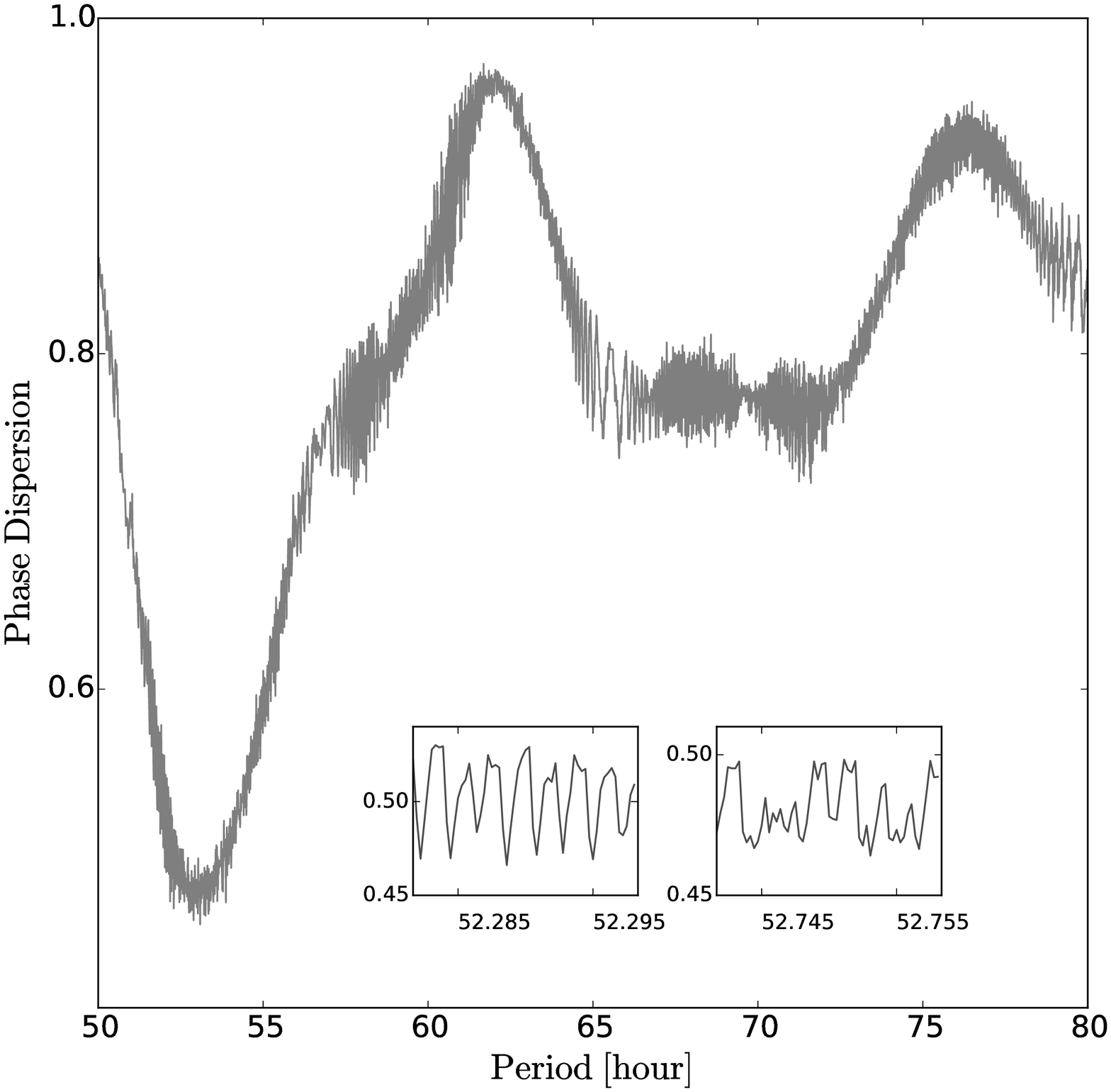}
\caption{Top panel: Search for periodicities for S2 with the Lomb-Scargle method. The periodogram is computed for five observations with $>$100 net source counts (Obsid 13813, 13812, 15496, 13814, 13815), and the power is given in the standard normalization ({\it astropy.stas}).
The black line represents a probability of 10$^{-10}$ for the periods to originate
from random fluctuations rather than true signal.
Bottom panel: phase dispersion calculated from the same five light-curves,
for candidate periods between 50 hr and 80 hr, with
a step of 0.01 hr. The two inserts show the result of finer searches around 52.29 hr and 52.75 hr.}
\label{method.fig}
\end{figure*}

\begin{figure*}
\center
\includegraphics[width=0.49\textwidth]{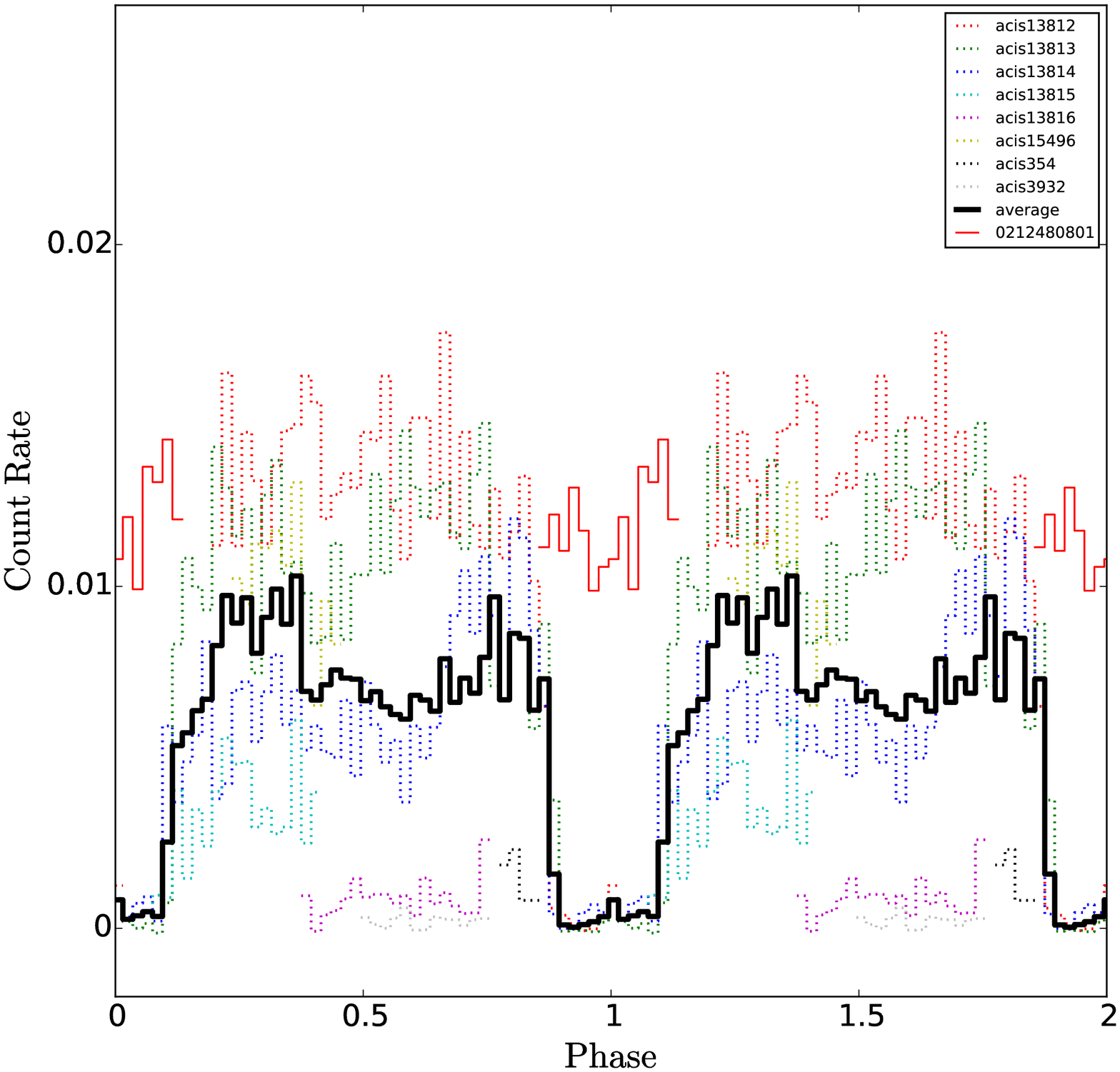}
\includegraphics[width=0.49\textwidth]{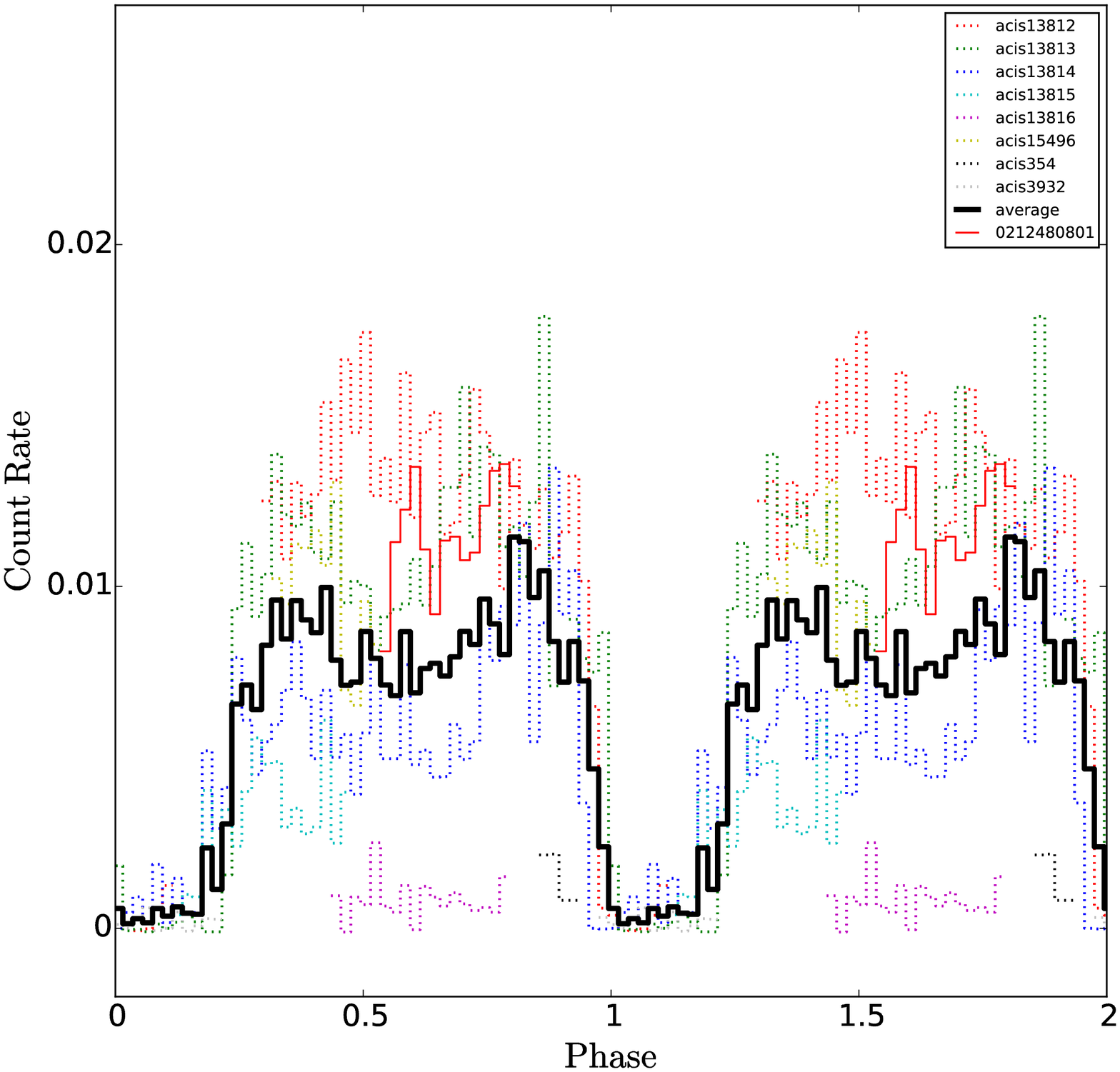}
\caption{Left panel: X-ray light-curves of S1, computed from the observations listed in the inset, and folded onto a candidate period of 52.2908 hr. The black solid histogram is the average light-curve from the {\it Chandra} observations; the red solid line is from the {\it XMM-Newton} observation. We converted all count rates to an equivalent {\it Chandra}/ACIS-S Cycle 13 count rate, using PIMMS with a power-law model (photon index $\Gamma = 1.7$) and Galactic line-of-sight absorption ($n_{\rm H} \approx 2\times 10^{-20}~{\rm cm}^{-2}$). For this choice of folding period, the {\it XMM-Newton} light-curve is inconsistent with the {\it Chandra} light-curve.
Right panel: as in the left panel, for a period of 52.7530 hr. This choice of period makes all the light-curves consistent with an eclipse at phase 0.}
\label{period.fig}
\end{figure*}

Having collected a sample of period candidates from our Lomb-Scargle and PDM searches, we folded the light-curves on each candidate period to compute averaged light-curves;
we compared them to the individual light-curves to check the phase preservation.
The most likely candidate periods are 52.2908 hr and 52.7530 hr.
For $P \approx{} 52.2908$ hr, the three egress profiles are consistent with the averaged {\it Chandra} light-curve; however, the light-curve of the {\it XMM-Newton} observation is clearly inconsistent with the folded profile, because it does not show an eclipse at a phase when it should have (Fig.~\ref{period.fig}, left panel).
For $P \approx 52.7530$ hr (Fig.~\ref{period.fig}, right panel), the three egress profiles from {\it Chandra} are still approximately consistent, and the light-curve from the {\it XMM-Newton} observation is also consistent with the folded light-curve (it does not cover the expected phase of the eclipse).
We conclude that the true period is $P \approx 52.7530$ hr. The small phase offsets of the three egress profiles can be used to estimate the uncertainty on the period.
The time span between ObsID 13813 and ObsID 13814 covers only 5 orbital cycles,
so we can only obtain a rough estimate of the period uncertainty; in other {\it Chandra} observations separated by a longer time interval ({\it e.g.}, ObsIDs 354 and 3932), the count rate is too low to provide meaningful constraints.
We estimate a preserved phase better than $\Delta \phi \approx 3/50$ over the 5 orbital cycles between ObsIDs 13813 and 13814; hence, the fractional error in the period is $\approx$($3/50$)/5 $\approx$0.012.
In conclusion, the binary period of the bright XRB S2 is $P = 52.75 \pm 0.63$ hr.

We determined a precise eclipse duration from {\it Chandra} ObsIDs 13813 and 13814; we defined the eclipse as the time when the count rate was less than 0.002 ct s$^{-1}$.
The two observations show an eclipse duration of $(4.170 \pm 0.128) \times 10^4$ s and $(4.270 \pm 0.168) \times 10^4$ s, respectively. Henceforth, we adopt the average of those two values, $T_{\rm ecl} = (4.220 \pm 0.149) \times 10^4$ s.
The eclipse fraction $\Delta_{\rm ecl}$ is then derived as the mean eclipse duration divided by the orbital period: we obtain $\Delta_{\rm ecl} = 0.222 \pm 0.008$.

\begin{table*}
\begin{center}
\scriptsize
\caption[]{X-ray spectral models and best-fitting parameters for S1. Errors are the 90\% confidence range for a single interesting parameter.}
\label{xspecS1.tab}
\begin{tabular}{llccccccc}
\hline\noalign{\smallskip}
Model  &    Data & $n_{H}${$^{a}$}             &    $kT_{\rm bb}/kT_{\rm in}$   &      $f_X$ [0.3--8 keV]$^b$   & $L_X$ [0.3--8 keV]$^c$  &  $L_{\rm bol}${$^{d}$}  & $\chi^2_{\nu}$/dof  & added models\\
       &         &(10$^{22}$ cm$^{-2}$) &    (eV) & (10$^{-14}$ erg cm$^{-2}$ s$^{-1}$)   & (10$^{39}$ erg s$^{-1}$) & (10$^{39}$ erg s$^{-1}$)  & \\
\noalign{\smallskip}\hline
\multirow{9}{*}{{\it bbody}}
&  354        & 0.11$^{+0.18}_{-0.01}$ & 70$^{+78}_{-32}$ & 6.5$^{+47.2}_{-2.7}$ & 2.0$^{+13.5}_{-1.6}$ & 1.4$^{+9.4}_{-1.1}$ & 0.97/5   & edge,mekal      \\
&  1622       & 0.22$^{+*}_{-0.01}$ & 107$^{+*}_{-15}$ & 3.1$^{+1.1}_{-0.4}$ & 1.4$^{+6.2}_{-1.1}$ & 1.5$^{+6.5}_{-1.2}$ & 1.14/6   & edge            \\
&  3932       & 0.02$^{+0.10}_{-0.01}$ & 130$^{+15}_{-27}$ & 7.3$^{+0.5}_{-0.7}$ & 0.63$^{+0.65}_{-0.07}$ & 0.8$^{+0.9}_{-0.1}$ & 1.38/35  & mekal           \\
&  13813      & 0.02$^{+0.08}_{-0.01}$ & 109$^{+16}_{-14}$ & 4.6$^{+0.4}_{-0.4}$ & 0.47$^{+0.05}_{-0.05}$ & 0.7$^{+0.1}_{-0.1}$ & 1.09/36  & edge,mekal      \\
&  13812      & 0.02$^{+0.10}_{-0.01}$ & 101$^{+10}_{-13}$ & 5.6$^{+0.5}_{-0.8}$ & 0.54$^{+0.56}_{-0.06}$ & 0.8$^{+0.8}_{-0.1}$ & 1.02/42  & 2*mekal         \\
&  15496      & 0.11$^{+*}_{-0.01}$ & 106$^{+20}_{-23}$ & 5.0$^{+1.7}_{-1.1}$ & 1.0$^{+5.2}_{-0.6}$ & 1.3$^{+6.6}_{-0.7}$ & 0.67/11  &                 \\
&  13814      & 0.19$^{+*}_{-0.12}$ & 100$^{+24}_{-15}$ & 7.0$^{+1.0}_{-0.5}$ & 2.6$^{+4.7}_{-1.6}$ & 2.0$^{+3.5}_{-1.2}$ & 1.16/51  & edge,2*mekal    \\
&  13815      & 0.02$^{+0.08}_{-0.01}$ & 125$^{+*}_{-26}$ & 6.7$^{+0.7}_{-0.6}$ & 0.60$^{+0.07}_{-0.06}$ & 0.8$^{+0.1}_{-0.1}$ & 1.11/20  & edge            \\
&  0212480801 & 0.07$^{+0.09}_{-0.01}$ & 102$^{+25}_{-19}$ & 9.8$^{+1.0}_{-0.7}$ & 1.5$^{+1.7}_{-0.6}$ & 1.4$^{+1.6}_{-0.5}$ & 0.96/48  & edge,mekal      \\
&  0303420201 & 0.11$^{+0.07}_{-0.06}$ & 93$^{+18}_{-13}$ & 6.8$^{+0.5}_{-0.4}$ & 1.4$^{+1.2}_{-0.6}$ & 1.5$^{+1.4}_{-0.7}$ & 0.83/46  & mekal           \\
\hline
\multirow{9}{*}{{\it diskbb}}
&  354        & 0.11$^{+*}_{-0.01}$ & 92$^{+*}_{-36}$ & 6.4$^{+4.2}_{-2.6}$ & 3.1$^{+1.0}_{-2.4}$ & 11.0$^{+3.5}_{-8.5}$ & 0.99/5   & edge,mekal      \\
&  1622       & 0.23$^{+*}_{-0.01}$ & 131$^{+*}_{-43}$ & 3.2$^{+1.4}_{-0.6}$ & 4.3$^{+30.9}_{-3.3}$ & 8.5$^{+61.1}_{-6.4}$ & 1.14/6   & edge            \\
&  3932       & 0.04$^{+0.09}_{-0.01}$ & 171$^{+27}_{-39}$ & 7.2$^{+0.7}_{-0.7}$ & 1.9$^{+1.8}_{-0.4}$ & 3.1$^{+3.0}_{-0.7}$ & 1.40/35  & mekal           \\
&  13813      & 0.03$^{+0.09}_{-0.01}$ & 142$^{+*}_{-29}$ & 4.7$^{+0.5}_{-0.7}$ & 1.3$^{+1.4}_{-0.2}$ & 2.9$^{+3.0}_{-0.5}$ & 1.14/36  & edge,mekal      \\
&  13812      & 0.03$^{+*}_{-0.01}$ & 99$^{+28}_{-26}$ & 5.6$^{+0.8}_{-0.8}$ & 1.8$^{+1.8}_{-0.5}$ & 4.2$^{+4.3}_{-1.1}$ & 1.03/42  & 2*mekal         \\
&  15496      & 0.22$^{+0.32}_{-0.01}$ & 113$^{+40}_{-31}$ & 4.7$^{+2.4}_{-0.9}$ & 8.3$^{+*}_{-6.7}$ & 25.7$^{+*}_{-20.7}$ & 0.73/11  &                 \\
&  13814      & 0.21$^{+0.13}_{-0.10}$ & 128$^{+31}_{-24}$ & 7.0$^{+0.9}_{-0.5}$ & 8.6$^{+16.5}_{-4.7}$ & 22.7$^{+43.5}_{-12.4}$ & 1.17/51  & edge,2*mekal    \\
&  13815      & 0.02$^{+0.09}_{-0.01}$ & 175$^{+*}_{-48}$ & 7.0$^{+0.7}_{-0.6}$ & 1.7$^{+1.7}_{-0.1}$ & 2.9$^{+3.0}_{-0.2}$ & 1.14/20  & edge            \\
&  0212480801 & 0.08$^{+0.08}_{-0.01}$ & 140$^{+40}_{-32}$ & 9.9$^{+1.0}_{-0.7}$ & 4.2$^{+4.3}_{-1.6}$ & 9.2$^{+9.4}_{-3.5}$ & 0.92/48  & edge,mekal      \\
&  0303420201 & 0.13$^{+0.07}_{-0.05}$ & 114$^{+21}_{-18}$ & 6.7$^{+0.5}_{-0.4}$ & 5.0$^{+4.3}_{-2.1}$ & 15.2$^{+13.0}_{-6.3}$ & 0.85/46  & mekal           \\
\hline
\end{tabular}
\end{center}
\begin{flushleft}
{$^a$}$n_{H}$ is the total column density, including the fixed Galactic foreground absorption ($1.57 \times 10^{20} $cm$^{-2}$) and the local component derived from spectral fitting.\\
{$^b$}$f_{\rm X}$ is the observed flux (not corrected for absorption).\\
{$^c$}$L_{\rm X}$ is the emitted luminosity (corrected for absorption). 
$L_{\rm X} \equiv 4\pi \, d^2 \, f^{\rm em}_{\rm X}$ for the {\it bbody} model, and
$L_{\rm X} \equiv 2\pi \, d^2 \, f^{\rm em}_{\rm X}/\cos \theta$ for the {\it diskbb} model. Here, we assumed $\theta = 80^{\circ}$ (see text for details).\\
$^d$For the {\it bbody} model, $L_{\rm bol} = \sqrt{N} \, d_{10{\rm kpc}} \, 10^{39}$ erg s$^{-1}$, where $N$ is the {\small {XSPEC}} normalization parameter; for the {\it diskbb} model, $L_{\rm bol} \approx 2\pi (d^2/{\cos \theta}) f[0.01$--20 keV], with $\theta = 80^{\circ}$.
\end{flushleft}
\end{table*}

\section{X-ray spectral results}
\label{xspec.sec}


\subsection{Spectral models for S1}
\label{sfits1.sec}

\begin{figure}
\center
\includegraphics[width=0.49\textwidth]{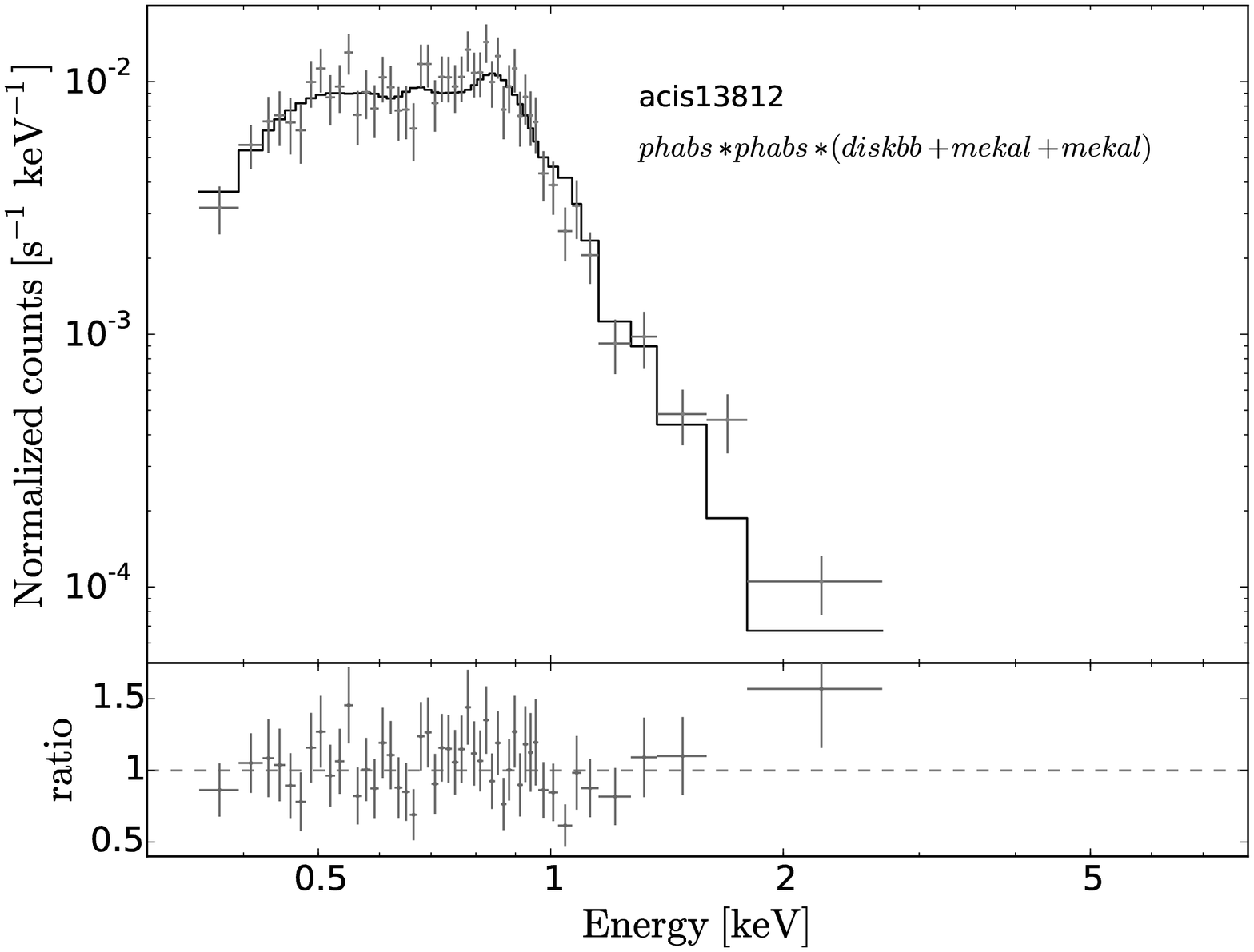}\\
\includegraphics[width=0.49\textwidth]{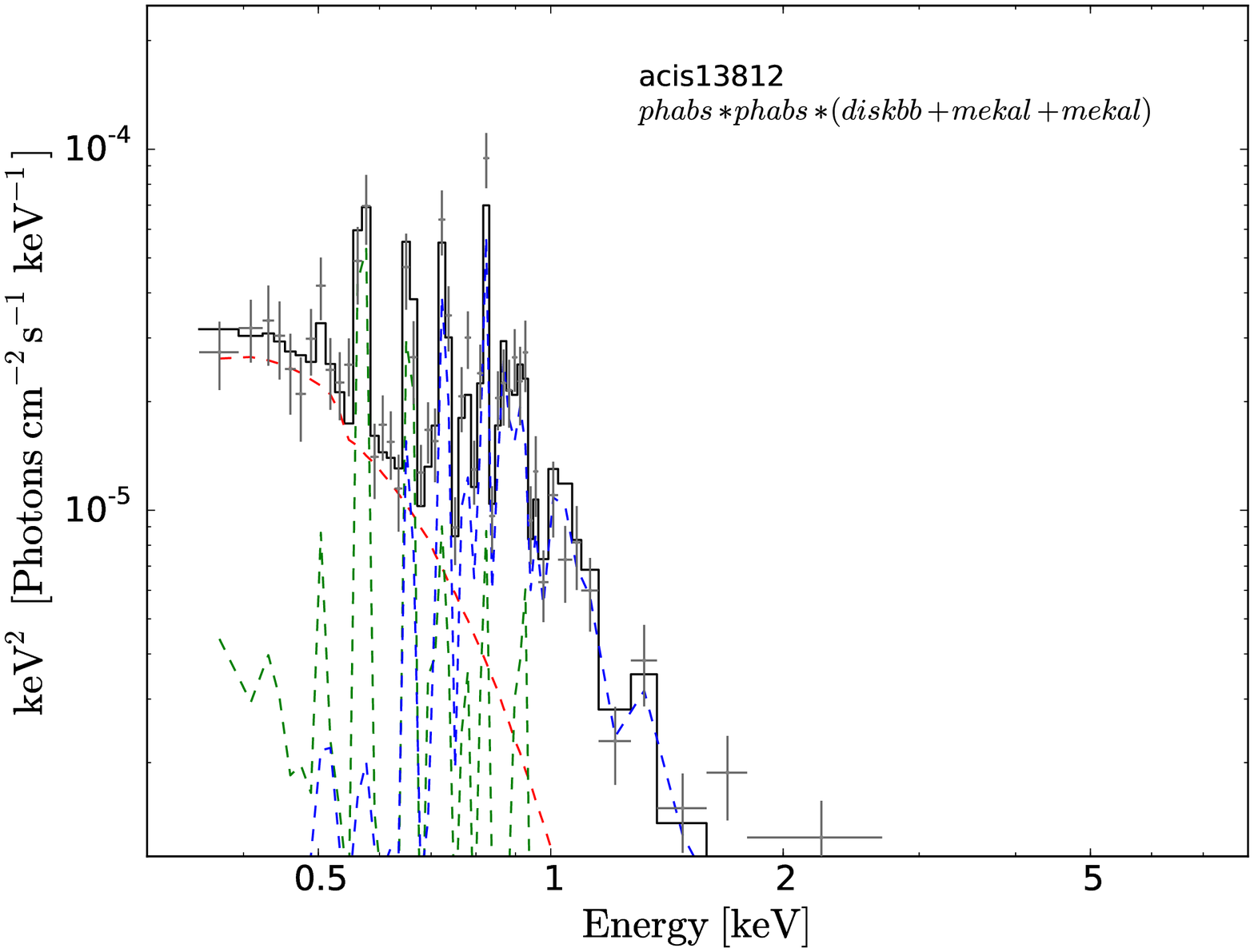}
\caption{Top panel: best-fitting disk-blackbody plus thermal plasma model to the spectrum of S1 from {\it Chandra} Obsid 13812. Datapoints and model values are plotted in the top sub-panel; data/model ratios in the bottom sub-panel. Error bars denote 68.3\% uncertainties. See Table \ref{xspecS1.tab} and Section \ref{sfits1.sec} for the fit parameters and their interpretation.
Bottom panel: unfolded best-fitting model to the spectrum of S1 from {\it Chandra} Obsid 13812, with individual model components. The dashed red line represents the {\it diskbb} component, and the dashed blue and green lines represent two {\it mekal} components; the solid black histogram represents the combined model.}
\label{specS1.fig}
\end{figure}

\begin{figure}
\center
\includegraphics[width=0.48\textwidth]{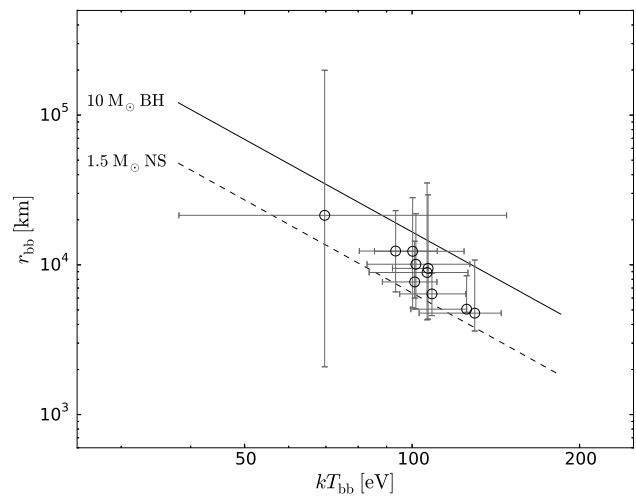}
\caption{Best-fitting blackbody radius versus temperature for S1. The solid and dashed lines represent the expected relations for optically-thick photospheres around a 10-$M_{\odot}$ BH and a 1.5-$M_{\odot}$ NS, respectively \citep{Soria2016,Urquhart2016a}.}
\label{S1model.fig}
\end{figure}

We tried two single-component models to fit the background-subtracted spectra of S1:
{\it phabs*phabs*bbody} and {\it phabs*phabs*diskbb}.
The first photoelectric absorption component was fixed at the line-of-sight value for M\,51
($N_{\rm H} = 1.57 \times 10^{20} $cm$^{-2}$: \citealt{Dickey1990}), and the second component was left free.
We find (Table \ref{xspecS1.tab}) that a single thermal component at $T \sim 0.1$ keV is a good fit, with no additional power-law at high energies; blackbody and disk-blackbody models are statistically equivalent, as expected given the low temperature. However, in most epochs (for example in {\it Chandra} ObsID 13812: Figure~\ref{specS1.fig}), as previously noted by \citet{Urquhart2016a}, we confirm the presence of significant soft x-ray residuals. In those cases, the fit is improved by the addition of one or two thermal-plasma components ({\it mekal} model in {\sc xspec}) and/or an absorption edge ({\it edge} in {\sc xspec}). This is also in good agreement with the findings of \citet{Urquhart2016a}. We also recover the inverse trend between characteristic radius and temperature of the thermal component (Figure \ref{S1model.fig}), noted and discussed in previous work \citep{Soria2016,Urquhart2016a}; the Spearman's rank correlation coefficient is $-$0.92.

Although the observed and de-absorbed fluxes in the X-ray band are in broad agreement between the {\it bbody} and {\it diskbb} models, the two models predict different bolometric luminosities $L_{\rm {bol}}$, because they have a different behaviour in the UV band, and because they have a different dependence on the viewing angle; in general, the blackbody model represents the idealized case of isotropic emission while the disk-blackbody model represents the equally idealized case of disk-like emission.
For the {\it bbody} model,
$L_{\rm bol} \equiv N_{\rm{bb}} \, d^2_{10{\rm kpc}} \, 10^{39}$ erg s$^{-1}$,
where $N_{\rm{bb}}$ is the {\sc xspec} model normalization and ${d}_{10\mathrm{kpc}}$ the distance in units of 10 kpc.
For the {\it diskbb} model,
$L_{\rm bol} \approx 4\pi r_{\rm in}^2 \sigma T_{\rm in}^4$,
where $r_{\rm in}$ is the apparent inner-disk radius, defined as $r_{\rm in} = \sqrt{N_{\rm {dbb}}} \, d_{10{\rm kpc}}/\sqrt{\cos \theta}$ km,
and $\theta$ the viewing angle to the disk plane (with $\theta = 0^{\circ}$ corresponding to a face-on view).
All the bolometric luminosities are estimated in the 0.01--20 keV band, and do not include the mekal components.
For the conversion of band-limited fluxes $f_{\rm X}$ to X-ray luminosities, first we calculated the de-absorbed fluxes $f^{\rm em}_{\rm X}$ and their errors with the task {\it cflux} within {\small XSPEC}; then, we calculated the emitted luminosities, defined as $L_{\rm X} \equiv 4\pi \, d^2 \, f^{\rm em}_{\rm X}$ for the {\it bbody} model, and
$L_{\rm X} \equiv 2\pi \, d^2 \, f^{\rm em}_{\rm X}/\cos \theta$ for the {\it diskbb} model.
The presence of dips and a likely eclipse suggest a high viewing angle with respect to the orbital plane. On the other hand, the spin axis of the compact object may be misaligned with respect to the spin axis of the binary system \citep{Fragos2010}. In that case, the inner disk is expected to be aligned with the equatorial plane of the compact object rather than with the binary plane \citep{Bardeen1975}. However, the degree of warping and the location of the transition radius between inner and outer disk (suggested to be at a radius $<$300 gravitational radii by \citealt{Fragile2001}) are still a matter of debate \citep[e.g.,][]{Fragile2007,Fragile2009, McKinney2013,Sorathia2013,Nixon2014,Zhuravlev2014,King2016,Motta2018,Middleton2018}. In the case of M\,51 S1, for the disk scenario, the characteristic radius of the emitting region would be $\approx$5,000--10,000 km (Figure~\ref{S1model.fig}), that is several 100s gravitational radii for a stellar-mass BH. It is unclear whether this region of the disk would still be aligned with the binary plane, especially in the case of highly super-Eddington accretion, with a spherization radius \citep{Shakura1973,Poutanen2007} also located at a similar distance from the compact object. In Table~\ref{xspecS1.tab}, for the luminosity estimates in the disk scenario, we have arbitrarily assumed $\theta = 80^{\circ}$ as an extreme upper limit to the plausible luminosity range. By comparison, we note that all other ULSs \citep{Urquhart2016a} have isotropic luminosities of the soft thermal component $\sim$10$^{39}$ erg s$^{-1}$.

\begin{table*}
\begin{center}
\setlength{\tabcolsep}{4pt}
\scriptsize
\caption[]{X-ray spectral models and best-fitting parameters for S2. Errors are the 90\% confidence range for a single interesting parameter.}
\label{xspecS2.tab}
\begin{tabular}{lcccccccccc}
\hline\noalign{\smallskip}
  Data & $n_{H}${$^{a}$} &    $kT${$^{b}$} & $kT_{bb}$ & $kT_m$  & $N_{db}${$^{c}$}  & $N_{bb}${$^{d}$}  &  $\Gamma$   &     $f_X$ [0.3--8 keV]$^e$   & $L_X$ [0.3--8 keV]$^f$ & $\chi^2_{\nu}$/dof \\
   &(10$^{22}$  &     &   &  &  &   &  & (10$^{-14}$   & (10$^{39}$  &  \\
   & cm$^{-2}$) &    (keV) &  (keV) & (keV) &  &   &  & erg cm$^{-2}$ s$^{-1}$)  & erg s$^{-1}$) &  \\
\noalign{\smallskip}\hline
\multicolumn{11}{c}{{\it phabs} $*$ {\it phabs} $*$ {\it powerlaw}}\\
\hline
13813      & 0.10$^{+0.04}_{-0.04}$ & - & - & - & - & - & 2.34$^{+0.20}_{-0.18}$ & 5.5$^{+0.4}_{-0.4}$ & 0.63$^{+0.10}_{-0.07}$ & 1.32/71  \\
13812      & 0.07$^{+0.12}_{-0.01}$ & - & - & - & - & - & 2.51$^{+0.19}_{-0.17}$ & 6.8$^{+0.5}_{-0.4}$ & 0.88$^{+0.14}_{-0.11}$ & 1.07/73  \\
13814      & 0.05$^{+0.06}_{-0.01}$ & - & - & - & - & - & 2.21$^{+0.29}_{-0.22}$ & 3.8$^{+0.5}_{-0.4}$ & 0.35$^{+0.07}_{-0.04}$ & 1.09/42  \\
0212480801 & 0.12$^{+0.08}_{-0.06}$ & - & - & - & - & - & 2.86$^{+0.51}_{-0.38}$ & 5.5$^{+0.7}_{-0.6}$ & 0.82$^{+0.53}_{-0.20}$ & 1.12/54  \\
\hline
\multicolumn{11}{c}{{\it phabs} $*$ {\it phabs} $*$ ({\it powerlaw} $+$ {\it mekal})}\\
\hline
13813      & 0.04$^{+0.05}_{-0.02}$ & - & - & 1.08$^{+0.26}_{-0.25}$ & - & - & 2.00$^{+0.24}_{-0.19}$ & 5.9$^{+0.5}_{-0.4}$ & 0.52$^{+0.07}_{-0.05}$ & 1.13/69  \\
13812      & 0.08$^{+0.04}_{-0.04}$ & - & - & 1.10$^{+0.35}_{-0.25}$ & - & - & 2.28$^{+0.22}_{-0.19}$ & 6.9$^{+0.5}_{-0.5}$ & 0.73$^{+0.12}_{-0.09}$ & 0.96/72  \\
13814      & 0.02$^{+0.03}_{-0.01}$ &- &  - &   1.12$^{+0.30}_{-0.26}$ &- &   - & 2.00$^{+0.10}_{-0.19}$ & 3.9$^{+0.3}_{-0.3}$ & 0.32$^{+0.02}_{-0.02}$ & 0.86/40  \\
0212480801 & 0.02$^{+*}_{-0.01}$ & - & - & 0.79$^{+0.24}_{-0.20}$ & - & - & 2.18$^{+0.15}_{-0.14}$ & 6.3$^{+0.5}_{-0.5}$ & 0.51$^{+0.04}_{-0.03}$ & 0.85/52  \\
\hline
\multicolumn{11}{c}{{\it phabs} $*$ {\it phabs} $*$ ({\it diskpbb} $+$ {\it mekal}) with $p = 0.75$}\\
\hline
13813      & 0.02$^{+0.01}_{-0.01}$ & 0.97$^{+0.14}_{-0.12}$ & - & 0.63$^{+0.07}_{-0.17}$ & 2.3$^{+1.3}_{-0.9}$$\times$$10^{-3}$ & - & - & 4.8$^{+0.3}_{-0.3}$ & 0.99$^{+0.06}_{-0.06}$ & 1.90/69  \\
13812      & 0.02$^{+0.00}_{-0.01}$ & 0.87$^{+0.10}_{-0.10}$ & - & 0.80$^{+0.09}_{-0.19}$ & 4.5$^{+2.7}_{-1.7}$$\times$$10^{-3}$ & - & - & 5.9$^{+0.3}_{-0.3}$ & 1.2$^{+0.1}_{-0.1}$ & 1.66/72  \\
13814      & 0.02$^{+0.01}_{-0.01}$ & 0.99$^{+0.27}_{-0.21}$ & - & 0.80$^{+0.29}_{-0.22}$ & 1.3$^{+2.1}_{-0.8}$$\times$$10^{-3}$ & - & - & 3.0$^{+0.3}_{-0.3}$ & 0.60$^{+0.06}_{-0.06}$ & 2.03/40  \\
0212480801 & 0.02$^{+0.02}_{-0.01}$ & 0.39$^{+0.05}_{-0.04}$ & - & [0.80] & 0.11$^{+0.06}_{-0.04}$                  & - & - & 4.4$^{+0.3}_{-0.3}$ & 0.99$^{+0.07}_{-0.07}$ & 1.50/53\\
\hline
\multicolumn{11}{c}{{\it phabs} $*$ {\it phabs} $*$ ({\it diskpbb} $+$ {\it mekal}) with $p = 0.6$}\\
\hline
13813      & 0.02$^{+0.01}_{-0.01}$ & 1.33$^{+0.29}_{-0.21}$ & - & 0.83$^{+0.24}_{-0.21}$ & 3.5$^{+3.4}_{-1.9}$$\times$$10^{-4}$ & - & - & 5.3$^{+0.4}_{-0.4}$ & 1.1$^{+0.1}_{-0.1}$ & 1.48/69  \\
13812      & 0.02$^{+0.01}_{-0.01}$ & 1.10$^{+0.17}_{-0.13}$ & - & 1.01$^{+0.12}_{-0.26}$ & 9.4$^{+6.7}_{-4.1}$$\times$$10^{-4}$ & - & - & 6.4$^{+0.4}_{-0.4}$ & 1.4$^{+0.1}_{-0.1}$ & 1.21/72  \\
13814      & 0.02$^{+0.01}_{-0.01}$ & 1.32$^{+0.52}_{-0.31}$ & - & 0.84$^{+0.41}_{-0.19}$ & 2.3$^{+4.4}_{-1.6}$$\times$$10^{-4}$ & - & - & 3.4$^{+0.4}_{-0.4}$ & 0.70$^{+0.07}_{-0.06}$ & 1.48/40  \\
0212480801 & 0.02$^{+0.02}_{-0.01}$ & 0.49$^{+0.10}_{-0.07}$ & - & [0.80] & 0.02$^{+0.02}_{-0.01}$                  & - & - & 4.6$^{+0.3}_{-0.3}$ & 1.0$^{+0.1}_{-0.1}$ & 1.38/53  \\
\hline
\multicolumn{11}{c}{{\it phabs} $*$ {\it phabs} $*$ ({\it diskbb} $+$ {\it powerlaw} $+$ {\it mekal})}\\
\hline
13813      & \multicolumn{10}{c}{diskbb component is not significant}\\
13812      & \multicolumn{10}{c}{diskbb component is not significant}\\
13814      & \multicolumn{10}{c}{diskbb component is not significant}\\
0212480801 & 0.05$^{+0.13}_{-0.03}$ & 0.25$^{+0.08}_{-0.08}$ & - & 1.10$^{+7.23}_{-0.44}$ & 0.73$^{+9.27}_{-0.54}$ & -& 1.19$^{+0.56}_{-0.61}$ & 6.8$^{+0.8}_{-0.8}$ & 1.2$^{+1.0}_{-0.2}$ & 0.67/50  \\
\hline
\multicolumn{11}{c}{{\it phabs} $*$ {\it phabs} $*$ ({\it diskbb} $+$ {\it bbodyrad} $+$ {\it mekal}) with $kT_{in} < kT_{bb}$}\\
\hline
13813      & 0.02$^{+0.07}_{-0.01}$ & 0.43$^{+0.07}_{-0.12}$ & 1.33$^{+0.67}_{-0.35}$ & 1.12$^{+0.46}_{-0.29}$ & 0.05$^{+0.21}_{-0.02}$ & 8.7$^{+19.2}_{-6.1}$$\times$$10^{-4}$ & - & 4.5$^{+1.8}_{-*}$ & 0.91$^{+0.16}_{-0.07}$ & 1.18/67  \\
13812      & 0.04$^{+0.10}_{-0.02}$ & 0.33$^{+0.10}_{-0.09}$ & 0.86$^{+0.23}_{-0.13}$ & 1.30$^{+*}_{-0.40}$ & 0.21$^{+1.07}_{-0.15}$ & 53$^{+50}_{-34}$$\times$$10^{-4}$ & - & 5.6$^{+1.5}_{-*}$ & 1.2$^{+0.5}_{-0.1}$ & 0.98/70  \\
13814      & 0.02$^{+0.09}_{-0.01}$ & 0.33$^{+0.10}_{-0.04}$ &   1.13$^{+0.47}_{-0.25}$ &   1.33$^{+0.77}_{-0.42}$ & 0.09$^{+0.06}_{-0.06}$ &   10.9$^{+5.7}_{-8.0}$$\times$$10^{-4}$ &- & 3.9$^{+0.5}_{-0.4}$ & 0.58$^{+0.07}_{-0.05}$ & 0.89/38  \\
0212480801 & 0.03$^{+0.10}_{-0.01}$ & 0.29$^{+0.05}_{-0.09}$ &   1.43$^{+0.63}_{-0.36}$ &   1.09$^{+4.68}_{-0.44}$ & 0.34$^{+2.21}_{-0.19}$ &   7.6$^{+12.7}_{-5.0}$$\times$$10^{-4}$ &- & 6.8$^{+0.8}_{-0.8}$ & 1.2$^{+0.6}_{-0.2}$ & 0.64/50  \\
\hline
\multicolumn{11}{c}{{\it phabs} $*$ {\it phabs} $*$ ({\it diskbb} $+$ {\it bbodyrad} $+$ {\it mekal}) with $kT_{in} > kT_{bb}$}\\
\hline
13813      & 0.02$^{+0.09}_{-0.01}$ & 1.86$^{+0.99}_{-0.46}$ & 0.22$^{+0.03}_{-0.03}$ & 1.27$^{+1.01}_{-0.41}$ & 1.60$^{+3.20}_{-1.26}$$\times$$10^{-4}$ & 0.74$^{+0.43}_{-0.24}$ & - & 5.7$^{+0.4}_{-0.4}$ & 0.96$^{+0.07}_{-0.07}$ & 1.20/67  \\
13812      & 0.02$^{+0.09}_{-0.01}$ & 1.26$^{+0.28}_{-0.19}$ & 0.20$^{+0.03}_{-0.03}$ & 1.34$^{+*}_{-0.40}$ & 8.67$^{+8.83}_{-8.65}$$\times$$10^{-4}$ & 1.21$^{+0.93}_{-0.43}$ & - & 6.7$^{+0.4}_{-0.4}$ & 1.2$^{+0.1}_{-0.1}$ & 0.99/70  \\
13814      & 0.02$^{+0.18}_{-0.01}$ & 1.85$^{+2.35}_{-0.64}$ & 0.18$^{+0.04}_{-0.09}$ & 1.38$^{+1.32}_{-0.35}$ & 1.07$^{+4.13}_{-1.01}$$\times$$10^{-4}$ & 1.04$^{+23.17}_{-0.54}$ & - & 3.8$^{+0.5}_{-0.4}$ & 0.64$^{+0.08}_{-0.07}$ & 1.00/38  \\
0212480801 & 0.02$^{+0.07}_{-0.01}$ & 3.0$^{+10.4}_{-1.3}$ & 0.19$^{+0.02}_{-0.02}$ & 1.32$^{+7.18}_{-0.57}$ & 0.28$^{+1.42}_{-0.27}$$\times$$10^{-4}$ & 2.47$^{+1.48}_{-0.78}$ & - & 6.7$^{+0.8}_{-0.8}$ & 1.0$^{+0.1}_{-0.1}$ & 0.66/50  \\
\hline
\multicolumn{11}{c}{{\it phabs} $*$ {\it phabs} $*$ ({\it simpl} $*$ {\it diskbb} $+$ {\it mekal})}\\
\hline
13813      & 0.19$^{+0.22}_{-0.17}$ & 0.17$^{+0.18}_{-0.06}$ & - & 1.18$^{+0.82}_{-0.30}$ & 6.77$^{+*}_{-6.66}$ & -& 1.90$^{+0.27}_{-0.72}$ & 5.7$^{+0.6}_{-0.4}$ & 0.83$^{+2.24}_{-0.37}$ & 1.12/67  \\
13812      & 0.12$^{+0.19}_{-0.07}$ & 0.15$^{+0.16}_{-0.08}$ & - & 1.28$^{+*}_{-0.34}$ & 10.49$^{+*}_{-10.24}$ & -& 2.22$^{+0.23}_{-0.29}$ & 6.8$^{+0.5}_{-0.4}$ & 0.86$^{+1.66}_{-0.24}$ & 0.98/70  \\
13814      & 0.05$^{+0.19}_{-0.03}$ & 0.08$^{+0.14}_{-0.06}$ &  - &   1.07$^{+0.29}_{-0.23}$ & [unconstrained]  & - & 1.88$^{+0.34}_{-0.26}$ & 4.1$^{+0.6}_{-0.5}$ &
0.41$^{+0.80}_{-0.08}$ & 0.82/38  \\
0212480801 & 0.05$^{+0.13}_{-0.03}$ & 0.25$^{+0.07}_{-0.08}$ & - & 1.10$^{+1.90}_{-0.42}$ & 0.93$^{+9.85}_{-0.63}$ & -& 1.19$^{+0.57}_{-0.18}$ & 6.8$^{+0.7}_{-0.8}$ & 0.63$^{+0.51}_{-0.12}$ & 0.67/50  \\
\hline
\multicolumn{11}{c}{{\it phabs} $*$ {\it phabs} $*$ ({\it bbodyrad} $+$ {\it bremss} $+$ {\it mekal})}\\
\hline
13813      & 0.02$^{+0.17}_{-0.01}$ & $> 5.2$ & 0.22$^{+0.03}_{-0.05}$ & 1.18$^{+0.55}_{-0.32}$ & - & 0.54$^{+0.46}_{-0.25}$ & - & 5.9$^{+0.4}_{-0.4}$ & 0.47$^{+0.03}_{-0.03}$ & 1.15/67  \\
13812      & 0.04$^{+0.14}_{-0.01}$ & 3.8$^{+2.4}_{-1.0}$ & 0.19$^{+0.05}_{-0.07}$ & 1.29$^{+5.97}_{-0.38}$ & - & 1.11$^{+26.08}_{-0.86}$ & - & 6.8$^{+0.5}_{-0.5}$ & 0.61$^{+0.38}_{-0.05}$ & 0.97/70  \\
13814      & 0.02$^{+0.20}_{-0.01}$ & 6.0$^{+11.1}_{-2.5}$ & 0.09$^{+0.11}_{-0.03}$ & 1.08$^{+0.35}_{-0.20}$ & - & 26.4$^{+220.9}_{-25.7}$ & - & 4.1$^{+0.5}_{-0.4}$ & 0.33$^{+0.04}_{-0.04}$ & 0.89/38  \\
0212480801 & 0.02$^{+0.09}_{-0.01}$ & $> 7.1$              & 0.19$^{+0.02}_{-0.03}$ & 1.27$^{+2.13}_{-0.55}$ & - & 2.4$^{+1.6}_{-0.8}$ & - & 6.8$^{+0.7}_{-0.7}$ & 0.55$^{+0.05}_{-0.06}$ & 0.66/51  \\
\hline
\end{tabular}
\end{center}
\begin{flushleft}
{$^a$}$n_{H}$ is the total column density, including the fixed Galactic foreground absorption ($1.57 \times 10^{20} $cm$^{-2}$) and the local component derived from spectral fitting.\\
$^b$For the {\it diskbb} and {\it diskpbb} models, $kT$ is the colour temperature at the inner radius ($kT_{in}$); for the {\it bremss} model, $kT$ is the plasma temperature.\\
$^c${\it diskbb} and {\it diskpbb} normalizations are in units of $(r_{\rm in}/{\rm km})^2\cos{\theta}\,(d/10\,{\rm kpc})^{-2}$, where $r_{\rm in}$ is the apparent inner-disk radius.\\
$^d$The {\it bbodyrad} normalization is in units of $(r/{\rm km})^2(d/10\,{\rm kpc})^{-2}$, where $r$ is the source radius.\\
{$^e$}$f_{\rm X}$ is the observed flux (not corrected for absorption).\\
{$^f$}$L_{\rm X}$ is the emitted luminosity (corrected for absorption). 
$L_{\rm X} \equiv 2\pi \, d^2 \, f^{\rm em}_{\rm X}/\cos \theta$ for the {\it diskbb} and {\it diskpbb} components of the models, and $L_{\rm X} \equiv 4\pi \, d^2 \, f^{\rm em}_{\rm X}$ for the other model components. We assumed $\theta = 80^{\circ}$.\\
\end{flushleft}
\end{table*}

\subsection{Spectral models for S2}
\label{sfits2.sec}

Unlike for S1, the spectra of S2 have significant emission $>$ 2 keV. First, we tried single-component models: power-law and disk-blackbody; for the disk model, we tried both a standard disk and a $p$-free disk, the latter choice justified by the near-Eddington regime \citep{Sutton2017}. The power law is a good model (Table \ref{xspecS2.tab}) for the {\it Chandra} spectra (with very low sensitivity below 0.5 keV and above 7 keV) but does not account for the broad-band curvature detected only in the {\it XMM-Newton} spectrum, because of its larger band coverage. Instead, both disk models fit poorly, for the opposite reason---too much curvature. Predictably, the $p=0.6$ model is a better fit than the standard disk ($p=0.75$) because it has a broader shape.
Both the power-law and the disk model show significant residuals around 1 keV; such residuals are well accounted for by an additional thermal-plasma component ({\it mekal} model). Once the thermal plasma emission is included, the best-fitting power-law photon index is $\approx$2.0--2.3 for all four epochs.

Next, we tried a set of two-component models (Table \ref{xspecS2.tab}): {\it diskbb} $+$ {\it powerlaw} (standard phenomenological models of XRBs); a Comptonization model, {\it simpl} $*$ {\it diskbb} \citep{Steiner2009}; a double-thermal model, {\it diskbb} $+$ {\it bbodyrad}; and a thermal model with a bremsstrahlung model, {\it bbodyrad} $+$ {\it bremss}. In all four cases, we also added a {\it mekal} component to account for the line emission around 1 keV.
Physically, the power-law plus thermal component model, and the Comptonization model, are applicable to a variety of physical scenarios for BH or NS accretors (especially at the low resolution of CCD spectra): either lower-temperature thermal emission from the disk, up-scattered in a hotter corona; or direct emission of hard X-ray photons, partly down-scattered in a cooler outflow. The double-thermal model may represent the two-component emission from the inner disk and the surface or boundary layer of a NS (in which case we expect the disk component to be cooler than the surface blackbody component); it may also represent the thermal emission from a disk plus that from the photosphere of a dense outflow (in which case we expect the disk emission to be hotter than the down-scattered blackbody emission).
We find that all those two-component models provide a better fit (an improvement $|\Delta \chi^2_{\nu}| \approx 11$--12 for the loss of two degrees of freedom) to the {\it XMM-Newton} spectrum than the power-law model; however, they are equivalent to each other, so we cannot rule out any physical scenario, or distinguish between a BH or NS accretor, from spectral fitting. For the {\it Chandra} spectra, as noted above, there is no advantage of multi-component models over a simple power-law. In the rest of this Section, we briefly examine the best-fitting values of the main parameters for the alternative models, to determine whether they are physically plausible and whether they resemble typical spectral parameters of other ULXs and XRBs.

\begin{figure*}
\center
\includegraphics[width=0.49\textwidth]{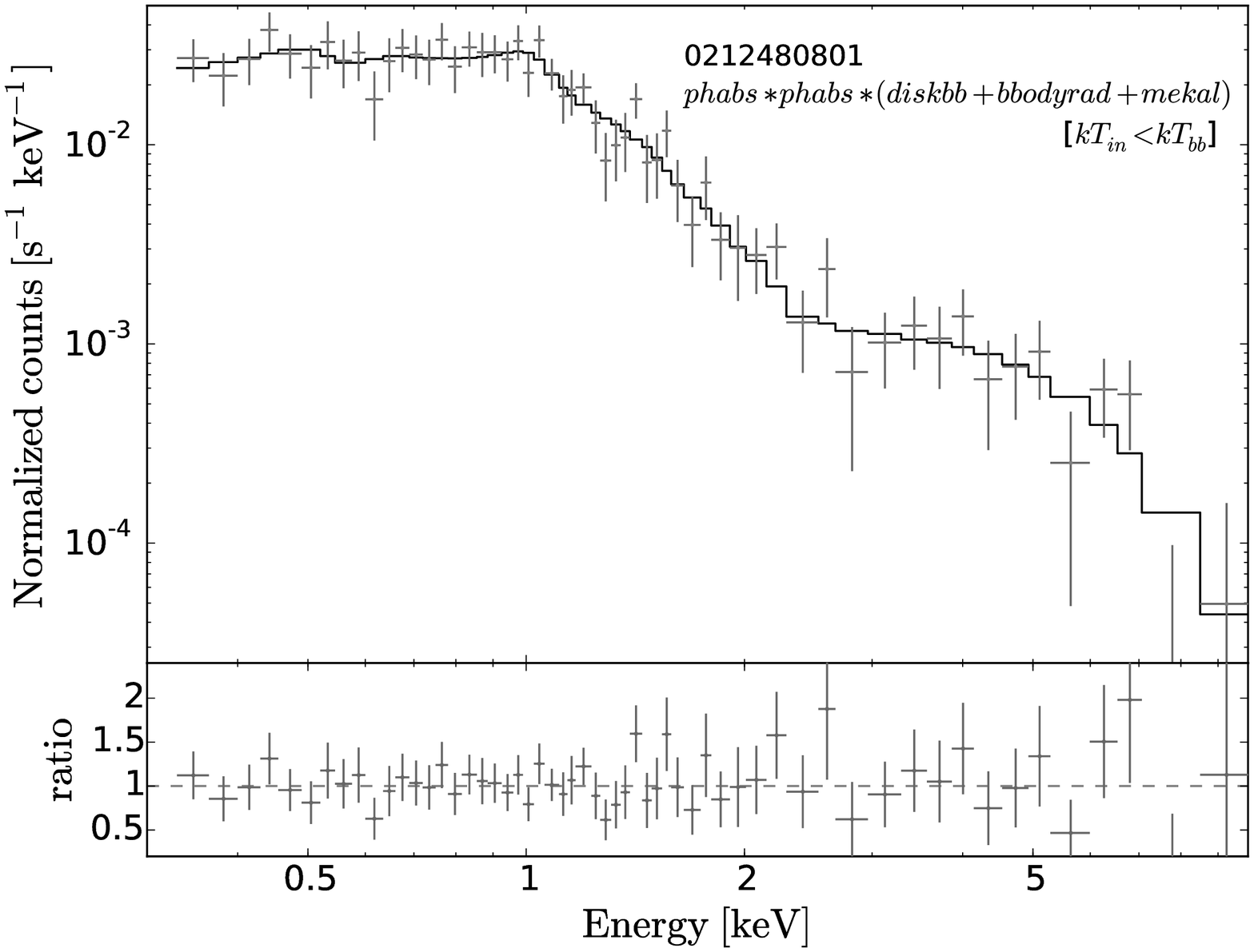}
\includegraphics[width=0.49\textwidth]{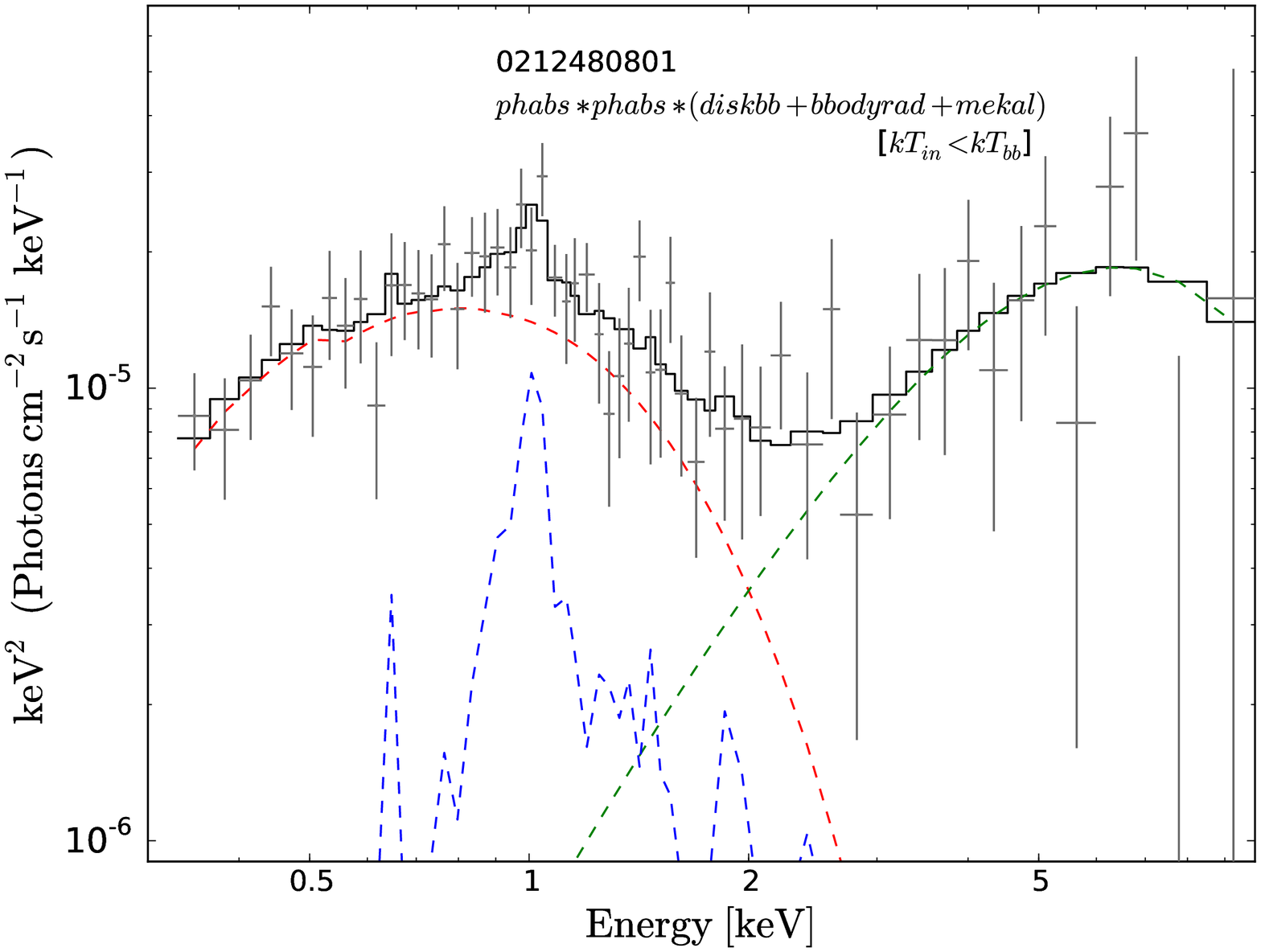}
\caption{Left panel: best-fitting double-thermal (blackbody plus disk-blackbody; $kT_{bb} > kT_{in}$) model to the spectrum of S2 from {\it XMM-Newton} Obsid 0212480801; a thermal plasma component is also present in the model, to account for systematic residuals around 1 keV. Datapoints and model values are plotted in the top sub-panel; data/model ratios in the bottom sub-panel. Error bars denote 68.3\% uncertainties. See Table \ref{xspecS2.tab} and Section \ref{sfits2.sec} for the fit parameters and their interpretation.
Right panel: unfolded spectrum, assuming the best-fitting double-thermal model.
Red, green, and blue dashed lines represent the {\it diskbb}, {\it bbodyrad}, and {\it mekal} components, respectively; the solid black histogram represents the combined model.}
\label{specS2.fig}
\end{figure*}

\begin{figure*}
\center
\includegraphics[width=0.49\textwidth]{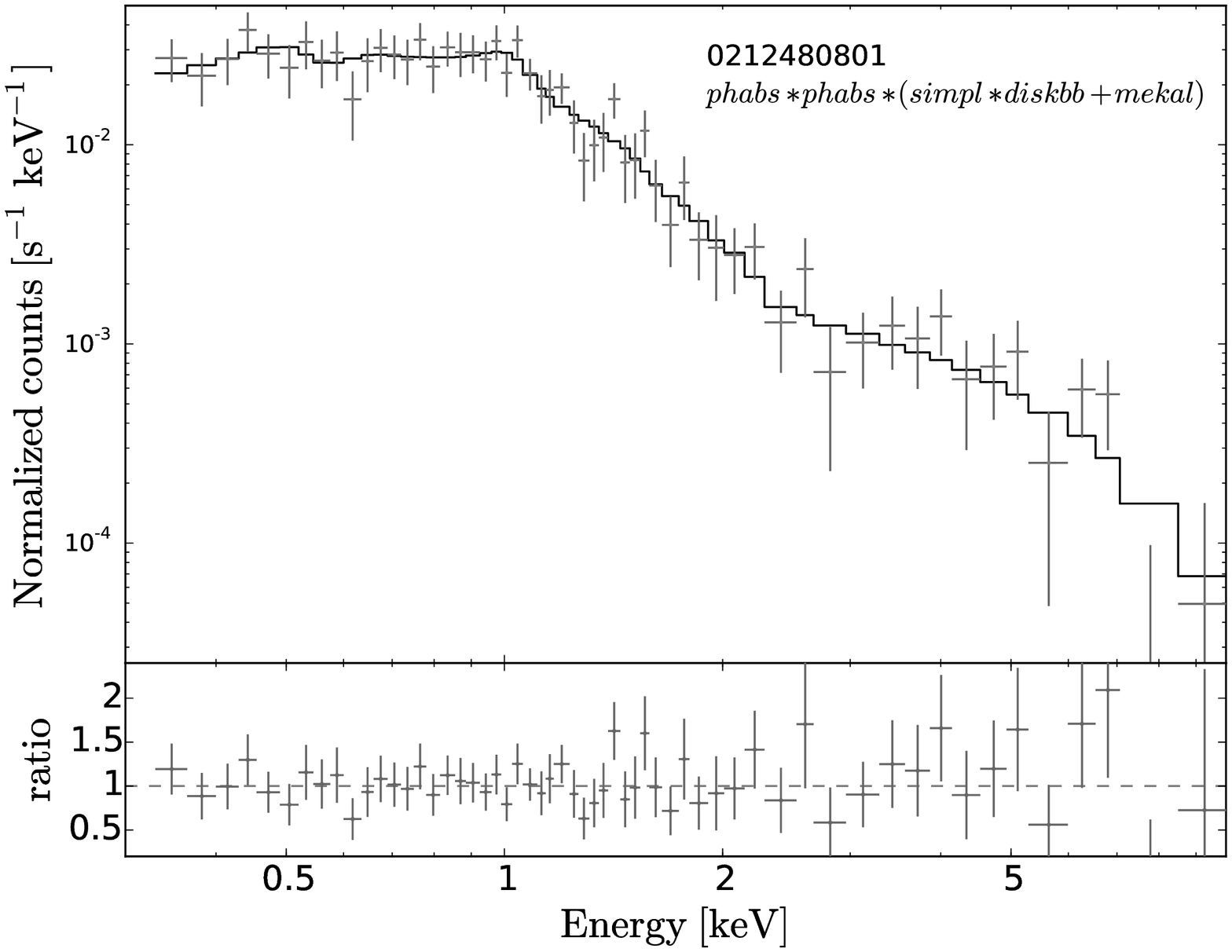}
\includegraphics[width=0.49\textwidth]{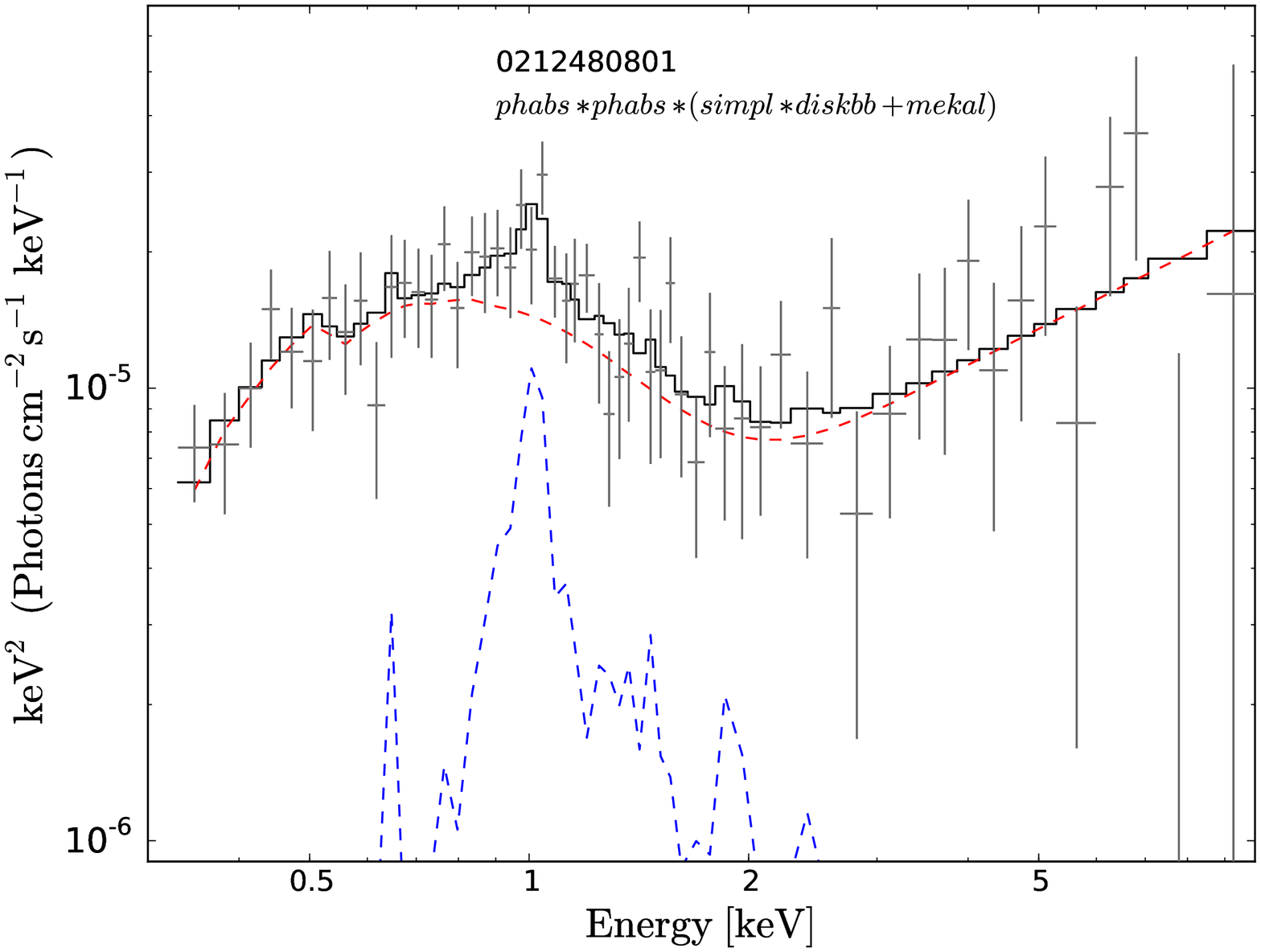}
\caption{Left panel: as in the left panel of Figure \ref{specS2.fig}, for the best-fitting Comptonization model plus thermal plasma (see Table \ref{xspecS2.tab} and Section \ref{sfits2.sec}).
Right panel: unfolded spectrum, assuming the best-fitting Comptonization model.
Red and blue dashed lines represent the {\it simpl*diskbb} and {\it mekal} components, respectively; the solid black histogram represents the combined model.}
\label{specS2b.fig}
\end{figure*}

\begin{figure*}
\center
\includegraphics[width=0.32\textwidth]{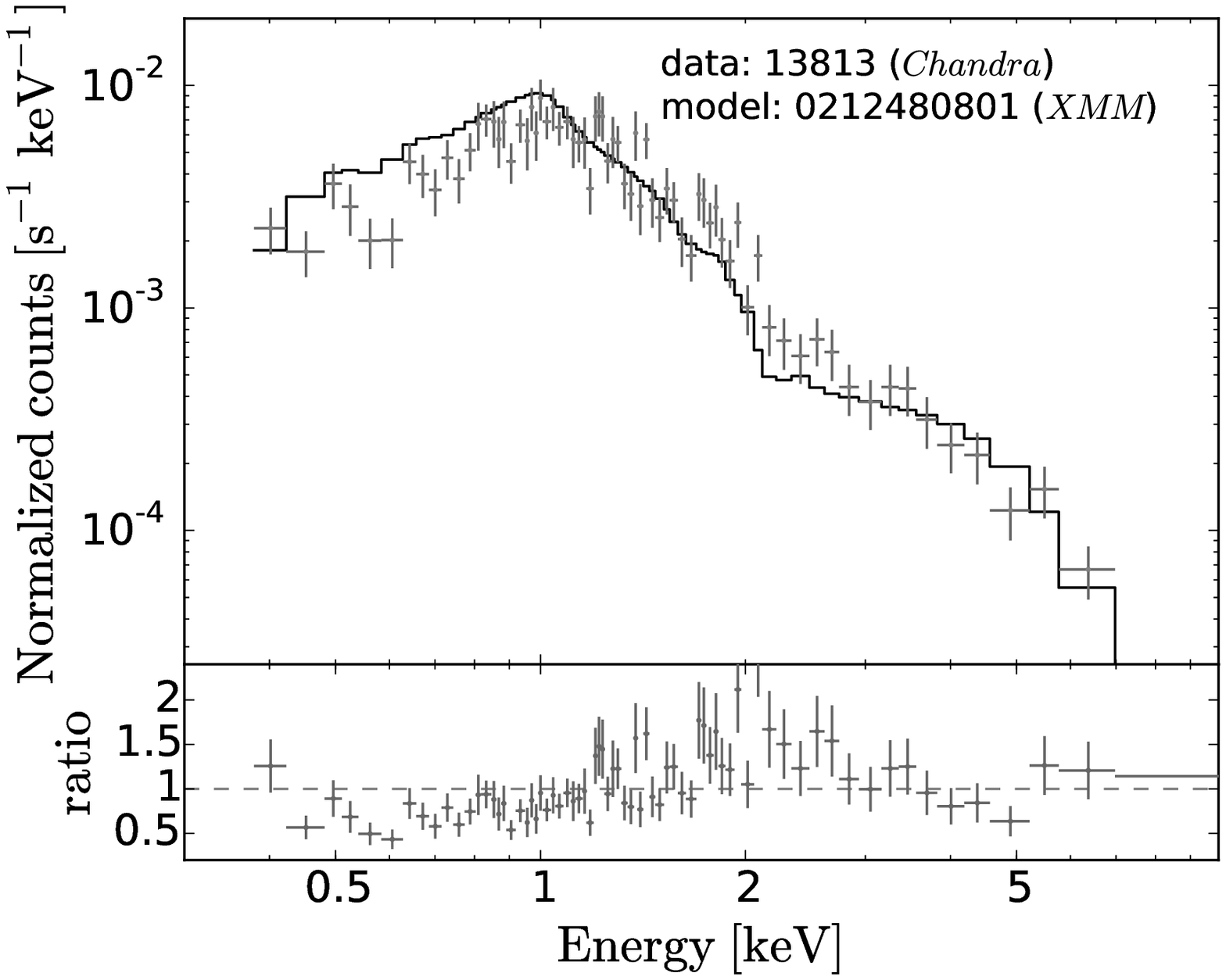}
\includegraphics[width=0.32\textwidth]{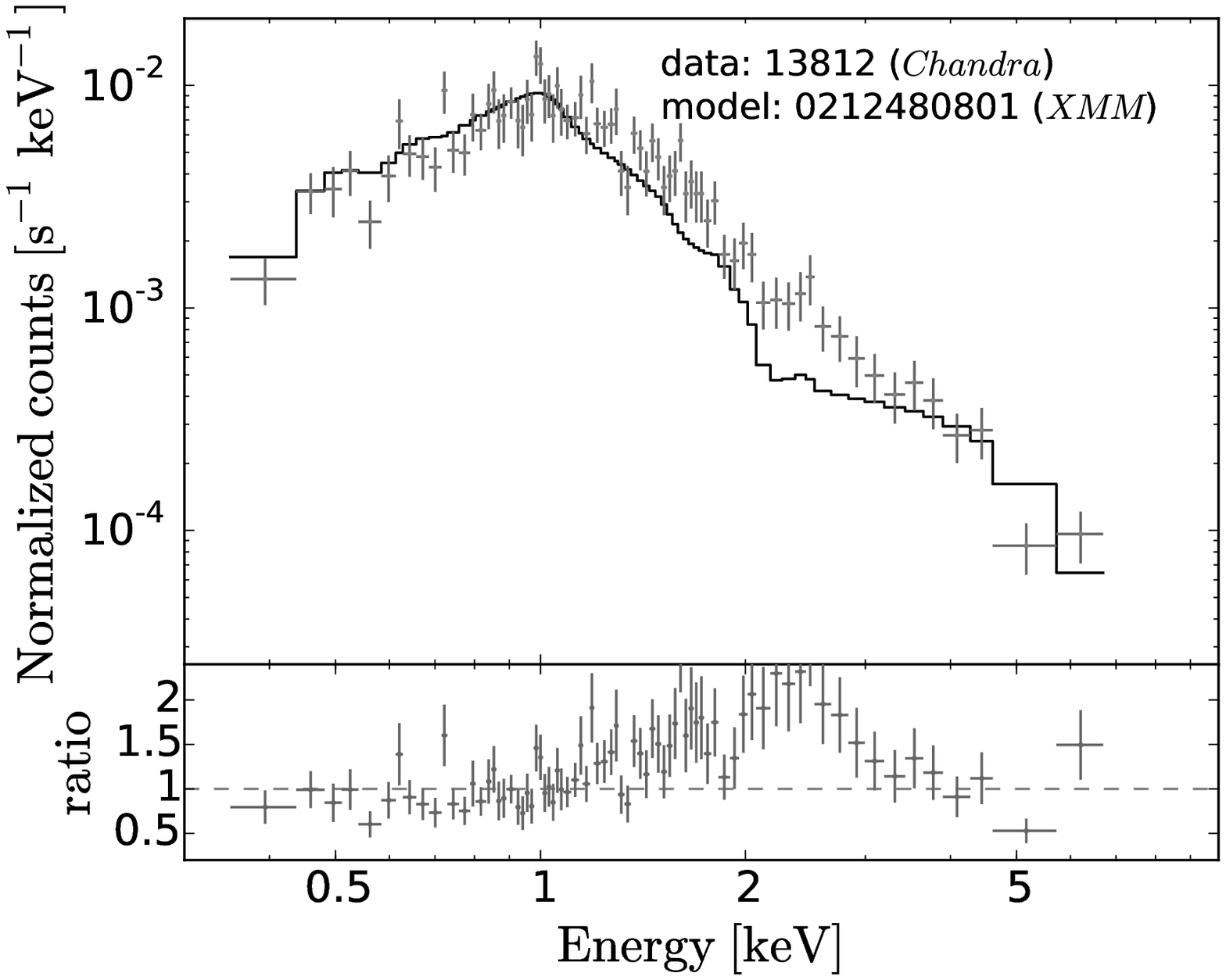}
\includegraphics[width=0.32\textwidth]{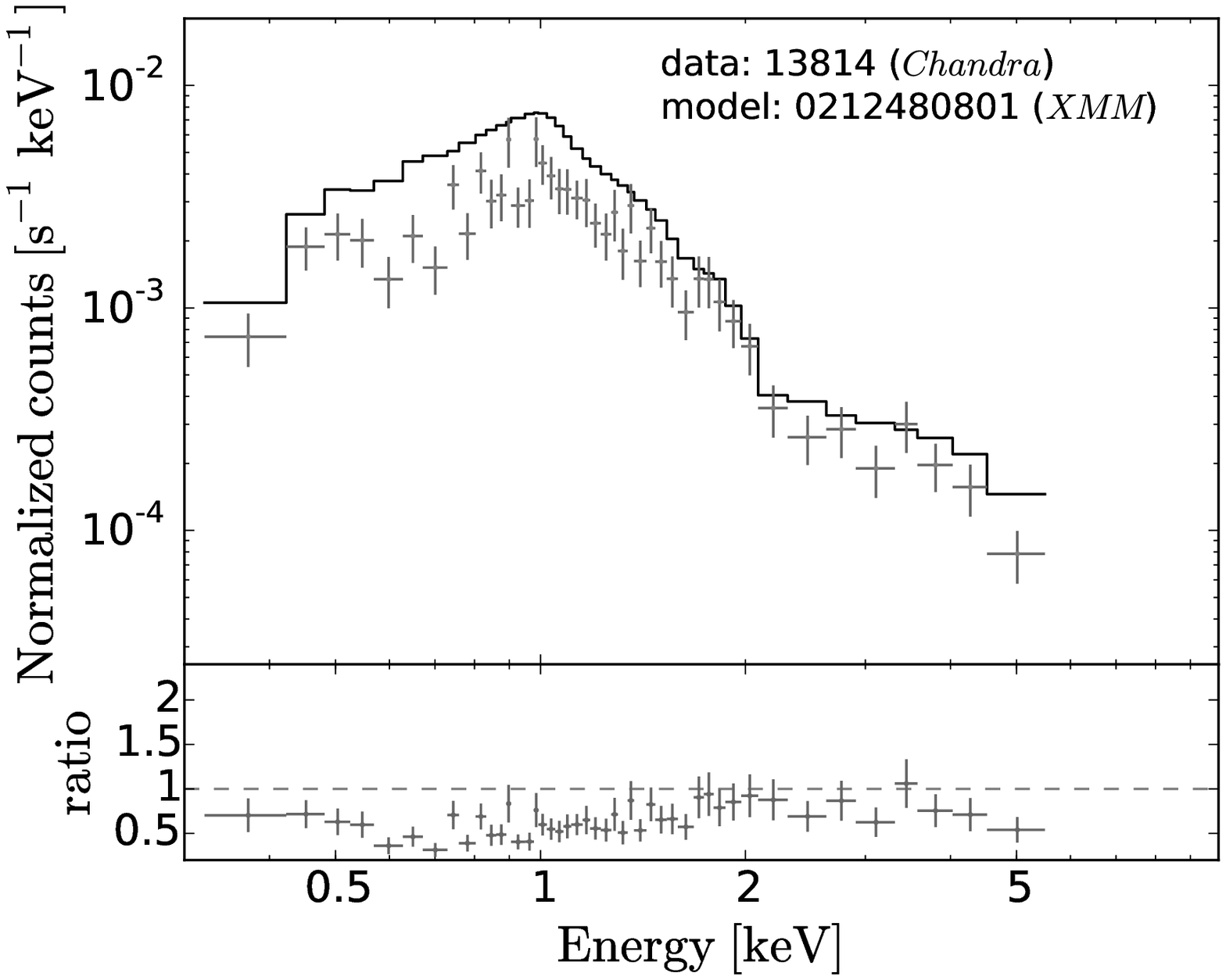}
\caption{Left panel: comparison between the best-fitting double-thermal model to the {\it XMM-Newton} spectrum of S2 from 2005 July 1 and the datapoints from {\it Chandra} ObsID 13813 (2012 September 9).
Central panel: as in the left panel, for {\it Chandra} ObsID 13812 (2012 September 12).
Right panel: as in the left panel, for {\it Chandra} ObsID 13814 (2012 September 20).
}
\label{specS2c.fig}
\end{figure*}

First, we consider the best-fitting double-thermal model with $kT_{\rm {bb}} > kT_{\rm {in}}$ (Figure \ref{specS2.fig}), which is more suited to a NS accretor. The characteristic size of the blackbody emitter is $r_{\rm {bb}} = 22^{+14}_{-9}$ km (90\% confidence level), and its temperature is $T_{\rm{bb}} = 1.4^{+0.6}_{-0.4}$ keV, in the {\it XMM-Newton} spectrum (Table \ref{xspecS2.tab}); both values are consistent with thermal emission from the surface of a NS or from a boundary layer between disk and surface. The best-fitting peak colour temperature of the {\it diskbb} component is $\approx$0.3 keV. The apparent inner-disk radius is $r_{\rm in} = 470^{+810}_{-160} \, (\cos \theta)^{-1/2}$ km; the physical inner disk radius $R_{\rm {in}}$ is generally estimated as $R_{\rm in} \approx 1.19 r_{\rm in}$ \citep{Kubota1998} for a standard thin disk. In this scenario, the inner disk is truncated very far from the innermost stable circular orbit (or from the NS surface), which inevitably reduces the radiative efficiency $\eta$ of the disk to $\eta = GM/(2R_{\rm in}c^2) \sim 10^{-3}$, two orders of magnitude lower than the efficiency of the boundary layer/surface emission. If so, it becomes difficult to explain why the disk and the surface emission contribute a similar luminosity ($\approx$ a few $10^{38}$ erg s$^{-1}$) to the X-ray spectrum, when we would expect the X-ray emission from the disk to be negligible compared with the total emission. For this reason, we conclude that the simplistic interpretation of the double thermal model as ``standard disk plus surface emission from the NS'' is physically not self-consistent.

Then, we consider the other double-thermal model ($kT_{bb} < kT_{in}$), which is equally applicable to BHs or weakly magnetized NSs.
The best-fitting colour temperature of the {\it diskbb} component is $\approx$3 keV, and the apparent inner-disk radius is $r_{\rm in} \sqrt{\cos \theta} = 4.2^{+7.0}_{-3.5}$ km; for example, for $i = 80^{\circ}$, $r_{\rm in} = 24^{+40}_{-20}$ km. In this case, it makes no sense to use the standard-disk conversion of $R_{\rm in} \approx 1.2 r_{\rm in}$, because a disk with a peak temperature of $\approx$3 keV is not consistent with a standard, sub-Eddington thin disk. In other words, the {\it diskbb} model is not self-consistent, and the only information we should take away from this fit is that the hottest thermal component is consistent with an emitting region close to the innermost stable circular orbit or to the surface of a weakly magnetized NS ($B \la 10^{10}$ G), at near-Eddington mass accretion rates \citep{Takahashi2017a}. The characteristic radius of the second (cooler) thermal component ({\it bbodyrad}) is $r_{\rm bb} \approx 1300^{+300}_{-200}$ km. Thus, if this second component comes from reprocessing of hotter photons, its size suggests the photosphere of a thick disk outflow, analogous to the explanation invoked for ULSs \citep{Soria2016}, or the walls of a polar funnel. However, a geometrically and optically thick outflow should prevent our direct view of the innermost disk, if we are looking at a high-inclination system (as suggested by the presence of long eclipses.
Thus, the co-existence of two optically thick thermal components with similar intensities but different temperature and radii requires a degree of misalignment, so that we see at the same time part of the direct inner-disk emission, and reprocessed emission for example from the outflow funnel walls.
Alternatively, a double thermal model may be taken as a purely phenomenological way to approximate more complex spectra of super-critical flows (including both thermal and bulk-motion Comptonization), such as those predicted by the MHD simulations of \citet{Kitaki2017} for super-critical BH accretion, or the accretion disk and magnetospheric envelope structure proposed by \citet{Mushtukov2017} for super-critical NS accretion.

We then used the $simpl*diskbb$ model (Figure \ref{specS2b.fig}) to represent Comptonized emission of an input disk spectrum \citep{Steiner2009}. For the {\it XMM-Newton} spectrum, the best-fitting parameters (Table \ref{xspecS2.tab}) include: a power-law photon index $\Gamma = 1.19^{+0.57}_{-0.18}$; a fraction of scattered photons of $0.21^{+0.19}_{-0.07}$; for the seed photon component, an inner-disk colour temperature $T_{\rm in} = 0.25^{+0.07}_{-0.08}$ keV and a characteristic inner-disk radius $r_{\rm in} \sqrt{\cos \theta} \approx 700$ km. We also tried other Comptonization models, for example {\it diskir} and {\it comptt}, which offer a more physical treatment of the Comptonized component when the electron temperature in the scattering region is $\la$ a few keV; for all those models, we obtained a similar size and temperature of the seed thermal component and a similar power-law photon index. Such parameters are typical of ULX spectra \citep{Gladstone2009,Mukherjee2015,Sutton2015}. 

We also searched for what is perhaps the main feature that identifies ULX spectra: the presence in most of those sources of a spectral rollover or downturn at energies around 5--10 keV ({\it e.g.}, \citealt{Gladstone2009,Sutton2013b,Bachetti2016,Brightman2016,Kaaret2017,Walton2018}); 
however, in our case, the small number of counts and the limited energy band of {\it Chandra} makes it impossible to confirm or exclude the spectral break.


Finally, the hard X-ray component can also be fitted with a bremsstrahlung model (Table \ref{xspecS2.tab}), with a hot gas temperature $\ga$ a few keV (not well constrained because of the limited band coverage of our spectra). Bremsstrahlung emission from hot, shocked gas may come from magnetically confined accretion columns above the surface of a NS \citep{Basko1976,Mushtukov2015}, while the lower-temperature thermal component may come from the truncated accretion disk or the magnetospheric curtain near the magnetospheric radius.

Regardless of the choice of continuum models, there are significant soft X-ray residuals around 1 keV in both the {\it Chandra} and {\it XMM-Newton} spectra of S2; they are well modelled with a {\it mekal} component (Table \ref{xspecS2.tab}). Such residuals are analogous to those seen in CCD-resolution spectra of several ULXs \citep{Middleton2015,Sutton2015,Feng2016}. Grating-resolution spectra of a few of those ULXs showed \citep{Pinto2016,Pinto2017,Kosec2018} that such features are absorption and emission lines from fast, thick, radiatively driven outflows, a hallmark of super-Eddington accretion. Instead, such residuals are usually absent from the spectra of sub-Eddington BH XRBs.



As already discussed in Section \ref{sfits1.sec}, we calculated the unabsorbed X-ray luminosity of disk-like emission components as $2\pi\,d^2/(\cos \theta)$ times the absorption-corrected flux, and the unabsorbed luminosity of the other components (including the {\it mekal} emission) as $4\pi\,d^2$ times the absorption-corrected flux (Table \ref{xspecS2.tab}). This is standard practice in the study of sub-Eddington XRBs, where the power-law emission comes from a geometrically thick corona and is assumed to be approximately isotropic. For super-critical accretion, this is probably not the case: if the power-law component represents the harder X-ray photons, it is preferentially emitted along the polar direction, while softer photons emerge at higher inclinations. In the absence of strong constraints for the geometry of the accretion flow, the intrinsic luminosities listed in Table \ref{xspecS2.tab} must be taken as rough estimates. We find characteristic 0.3--8 keV luminosities  $L_{\rm X} \sim (1$--$2) \times 10^{39}$ erg s$^{-1}$ if we use the disk-like geometric collimation $\propto 1/\cos \theta$ (assuming $\theta \approx 80^{\circ}$ from the presence of long eclipses), and $L_{\rm X} \sim (0.5$--$1) \times 10^{39}$ erg s$^{-1}$ for perfectly isotropic emission, over the four epochs studied in details (corresponding to the brightest level observed for S2).  The isotropic contribution of the thermal-plasma component $L_{\rm X,mekal} \approx$ (3--5) $\times 10^{37}$ erg s$^{-1}$.

In addition, we directly compared the shape of the spectrum in the four observations, to illustrate the relative change. We have overplotted (Figure \ref{specS2c.fig}) the best-fitting model to the {\it XMM-Newton} spectrum from 2005 July 1 onto the datapoints for the three {\it Chandra} spectra extracted during the 2012 September outburst (see also Figure 3, bottom right panel). The main difference is that the 2012 spectra have a factor-of-2 excess of flux in the $\approx$1--3 keV band, relative to the flux at lower and higher energies, even when the integrated 0.3--8 keV flux is the same.


In summary, the outburst luminosity of S2 in 2005 and 2012 peaks at $\approx$10$^{39}$ erg s$^{-1}$. Its spectral properties are different from the simple canonical states of stellar-mass BHs expected at near-Eddington luminosities \citep{Remillard2006}, that is the top of the high/soft state or the steep power-law state. The two-component nature of the best-fitting models, and the presence of line features around 1 keV, point to a system in the super-Eddington regime (even though we cannot say anything about the presence of absence of a high-energy rollover). There is no reason to assume that the compact object is a stellar-mass BH (which would barely reach the Eddington limit at such moderate luminosities). Instead, it could be a NS reaching a peak luminosity of several times Eddington in the two recorded outbursts, while hovering around $L_{\rm Edd} \approx 2 \times 10^{38}$ erg s$^{-1}$ in most of the other {\it Chandra} and {\it XMM-Newton} observations (Figure 3, bottom right panel). We will now show how optical studies discriminate between the BH and NS scenario.

\section{Optical imaging of the S2 field}
\label{hst.sec}

Knowing the orbital period and the eclipse fraction
gives us an exceptional opportunity to constrain the binary system parameters of S2, using only photometric data (following the method of \citealt{Porquet2005}); if the properties of the donor star are also constrained from optical photometry, we can then constrain the mass of the accreting object.

\begin{figure*}
\center
\includegraphics[width=0.48\textwidth]{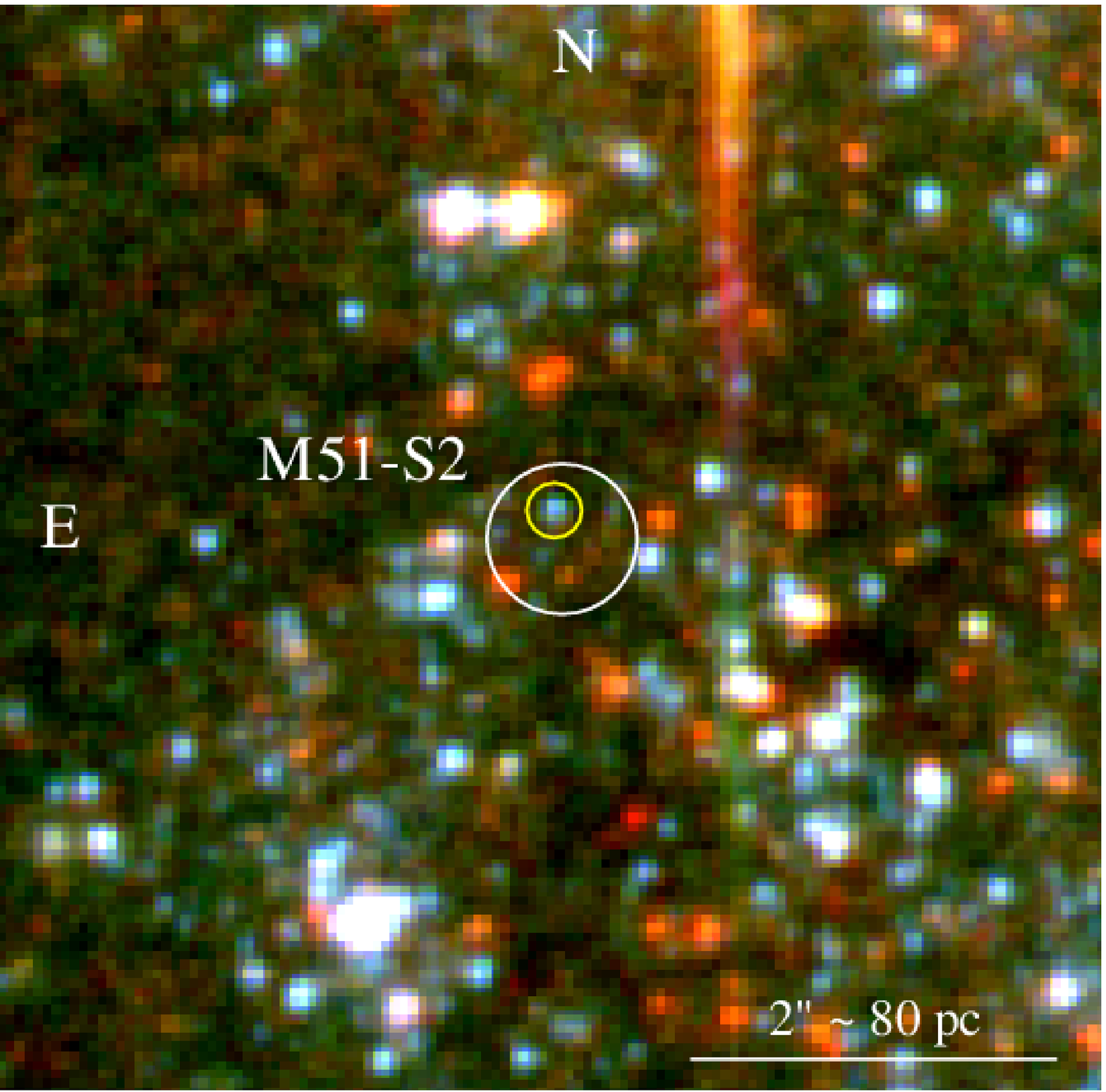}
\includegraphics[width=0.48\textwidth]{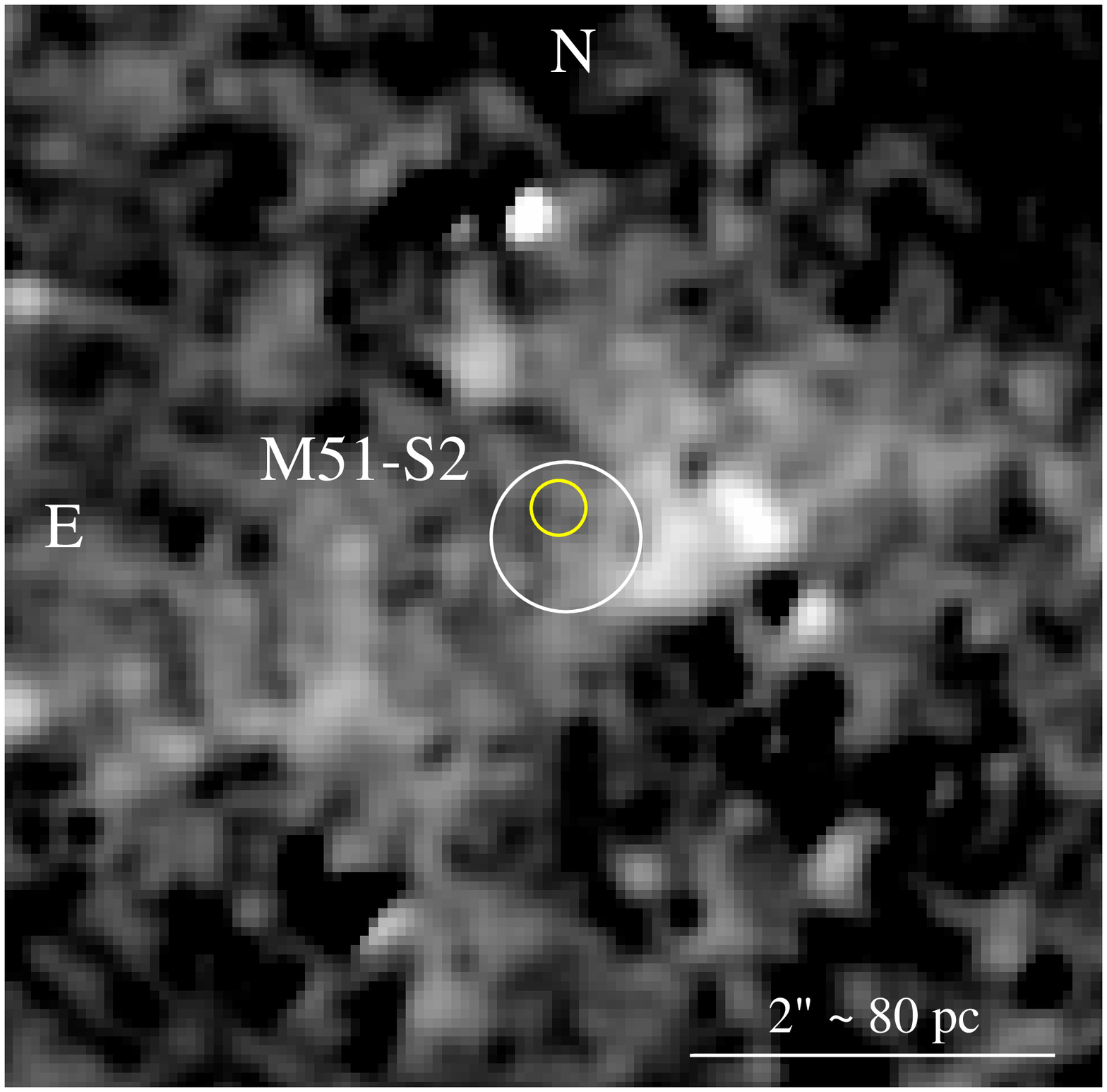}
\caption{Left panel: true-colour optical image of the stellar field around S2, using {\it HST}/ACS-WFC images in the F435W filter (blue), F555W filter (green), and F814W filter (red).
The white circle is the astrometry-corrected position of the X-ray source (error radius $\approx 0''.4$). The small yellow circle marks the blue point-like source that we regard as the most likely counterpart of S2.
Right panel: continuum-subtracted, smoothed image in the F658W filter (H$\alpha$ plus [N {\sc ii}] emission) for the same field of view; it does not show any emission-line enhancement at the location of the X-ray source.}
\label{S2ds9.fig}
\end{figure*}

To localize and identify the optical counterpart of S2, we used several X-ray/optical coincidences between point-like sources in {\it Chandra}, {\it HST} and Sloan Digital Sky Survey (SDSS) observations \citep{Wang2015}.
For the preliminary X-ray astrometry, we used the {\it Chandra}/ACIS image from Obsid 13812, in which S2 has the highest number of counts.
{\it HST} images of the field were taken on Jan 18--22, 2008 with the Advanced Camera for Surveys (ACS), Wide Field Camera (WFC), in several filters; for the purpose of astrometric alignment, we used the drizzled images, produced by the standard pipeline calibration, in the F814W filter (Dataset j97c61hcq, exposure time 340s).
%
We also used images from the SDSS Data Release 12 as an intermediate step, to match the optical and X-ray positions of more sources. Although the ACS-WFC has a relatively large field of view ($202''\times202''$) for an {\it HST} instrument, it is still hard to find matches with strong X-ray sources for a direct registration of {\it Chandra} and {\it HST} images. Thus, first we matched the {\it Chandra} and SDSS astrometry using five X-ray/optical point-like sources, and then we matched the SDSS and {\it HST} astrometry using twelve common optical sources.
We estimate that the relative offset between the original astrometry of the public-archive {\it Chandra} and {\it HST} images is $\Delta {\rm RA} = 0''.32 \pm 0''.14$ and $\Delta {\rm Dec} = -0''.39 \pm 0''.36$ near the position of S2. 
We show the corrected position of S2 on the {\it HST} image in Figure \ref{S2ds9.fig}; the error circle (at the 90\% confidence level) has a radius of $0''.4$, which is the quadratic sum of the {\it Chandra} position uncertainty of S2 and the random scatter between {\it Chandra} and {\it HST} positions of the common sources after the systematic offset has been removed.

We double-checked the X-ray optical alignment with another independent test. We used three radio/X-ray associations identified by \cite{Rampadarath2015} (listed in their Table 4) to correct the absolute astrometry of the {\it Chandra} images. We then used the Gaia Data Release 1 source catalogue \citep{Gaia2016} to correct the absolute astrometry of the {\it HST} images. This method gave us essentially the same result in terms of X-ray/optical offset and uncertainty as the other method.

In summary, none of the $\approx$10 reliable X-ray/optical and X-ray/radio associations available for astrometric registration has a discrepancy $>$0$''.4$ from the best-fitting astrometric solution. Thus, we argue that the optical counterpart of S2 is located inside that circle, with a conservative confidence level of at least 90\% (limited only by the small number of reliable test associations). We shall then assume that the most likely counterpart is the brightest star inside that circle. There are two possible objections to this assumption. The first possibility is that the true counterpart of S2 could be a fainter optical source also inside the error circle. In that case, the arguments about the donor star mass that we shall discuss in Section \ref{fraction.sec} will still stand: in fact, in that case the upper limit to the mass of the donor star (and, hence, to the mass of the compact object) will be even more constraining. The second possibility is that the true counterpart is the bright blue star located just at the outside edge (to the west) of the error circle (Figure~\ref{S2ds9.fig}); that star is slightly brighter than our assumed counterpart inside the circle. Although we consider that association very unlikely, we also briefly mention the corresponding mass limits for that counterpart, at the end of Section 7.1.

We performed aperture photometry with standard {\sc DAOPHOT} tasks implemented in the Image Reduction and Analysis Facility ({\sc IRAF}) software package \citep{Tody1993}.
For the assumed optical counterpart, we used a small aperture radius ($0''.125$) to avoid contamination from nearby objects in this crowded region.
For the local background, we used an annulus with inner radius of $0''.175$ and outer radius of $0''.325$ around the source. We then used point-spread function models from Tiny Tim
\footnote{http://tinytim.stsci.edu/cgi-bin/tinytimweb.cgi.}
to determine and apply the aperture correction from an aperture of radius $0''.125$ to one with a radius of $0''.5$.
Finally, we corrected the net count rates from a $0''.5$ aperture radius to an ``infinite'' aperture radius using the tables by \citet{Sirianni2005}.
We converted the infinite-aperture count rates to Vega magnitudes using the online tables of zeropoints for ACS-WFC (see also \citealt{Sirianni2005}).
We summarize our results in Table \ref{S2photo.tab}.\\

\begin{table}
\begin{center}
\caption[]{{\it HST} ACS/WFC Observations and Brightness of the Optical Counterpart of S2.}
\label{S2photo.tab}
\begin{tabular}{ccccc}
\noalign{\smallskip}\hline\noalign{\smallskip}
Data Set  &  MJD & Exp. Time &   Filter &  $m_{\rm Filter}$    \\
          &      &  (s)      &            &       (mag)       \\[5pt]
\hline\noalign{\smallskip}
j97c61h9q &  53388.778 &    680 &  F435W &          24.59$\pm$0.06    \\
j97c62xzq &  53392.909 &    680 &  F435W &          24.65$\pm$0.06    \\
j97c63rbq &  53390.844 &    680 &  F435W &          24.77$\pm$0.08     \\
j97c64rjq &  53390.910 &    680 &  F435W &          24.59$\pm$0.06    \\
j97c61haq &  53388.788 &    340 &  F555W &          24.83$\pm$0.10    \\
j97c62y0q &  53392.919 &    340 &  F555W &          24.77$\pm$0.09    \\
j97c63rcq &  53390.854 &    340 &  F555W &          24.83$\pm$0.09    \\
j97c64rkq &  53390.920 &    340 &  F555W &          24.94$\pm$0.11    \\
j97c61hcq &  53388.794 &    340 &  F814W &          25.00$\pm$0.23    \\
j97c62y2q &  53392.925 &    340 &  F814W &          25.23$\pm$0.31    \\
j97c63req &  53390.859 &    340 &  F814W &          24.87$\pm$0.19    \\
j97c64rmq &  53390.926 &    340 &  F814W &          25.24$\pm$0.30    \\
\noalign{\smallskip}\hline
\end{tabular}
\end{center}
\end{table}

The average apparent brightness of the candidate optical counterpart over the 4 exposures in each filter is $m_{435} = 24.64$ mag, $m_{555} = 24.84$ mag, $m_{814} = 25.08$ mag. Converting to standard Johnson-Cousins photometric system \citep{Sirianni2005}, correcting for a line-of-sight reddening $E(B-V) \approx 0.030$ \citep{Schlafly2011}, and assuming a distance modulus of 29.50 mag to M\,51\footnote{From the NASA/IPAC Extragalactic Database.}, we obtain an intrinsic absolute brightness $M_B \approx -5.0 \pm 0.1$ mag, $M_V \approx -4.8 \pm 0.2$ mag, $M_I \approx -4.5 \pm 0.2$ mag.

\begin{figure}
\center
\includegraphics[width=0.48\textwidth]{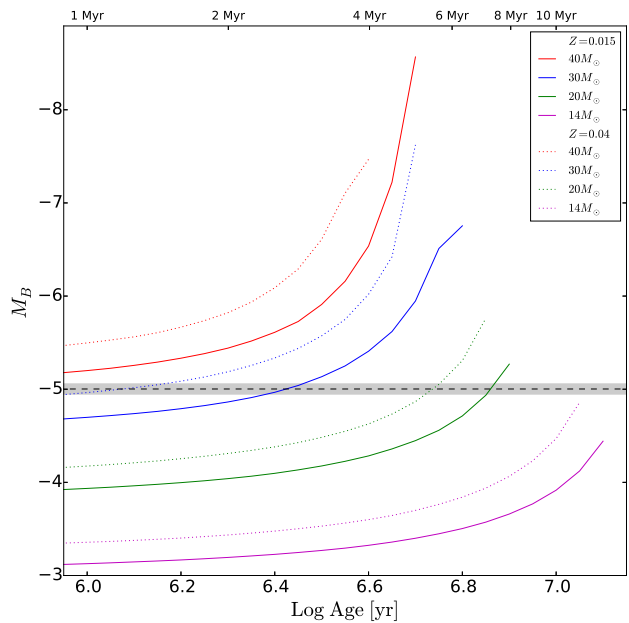}
\caption{Brightness evolution as a function of stellar age, for a range of initial masses (data from the Padova tracks, \citealt{Marigo2017}). Solid lines are for $Z = 0.015$ and dotted lines for $Z = 0.04$. The dashed black line represents the mean absolute magnitude of the brightest point-like source in the X-ray error circle for S2 (best candidate optical counterpart), with the shaded region being the 68\% uncertainty. Only stars with masses lower than $\approx$35 $M_{\odot}$ can be consistent with the optical counterpart of S2.}
\label{Mbageall.fig}
\end{figure}

The brightness evolution over time for stars with different initial masses is shown in Figure \ref{Mbageall.fig}, based on the Padova evolutionary tracks\footnote{Downloaded from http://stev.oapd.inaf.it/cgi-bin/cmd\_2.8.} \citep{Marigo2017}.
For approximately solar metallicity ($Z = 0.015$), as is now thought to be the case in the M\,51 disk \citep{Croxall2015,Bresolin2004}, stars more massive than $\approx$35 $M_{\odot}$ are {\it always} more luminous than $M_B \approx -5.0$ mag, even during their main-sequence phase; stars with masses $\approx$20--35 $M_{\odot}$ are initially fainter than $M_B \approx -5.0$ mag, and reach or exceed that luminosity during their giant or supergiant phases. As a result, we can put a conservative upper limit to the mass of the donor star in S2: $M_2 \la 35 M_{\odot}$.
An alternative calibration of the metal abundance in M\,51 \citep{Moustakas2010,Zaritsky1994} suggested a 2.5-times super-solar metallicity: in that case, the upper limit to the mass of the donor star would be $\approx$30 $M_{\odot}$ (Figure \ref{Mbageall.fig}).
Note that we have assumed that the donor star is the brightest star in the error circle, and that all the blue luminosity comes from the star itself. If some of the optical light comes from the accretion disk, and/or the true optical counterpart is one of the fainter objects, the true mass of the donor star would be even lower.  We will use the upper limit $M_2 \la 35 M_{\odot}$, together with the binary period and eclipse fraction, to constrain the mass of the accreting object.

\section{Discussion}
\label{discuss.sec}

In this work, we have done a detailed study of two bright, eclipsing X-ray sources in M\,51, a galaxy in which two other eclipsing X-ray sources were recently found \citep{Urquhart2016b}. There are two main reasons why we are looking for bright (ideally, super-Eddington) eclipsing XRBs. The first reason is that in principle, eclipsing systems offer us a good chance to constrain the binary period, mass ratio, and mass of the compact object with photometric observations alone, without the need for phase-resolved spectroscopy. The second reason is because we want to determine the effect of the viewing angle on the observational appearance of super-critical accreting stellar-mass objects---a problem similar to the unification scenario in AGN. It is generally very difficult to constrain the viewing angle and therefore also the geometry of emission; however, the presence of eclipses is strong evidence that we are viewing that particular system at high inclination. We can then compare the X-ray spectral properties of the sub-population of eclipsing sources with the properties of the non-eclipsing general population, to determine how the observed spectrum changes at higher inclination, and whether high-inclination sources have softer X-ray spectra, as predicted by MHD simulations \citep[{\it e.g.},][]{Narayan2017,Kawashima2012} and phenomenological models \citep{Sutton2013b}.

\subsection{Photometric identification of an accreting NS in S2}
\label{fraction.sec}

In Section \ref{xtimeS2.sec}, we identified the eclipse period ($P = 52.75 \pm 0.63$\,hr) of S2 as the orbital period of the binary system.
To proceed further, we need to make two simplifying assumptions. First of all, given the high luminosity of the system, we can plausibly assume that the donor star is filling its Roche lobe. Second, we also assume that the orbit has already been approximately circularized by tidal forces. The latter assumption is supported by the analytic solutions of \citet{Zahn1975} and \citet{Hurley2002}, and their application to high-mass XRBs \citep{Stoyanov2009}, which show that both BH and NS systems with a donor star mass $M_2 \ga 10 M_{\odot}$ and a binary period $P \la 10$ d will circularize in $\la$ 10$^6$ yrs, while systems with longer periods (usually including Be XRBs) may not circularize. Since S2 has a binary period $\approx$2.2 days, and a likely massive donor mass (as we discussed in Section \ref{hst.sec}), our assumption of tidal circularization is reasonable. Other NS ULXs with a supergiant donor, most notably NGC\,7793-P13, appear to have eccentric orbits \citep{Motch2014}. This may be due to their longer orbital period, which is $\approx$64 d for P13, if it corresponds to the observed X-ray/optical periodicity (although the identification of such periodicity as orbital or super-orbital remains in dispute: see, {\it e.g.}, \citealt{Furst2016} and \citealt{Hu2017}).
A detailed calculation of the circularization timescale across a more general parameter space for ULXs is well beyond the scope of this work.

Under the assumptions of a circularized, semi-detached system, the eclipse fraction is a function of the viewing angle $i$, of the binary separation $a$, and of the radius of the donor star $R_2$:
\begin{equation}
\Delta_{\rm ecl} = \frac{1}{\pi} \,
       \arccos\left[\, \frac{1}{\sin i} \,
           \sqrt{1-\left(\frac{R_2}{a}\right)^2} \right]
\end{equation}
\citep[{\it {e.g.}},][]{Chanan1976,Weisskopf2004,Porquet2005}. Using the standard Roche-lobe approximation of \citet{Eggleton1983} for $R_2/a$, we get:
%
\begin{equation}
\frac{\ln\left(1 + q^{1/3}\right)}{q^{2/3}} =
    \frac{0.49}{\sqrt{1-\left[\sin i \, \cos \left(\pi \Delta_{\rm ecl}\right)\right]^2}} - 0.60
\label{eq:qi}
\end{equation}
\citep{Porquet2005}. Eq.~(\ref{eq:qi}) can be solved numerically for $q(i;\Delta_{\rm ecl})$, and it also has a simple analytic solution for $i=90^{\circ}$.

\begin{figure}
\center
\includegraphics[width=0.48\textwidth]{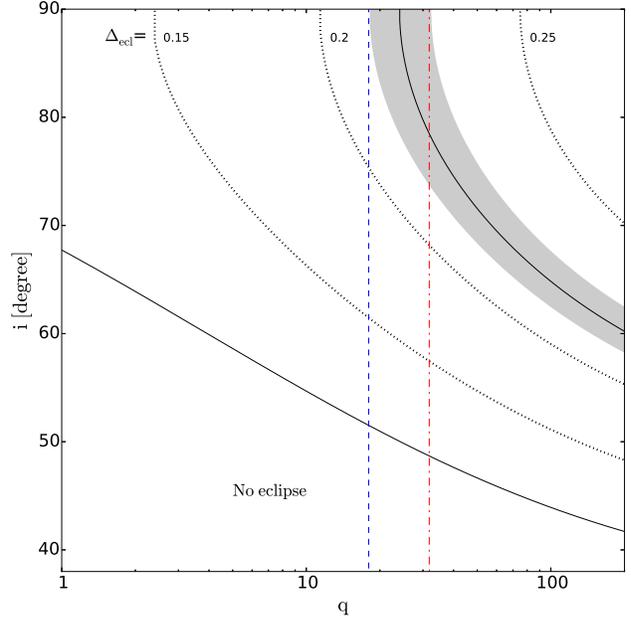}
\caption{Relation between viewing angle $i$ and mass ratio $q$, for various values of the primary eclipse fraction $\Delta_{\rm ecl}$ in semi-detached binary systems. The lower solid line represents the minimum angle for a grazing eclipse ($\Delta_{\rm ecl} \rightarrow 0$). Longer eclipses (top right sector of the plot) require higher values of $i$ and/or $q$. The grey band represents the observed eclipse fraction in S2 with its uncertainty range, $\Delta_{\rm ecl} = 0.222 \pm 0.008$. The dashed blue line and dashed-dotted red line represent the lower and upper limit of $q$ corresponding to this eclipse fraction, for $i=90^{\circ}$; for any other value of $i$, the acceptable range of $q$ is higher. Thus, $q\ga 18$ represents the minimum mass ratio in the system. Coupled with the knowledge that $M_2 < 35 M_{\odot}$, we conclude that the primary is in the NS mass range.}
\label{fraction.fig}
\end{figure}

For the observed eclipse fraction $\Delta_{\rm ecl} \approx 0.222 \pm 0.008$, we plot in Figure \ref{fraction.fig} the corresponding values of $q(i)$. The inclination angle is unknown, but in all cases, the value of $q(i=90^{\circ})$ provides the lower limit to the acceptable values of $q$. In our case, $q \ga 18$. We have already shown (Section \ref{hst.sec}) that the donor star has a mass $M_2 \la 35 M_{\odot}$.
We conclude that $M_1 \la 1.9 M_{\odot}$.
Hence, we have proved that the compact object in S2 is a NS rather than a BH.
If we also assume that the compact object cannot be lighter than $M_1 \approx 1.1~M_{\odot}$ (based on the observed range of NS and BH masses: \citealt{Kiziltan2013,Kreidberg2012}), we have $20  \la (M_2/M_{\odot}) \la 35$, and $18 \la q \la 32$.

\begin{figure}
\center
\includegraphics[width=0.48\textwidth]{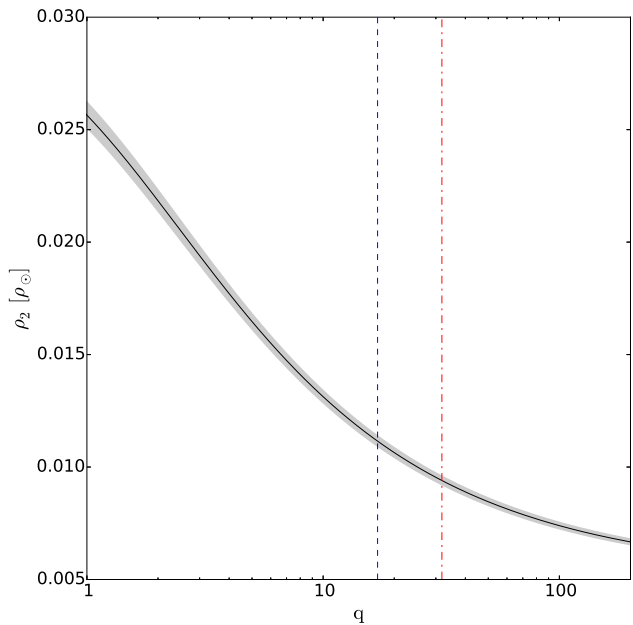}
\caption{Mean density of the secondary as a function of mass ratio, assuming a semi-detached system with binary period $P = 52.75$ hr. The dashed blue line and dashed-dotted red line represent the lower and upper limit of $q$ allowed in the system. The lower limit ($q > 18$) comes from the eclipse fraction (Figure~\ref{fraction.fig}); the upper limit ($q < 32$) comes from the observed mass of the secondary ($M_2 < 35 M_{\odot}$) and the plausible lowest limit for the mass of a NS ($M_1 \ga 1.1~M_{\odot}$). The corresponding acceptable range for the stellar density is $9.2 \times 10^{-3} \la \left(\rho_2/\rho_{\odot}\right) \la 11.2 \times 10^{-3}$.}
\label{rho2M2.fig}
\end{figure}

Using the above constraint on $q$ from the eclipse fraction, we can further refine our estimate of the secondary mass and evolutionary stage. The mean density $\rho_2$ of a Roche-lobe-filling secondary is related to the binary period and mass ratio by the relation:
\begin{equation}
\frac{\rho_2}{\rho_{\odot}} = 0.66 \, \left( \frac{P}{10\,{\rm {hr}}} \right)^{-2} \frac{\left[0.6\,q^{2/3}+\ln{(1+q^{1/3})}\right]^3}{q(1+q)}
\label{eq:rho}
\end{equation}
\citep{Eggleton1983}, where the mean solar density ${\rho}_\odot = 1.41$ g cm$^{-3}$. Taking into account that $18 \la q \la 32$, we obtain $9.2 \times 10^{-3} \la \left(\rho_2/\rho_{\odot}\right) \la 11.2 \times 10^{-3}$, that is $\rho_2 \approx 0.013$--0.016 g cm$^{-3}$ (Figure \ref{rho2M2.fig}). This is a range of values typical of giant and supergiant stars, and some O-type main sequence stars (although very massive O-type stars were already ruled out by the luminosity constraint).

\begin{figure*}
\center
\includegraphics[width=0.48\textwidth]{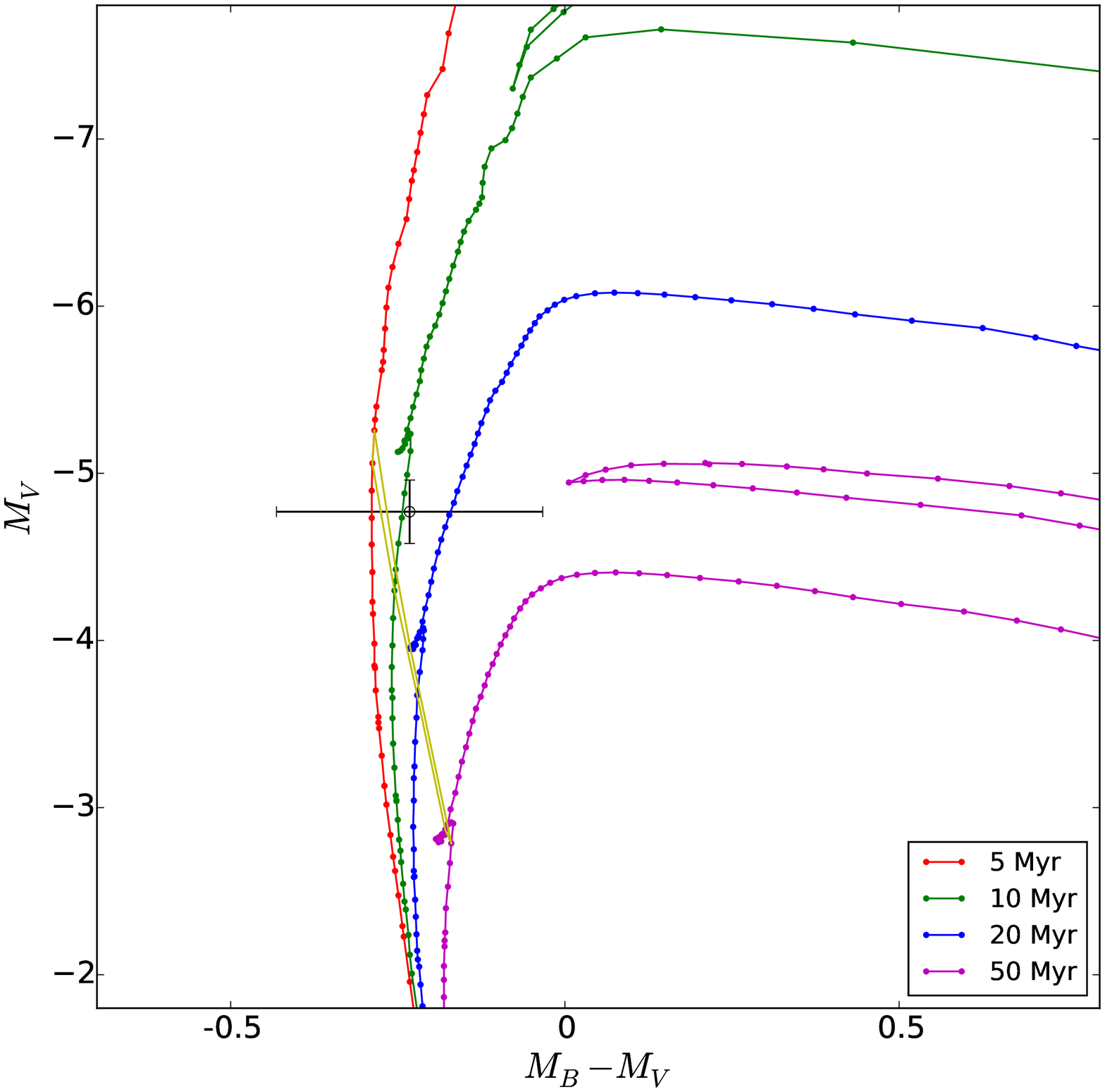}
\includegraphics[width=0.48\textwidth]{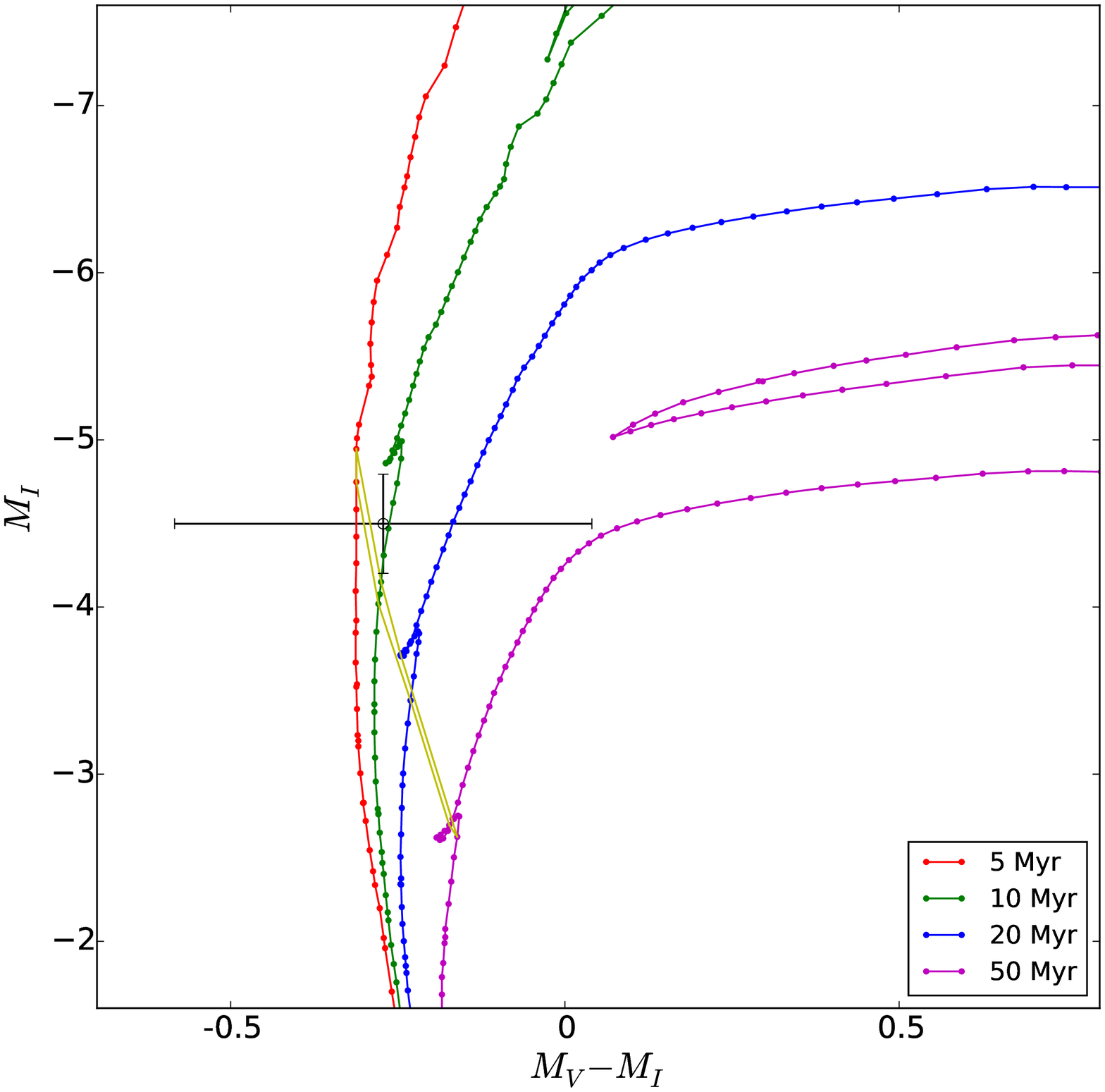}
\caption{Left panel: absolute brightness $M_V$ versus colour $M_B - M_V$ of the observed counterpart of S2, compared with a series of theoretical isochrones at $Z = 0.015$ (from the Padova database), and with the region of the isochrones (yellow band) that satisfies the mean density condition (Figure~\ref{rho2M2.fig}). Right panel: as in the left panel, for $M_I$ versus $M_V - M_I$.}
\label{ivis.fig}
\end{figure*}

\begin{figure}
\center
\includegraphics[width=0.48\textwidth]{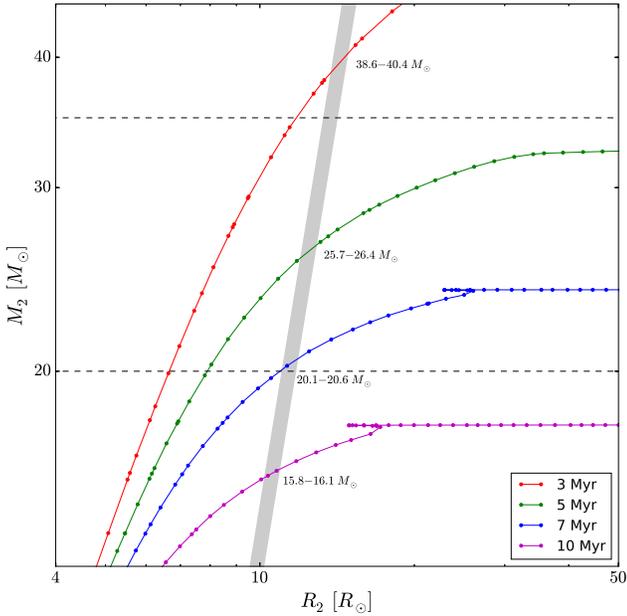}
\caption{Theoretical stellar mass $M_2$ versus radius $R_2$ along a set of isochrones ($Z = 0.015$), compared with the allowed density range in S2 (grey band: $\rho \approx 0.013$--$0.016$ g cm$^{-3}$).
The additional conditions that $q \ga 18$, and that the NS mass $M_1 \ga 1.1~M_{\odot}$, further constrain the donor star mass to $20 \la (M_2/M_{\odot}) \la 35$ (horizontal dashed lines).}
\label{M2R2.fig}
\end{figure}

We can now put together the constraints on the donor star derived from the eclipse properties and the geometry of the system ($20~M_{\odot} \la M_2 \la 35~M_{\odot}$, $\rho_2 \approx 0.013$--0.016 g cm$^{-3}$), with the inferred optical brightness of the counterpart ($M_V \approx -5.0 \pm 0.1$ mag, $M_V \approx -4.8 \pm 0.2$ mag). Using again the Padova stellar evolution tracks \citep{Bressan2012, Marigo2017}, we want to determine whether there are stars that satisfy all those properties at the same time.
In the colour-magnitude diagrams shown in Figure \ref{ivis.fig}, we have plotted the intrinsic brightness of the optical counterpart, together with a representative sample of Padova isochrones from 5 to 50 Myr.
Only stars located along the yellow band have a mean density consistent with the range derived for S2.
Both diagrams clearly show that the inferred brightness of the star (with its error range) overlaps the allowed density band only for stellar ages $\la$ 10 Myr and a narrow range of stellar masses along each isochrone. To further investigate this allowed range of masses, we have plotted (Figure \ref{M2R2.fig}) the same theoretical isochrones in a mass versus radius plot, with a grey band representing the permitted range of densities for the S2 donor. We find that only isochrones younger than 7 Myr allow solutions with $M_2 > 20 M_{\odot}$ and the correct density.
%
In summary, we find that at $Z = 0.015$, there are physical solutions consistent with all constraints in the narrow mass range $20  \la (M_2/M_{\odot}) \la 35$, with a radius $11 \la (R_2/R_{\odot}) \la 13$ and an age of $\approx$4--7 Myr.

\begin{figure}
\center
\includegraphics[width=0.48\textwidth]{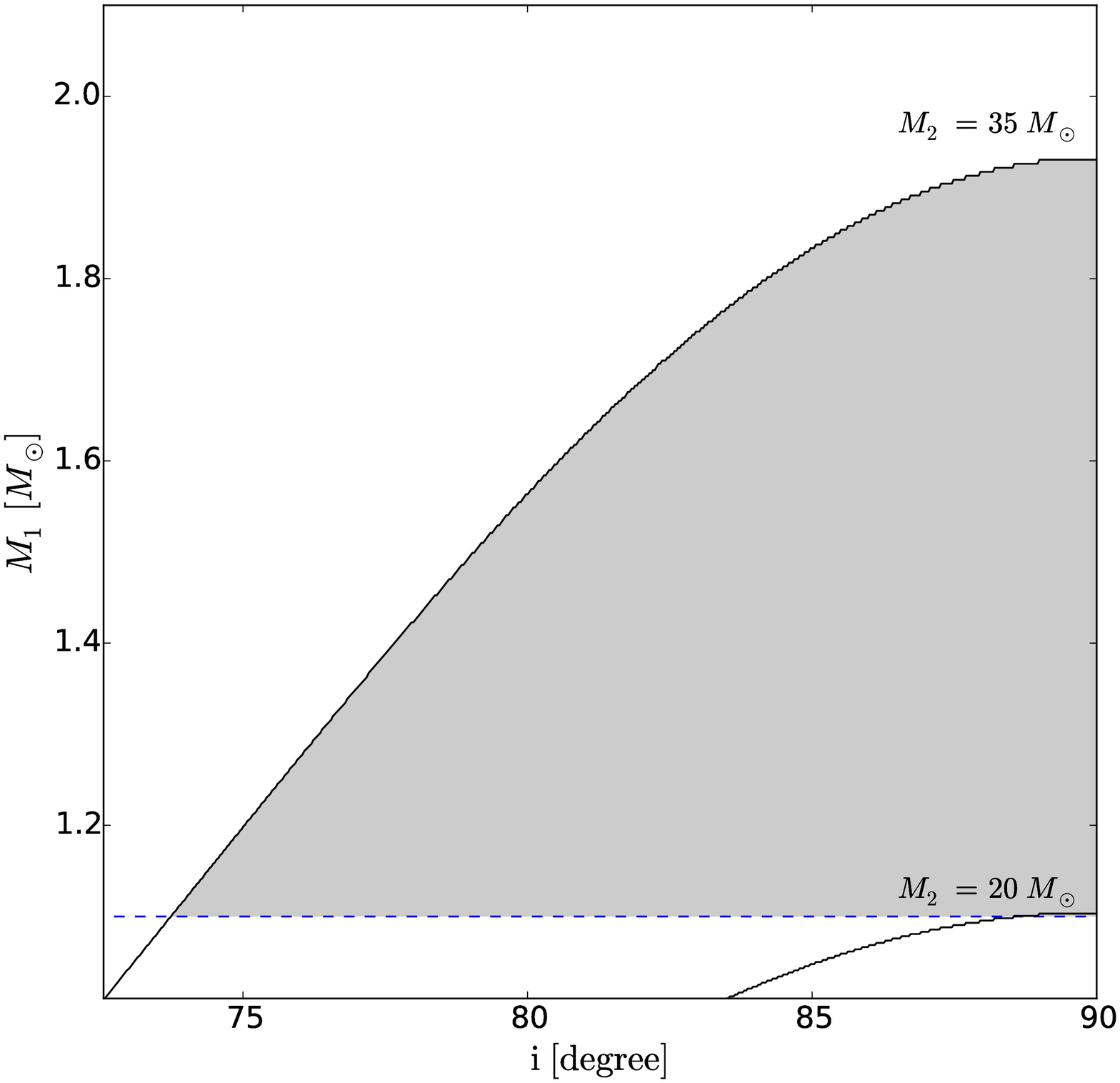}
\caption{Permitted mass range for the primary (grey shaded region) as a function of viewing angle $i$. This range is determined by the permitted range of $q(i)$, combined with the photometric upper limit to the mass of the donor star ($M_2 < 35 M_{\odot}$) and the empirical lower limit to the mass of a NS ($M_1 > 1.1 M_{\odot}$, dashed blue line).}
\label{M1i.fig}
\end{figure}

As an alternative way of presenting our main results, we also plotted the permitted range of $M_1$ as a function of inclination angle, for the acceptable range of donor-star masses (Figure \ref{M1i.fig}). Only high values of $i \ga 73^{\circ}$ are consistent with all constraints; more face-on viewing angles are inconsistent with the long eclipse fraction observed in S2. A "canonical" 1.4-$M_{\odot}$ NS is consistent with the observed eclipse fraction for all angles $i > 78^{\circ}$.
A compact object mass $M_1 \ga 2M_{\odot}$ is always inconsistent with the combined eclipse properties and secondary star mass range (because of the upper limit $M_2 \la 35 M_{\odot}$).

Finally, as we anticipated in Section 6, for the sake of completeness we also considered the (unlikely) possibility that the true counterpart is the bright blue star at the periphery (just to the west) of the error circle (Figure \ref{S2ds9.fig}). The average apparent brightness in each filter is $m_{435} = 24.01$ mag, $m_{555} = 24.05$ mag, and $m_{814} = 23.97$ mag, corresponding to an intrinsic (dereddened) absolute brightness $M_B \approx -5.6 \pm 0.1$ mag, $M_V \approx -5.5 \pm 0.1$ mag, and $M_I \approx -5.6 \pm 0.1$ mag. From the Padova evolutionary tracks, we derive an upper mass limit of $\approx$53 $M_{\odot}$ ($Z =$ 0.015) or $\approx$45 $M_{\odot}$ ($Z =$ 0.04). All our arguments based on the eclipse fraction remain unchanged: the lowest limit on the mass ratio is still $q> 18$ regardless of the assumed donor star. Therefore, the mass of the compact object must be $\lesssim (53/18)\,M_{\odot} \approx 2.9~M_{\odot}$. This is still substantially lower than the mass range observed in stellar-mass BHs. Thus, even for that alternative choice of optical counterpart, we stand by our identification of the compact object as a NS.









%




\subsection{Super-Eddington accretion in NSs}

The highlight of our study of M\,51 S2 is that we were able to identify the NS nature of a bright X-ray binary (with super-Eddington outbursts up to $\approx$0.5--1 $\times 10^{39}$ erg s$^{-1}$ in 2005 and 2012) simply from its X-ray lightcurve. This was done with a precise determination of the binary period, combined with constraints on the donor star mass from its optical luminosity, without the need for phase-resolved optical spectroscopy and without the detection of X-ray pulsations. Now we need to place this system in the context of other luminous types of NS high-mass XRBs.


In our Local Group, the most luminous NS outbursts are seen in transient Be XRBs. For example, the transient X-ray pulsar SMC X-3 also reached an X-ray luminosity of $\approx$10$^{39}$ erg s$^{-1}$ during the 2016--2017 outburst \citep{Weng2017,Townsend2017,Tsygankov2017}; the mass donor is a Be star with a mass of $\approx$10 $M_{\odot}$ \citep{McBride2008}. Similarly, the Be XRB A0538$-$66 in the Large Magellanic Cloud has reached an X-ray luminosity $\approx$10$^{39}$ erg s$^{-1}$ in outburst \citep{White1978,Corbet1997}. However, this class of systems is characterized by long orbital periods, $>$10d (for example, 44.9 d for SMC X-3, 16.7 d for A0538$-$66). Moreover, the peak outburst luminosity is three or four orders of magnitudes higher than the inter-outburst luminosity, contrary to what is seen in S2.

We then compare S2 to Local Group NSs with a supergiant donor. Within this class, outbursts are seen in supergiant fast X-ray transients \citep{Negueruela2006,Sidoli2008,Romano2014}. However, these wind-fed systems only reach X-ray luminosities $\la$10$^{37}$ erg s$^{-1}$ in outburst, and $\sim$10$^{33}$ erg s$^{-1}$ in between outbursts.

Thus, the closest Local Group analogs to S2 must be found among persistently luminous X-ray pulsars such as SMC X-1, LMC X-4, and Cen X-3, with a Roche-lobe-filling supergiant donor.
In those systems, the stellar masses $M_2 \approx$ 15--25 $M_{\odot}$, the periods are a few days, and the viewing angles are $\approx$70$^{\circ}$--80$^{\circ}$, which leads to observed eclipse fractions of $\approx$0.15--0.20 of the orbital periods \citep{Rawls2011}. Such systems reach X-ray luminosities between $\approx$1--4 $\times 10^{38}$ erg s$^{-1}$ in their high states \citep{Farinelli2016,Marcu2015,Hickox2004,Vrtilek2001}, at the Eddington limit for a NS. However, none of the Local Group systems in this class has shown super-Eddington outbursts. Thus, M\,51 S2 represents a new type of behaviour for luminous NS XRBs. It  is another indication of how much we still do not know about super-Eddington in NSs. Even more luminous examples of super-Eddington NSs were recently found in other galaxies, such as M\,82 X-2 \citep{Bachetti2014}, NGC\,7793 P13 \citep{Israel2017b,Furst2016}, and NGC\,5907 X-1 \citep{Israel2017a,Furst2017}, also likely to be fed by Roche-lobe-filling supergiant donors.

One important property of magnetized NS XRBs, which distinguishes them from BH systems, is the switch from accretor to propeller regime \citep{Cui1997, Campana2014,Christodoulou2017,Tsygankov2017}. As the accretion rate and luminosity decrease, the magnetospheric radius increases, until it becomes larger than the corotation radius in the disk. At that threshold, the magnetosphere creates a centrifugal barrier that halts accretion and switches off the X-ray source. The threshold luminosity constrains the magnetic field as a function of spin period. Unfortunately, the spin period is unknown for S2 (no X-ray pulsations detected); therefore, we cannot constrain the magnetic field. The lowest X-ray luminosity at which the source was still detected was $\approx$10$^{37}$ erg s$^{-1}$, at the end of the 2012 outburst (Table 2).

The X-ray spectra of near-Eddington, persistent X-ray pulsars in the Local Group are usually interpreted in terms of thermal and bulk Comptonization of bremsstrahlung and cyclotron emission \citep{Farinelli2016,Becker2007}, with most of the hard X-ray photons emerging from the lateral wall of the accretion column, and a soft excess below 1 keV caused by the reprocessing (down-scattering) of a fraction of the hard X-ray photons near the surface of the inner disk \citep{Hickox2004}. The resulting X-ray spectrum has a hard slope (photon index $\Gamma \approx 1$) in the $\approx$1--7 keV range. This is also consistent with the spectrum of S2, with $\Gamma \approx 1.2$ when the soft excess is properly taken into account ({\it i.e.}, in the {\it XMM-Newton} spectrum, with its broader energy coverage; see Table 3). A hard spectral slope below $\approx$7 keV also characterizes the few NS ULXs identified so far from their pulsations: for example, in the {\it Chandra} energy band, P13 in NGC\,7793 can be fitted with a photon index $\Gamma \approx 1.1$--1.2 and a soft excess at $kT_{\rm bb} \approx 0.2$ keV \citep{Motch2014,Pintore2017}, or with a double thermal model plus power-law \citep{Walton2018}. On the other hand, the softer continuum component and the thermal plasma emission are much stronger in S2 than in the NS ULXs; we do not have enough empirical evidence to determine whether this is due to the higher viewing angle and/or denser wind in S2.

In summary, we suggest that we are finally starting to detect a continuum range of NS luminosities, between $L_{\rm X} \approx$ a few  $10^{37}$ erg s$^{-1}$ to $L_{\rm X} \approx 10^{40}$ erg s$^{-1}$, as predicted 
\citep{Mushtukov2015}, likely characterized by the same basic physical structure (Roche lobe overflow, truncated inner disk, magnetized accretion columns), without any significant observable transition or threshold at the classical Eddington limit for a NS. In this context, M\,51 S2 represents a source in the middle of this range (See Table \ref{XlumS2.tab}). Searching for X-ray eclipses and for X-ray pulsations are two complementary techniques for finding more sources in this class. Determining the ratio between soft thermal component(s) and hard power-law in those sources will provide important clues for the accretion geometry and physical properties of super-Eddington NSs.






\subsection{Outflows in super-Eddington sources}
\label{super.sec}

The four eclipsing sources studied in this work and in \cite{Urquhart2016b} give us important insights on the observational appearance of super-critical compact objects seen at high inclination.  Although all four sources have comparable luminosity (within a factor of 2) and inclination, their spectral properties are significantly different. One is a ULS (S1), one has a ``soft ultraluminous" spectrum (ULX-1), one is well fitted by a slim disk (ULX-2) and one (S2) has a hard photon index similar to the ``hard ultraluminous" regime (using the classification of \citealt{Gladstone2009} and \citealt{Sutton2013b}), which has also been interpreted as a characteristic feature of super-Eddington NSs \citep{Pintore2017}. 
In other words, all four main phenomenological spectral types suggested for super-critical accreting sources are represented in this small sample. Clearly, this shows that the accretion rate and the nature of the compact object must have as much an effect on the X-ray spectral appearance as the viewing angle; we cannot simply assume that higher inclination means softer spectra. In particular, it is remarkable that we see hard X-ray emission in S2, considering that its inclination angle is $\ga 80^{\circ}$.

It is well known that near-Eddington NSs such as SMC X-1, LMC X-4, and Cen X-3 show hard X-ray spectra even for viewing angles $\approx$70$^{\circ}$--80$^{\circ}$ \citep{Hickox2004, Rawls2011}. M\,51 S2 is analogous to them or even more luminous, depending on the inclination angle. Since it exceeds the Eddington limit in outburst, we would naively expect dense, optically-thick outflows launched from the inner part of the disk \citep{Shakura1973,Poutanen2007,King2009}, shrouding both the surface of the NS and the base of the accretion column from our view.
MHD simulations of super-critical accretion disks in BHs ({\it e.g.}, \citealt{Ohsuga2011,Kawashima2012,Jiang2014,Narayan2017,Ogawa2017}) and in non-magnetic NSs \citep{Ohsuga2007,Takahashi2017b} generally predict the formation of a geometrically thick disk outflow and a lower-density polar funnel (around the spin axis of the compact object), with the hard X-ray photons visible only to observers looking into the funnel. This is the basis for the most common interpretation of ULXs \citep{King2001,King2009,Feng2011}, with harder spectra being associated with low-inclination sources, and softer spectra (with higher short-term variability) associated with high-inclination sources \citep{Sutton2013b,Middleton2015a}. In fact, the lack of eclipsing ULXs was used as an argument in support of the geometric beaming scenario \citep{Middleton2016}, in the sense that high-inclination sources would be much less likely to appear ultraluminous to us, for a given accretion rate.

However, for super-critical accretion onto magnetized NSs \citep{Kawashima2016,Takahashi2017a} the situation is more complicated than for non-magnetic accretors, for at least two reasons. Firstly, the inner disk is expected to be truncated at the magnetospheric radius, far from the surface of the NS. If the truncation radius is outside the spherization radius \citep{Shakura1973,Poutanen2007}, that is for relatively strong magnetic fields and/or relatively moderate accretion rates, the disk is still geometrically thin \citep{Chashkina2017}, with $H/R \la 0.1$, and may not occult the magnetized accretion columns. In that situation, most of the radiative energy release occurs in the polar accretion columns, and the radiatively-driven disk outflow is less powerful than in non-magnetic systems, for the same mass transfer rate at the outer disk boundary. Secondly, the viewing angle derived from the eclipse duration and mass ratio refers to the inclination of the orbital plane, but the magnetic dipole axis (and therefore the axis of the accretion columns) may be misaligned and not perpendicular to the orbital plane. This could give us a more direct view of the polar regions. It would also cause super-orbital cycles in the X-ray luminosity due to disk precession, similar to those proposed for example for SMC X-1 \citep{Priedhorsky1987}, LMC X-4 \citep{Heemskerk1989}, NGC\,7793 P13 \citep{Motch2014}. It is worth noting that the binary parameters of S2, namely $q \approx 20$ and $a = (1.5 \pm 0.1) \times 10^{12}$ cm $\approx 8 \times 10^6~r_{\rm g}$, correspond precisely to the sub-population of NS high-mass XRBs with observed super-orbital periodicities, interpreted as disk precession (\citealt{Ogilvie2001}, their Figure 7).
In summary, the relative contribution of direct (hard) X-ray emission and reprocessed (soft) component in the spectra of super-critical NSs is not a simple function of accretion rate and binary viewing angle, but depends also on magnetic field and angle of misalignment, and can have long-term variations due to precession \citep{Weng2018}.

By contrast, M\,51 S1 behaves like a standard ULS \citep{Urquhart2016a}. The lack of hard X-ray photons is consistent with complete reprocessing of the direct emission in an optically thick outflow at high inclination \citep{King2003,Poutanen2007,Soria2016}. The characteristic temperature ($\approx$85--140 eV) and radius ($\approx$4,000--20,000 km) are typical for this class of sources, and so is the $r_{\rm bb} \propto T_{\rm bb}^{-2}$ relation between different observations (Figure \ref{S1model.fig}). The high degree of short-term variability in its X-ray light-curves (Figure \ref{eachS1.fig}) and the detection of spectral residuals around 1 keV (Figure \ref{specS1.fig}) are also typical of this class of sources \citep{Urquhart2016a}. They can be naturally interpreted \citep{Middleton2014,Middleton2015} as evidence of clumpy disk outflows, as predicted by MHD simulations \citep{Takeuchi2013,Takeuchi2014,Jiang2014}. The sharp, isolated dip (total duration of $\approx$8 ks) seen in {\it Chandra} Obsid 13814 is quite puzzling. A distribution of cold clouds in a clumpy wind cannot produce a single isolated dip of such duration. More likely, by analogy with dips seen in Galactic sources, the dip in S1 is due to the obscuration of the emitting photosphere (characteristic radius $\sim$10$^9$ cm) by a thickened lump at the outer edge of the disk (characteristic radius $\ga$10$^{11}$ cm), caused by the interaction between the accretion stream and the outer disk \citep{White1982,Parmar1986,Balucinska-Church1999}. Similar dips have been found in the M\,81 ULS \citep{Swartz2002,Liu2008}. In the Local Group, this model has successfully explained the complex dipping behaviour of EXO 0748-676 \citep{Parmar1986,Cottam2001} and of several other NS XRBs ({\it e.g.}, \citealt{Oosterbroek2001,Smale2001,Diaz-Trigo2006,Gambino2017}).

\section{Conclusions}
\label{conclusion.sec}

We are carrying out systematic searches and studies of eclipsing super-critical sources in nearby galaxies. In particularly favourable circumstances, eclipse patterns can reveal the binary period and constrain the mass of the compact object, based only on X-ray and optical photometric observations, as we have illustrated in this paper.
X-ray hardness and spectral differences between the eclipsing and non-eclipsing ULX/ULS populations will test the empirical scenario \citep{Sutton2013b,Middleton2015a,Middleton2016} that for a given super-Eddington accretion rate, sources appear harder and more luminous at low inclination (geometric beaming of the X-ray emission inside a polar funnel) and softer and fainter at high inclination (down-scattering in a thick disk outflow). More generally, ULXs/ULSs with a known inclination angle provide important constraints to MHD simulations of super-critical inflows and outflows.

In this paper, we have focused on the face-on spiral galaxy M\,51, because of its extensive X-ray coverage over the years. Using {\it Chandra} and {\it XMM-Newton} archival data, we discovered eclipses in two more X-ray sources, which we have labelled S1 and S2. Together with the two eclipsing ULXs discovered by \citet{Urquhart2016b} (labelled ULX-1 and ULX-2), we now know four eclipsing sources, all of them with X-ray luminosities $\sim 10^{39}$ erg s$^{-1}$, in the same quarter of a single galaxy (covered by a single ACIS-S chip). This seems suspiciously unlikely (as also noted by \citealt{Urquhart2016b}), considering how rare such findings are. We did verify that the light-curves of other nearby sources in the same observations do not have eclipses or any other glitches; the four sources themselves enter and exit their eclipses at different times, which confirms that such events are not instrumental effects. We suggest that our luck in finding eclipsing binaries in M\,51 is mostly due to a sequence of long X-ray observations repeated every few days (on a similar timescale as the characteristic binary periods), in a galaxy rich with luminous X-ray sources. If the galaxy had been covered with shorter observations, scattered over a longer period of time, we might not have recognized flux changes and source disappearances as a repeated pattern of eclipses.

In one of the four eclipsing M\,51 sources (S2, discussed in this paper), we determined the binary period ($P = 52.75 \pm 0.63$ hr) and eclipse fraction ($\Delta_{\rm ecl} = 0.222 \pm 0.008$) accurately. This gave us a lower limit on the mass ratio $M_2/M_1$. We then used optical photometry to determine (from stellar evolution tracks) an upper limit to the mass of the donor star; it is an upper limit because the optical flux of the counterpart includes unknown contributions from the accretion disk. By combining the two constraints, we obtained an upper limit to the mass of the compact object, identified as a NS ($M_1 < 1.9~M_{\odot}$) with a high-mass donor ($20~M_{\odot} \la M_2 \la 35~M_{\odot}$), reaching an X-ray luminosity $L_{\rm X} \approx 10^{39}$ erg s$^{-1}$ in outburst.

Beyond the specific characterization of one luminous XRB, the identification of a NS in that system has a more general significance. One of the main unsolved problems in the study of XRB populations is the relative contribution of NSs and stellar-mass BHs to sources at $L_{\rm X} \ga 10^{39}$ erg s$^{-1}$. It was previously assumed that the high-luminosity end of the population consisted only of BHs; however, the recent identification of at least three NS ULXs \citep{Bachetti2014,Furst2016,Furst2017,Israel2017a,Israel2017b} has changed that view. Moreover, it is well known \citep{Swartz2004,Swartz2011,Roberts2011,Mineo2012} that there is no break or downturn in the luminosity function at the Eddington limit of a NS. This suggests that accreting NSs continue to power {\it the majority} of high-mass XRBs well above $10^{39}$ erg s$^{-1}$. Unfortunately, direct evidence that an accreting source is a NS is very difficult to obtain from phase-resolved spectroscopy of its optical counterpart, and only a small fraction of them may show X-ray pulsations. Here we have shown that there is a complementary way to make progress on this issue, based on X-ray and optical photometry.

Another result of our M\,51 work is that high-inclination, super-critical sources can have all types of X-ray spectra: disk-like, hard, soft, and supersoft. The difference between disk-like spectrum (ULX-2 in \citealt{Urquhart2016b}) and supersoft spectrum (S1 in this paper and in \citealt{Urquhart2016a}) can be explained by a much higher accretion rate in the ULS, and therefore a much denser and optically thick disk outflow. Instead, the hard spectrum of the eclipsing NS S2 is consistent with the different behaviour expected for super-critical BHs and NSs. The spectral classification scheme of \cite{Sutton2013b} (based on denser outflows and softer spectra at higher inclination) is valid only for BH and weakly magnetized NS ULXs (that is, those systems where the accretion disk reaches its spherization radius and gives rise to strong outflows). For NS ULXs with high enough dipole B field, instead, the magnetospheric radius is outside the spherization radius, and the truncated disk may not be able to launch strong radiatively-driven outflows. Moreover, the magnetic axis may be misaligned with the spin axis of the binary orbit, and may enable the view of the hottest part of the accretion column (an order of magnitude hotter than the peak colour temperature in an accretion disk). For both reasons, a hard X-ray spectrum may be seen also from NS ULXs that have a high orbital inclination, as proposed by \cite{Pintore2017}.

One of the unsolved issues in super-Eddington accretion is that MHD simulations consistently predict \citep{Kawashima2012,Narayan2017} a higher apparent luminosity (by an order of magnitude) for sources with harder X-ray spectra, because of geometric beaming in the polar funnel as opposed to down-scattering in the disk outflows. On the other hand, observations show \citep{Sutton2013b} that the luminosity distribution of ULXs with hard spectra largely overlaps the luminosity distribution of sources in the soft ultraluminous regime: there is no significant trend between different sources in the luminosity versus hardness plane. We suggest that this contradiction can be explained if the ULX population with hard spectra includes a substantial contribution from strongly magnetized NSs, especially at luminosities $\la$ a few $10^{39}$ erg s$^{-1}$, while the soft ultraluminous regime includes mostly BHs at much higher accretion rates but also with more down-scattered emission.

  \section*{Acknowledgments}
We especially thank the anonymous referee for his/her thorough report and helpful comments
and suggestions that have significantly improved the paper.
We thank Rosanne Di Stefano, James Miller-Jones, Christian Motch, Manfred Pakull, Yanli Qiu, Kinwah Wu for useful suggestions.
This work has made use of data obtained from the Chandra Data Archive,
and software provided by the Chandra X-ray Center (CXC) in the application packages CIAO.
We also used observations obtained with {\it XMM-Newton},
an ESA science mission with instruments and contributions directly funded by ESA Member States and NASA.
Part of our study relied on observations made with the NASA/ESA {\it Hubble Space Telescope}, obtained from the Data Archive at the Space Telescope Science Institute, which is operated by the Association of Universities for Research in Astronomy, Inc., under NASA contract NAS 5-26555.
We acknowledge use of the SIMBAD database and the VizieR catalogue access tool,
operated at CDS, Strasbourg, France, and of Astropy,
a community-developed core Python package for Astronomy (Astropy Collaboration, 2013).
SW is grateful for support from the National Science Foundation of China
(NSFC, Nos. 11273028, 11333004, and 11603035), and the National Astronomical Observatories,
Chinese Academy of Sciences, under the Young Researcher Grant.
RS acknowledges support from a Curtin University Senior Research Fellowship; he is also grateful for support, discussions and hospitality at the Strasbourg Observatory during part of this work.
RU acknowledges that this research is supported by an Australian Government Research Training Program (RTP) Scholarship. The International Centre for Radio Astronomy Research is a joint venture between Curtin University and the University of Western Australia, funded by the state government of Western Australia and the joint venture partners.


\end{document}